\documentclass[%
reprint,
superscriptaddress,
amsmath,amssymb,
aps
]{revtex4-2}
\usepackage{times}
\usepackage[a4paper, total={18cm, 25cm}]{geometry}
\usepackage{microtype}
\usepackage{graphicx}
\usepackage{wrapfig}
\usepackage{multirow}
\usepackage[table,svgnames]{xcolor}
\usepackage{bm}
\usepackage{soul} 
\usepackage[separate-uncertainty=true,number-mode=math,unit-mode=text]{siunitx}
\usepackage{tabularx}
\usepackage{physics}
\usepackage{hyperref}
\hypersetup{pdfborder={0 0 0}} 
\hypersetup{
	colorlinks=true, 
	linkcolor={red!65!black},
	citecolor={blue!75!black},
	urlcolor={blue!50!black}
}
\usepackage{titlesec}
\titleformat{\section}{\normalfont\bfseries}{}{1em}{}
\titlespacing*{\section}{0pt}{15pt}{5pt}

\begin{document}

\title{\large \textbf{Imaging scattering resonances in low-energy inelastic ND$_3$-H$_2$ collisions}}

\author{Stach E.\,J. Kuijpers}
\affiliation{Radboud University, Institute for Molecules and Materials, Heyendaalseweg 135, 6525 AJ Nijmegen, the Netherlands}
\author{David H. Parker}
\affiliation{Radboud University, Institute for Molecules and Materials, Heyendaalseweg 135, 6525 AJ Nijmegen, the Netherlands}
\author{J\'er\^ome Loreau}
\affiliation{KU Leuven, Department of Chemistry, Celestijnenlaan 200F, bus 2404, 3001 Leuven, Belgium}
\author{Ad van der Avoird}
\affiliation{Radboud University, Institute for Molecules and Materials, Heyendaalseweg 135, 6525 AJ Nijmegen, the Netherlands}
\author{Sebastiaan Y.\,T. van de Meerakker}
\thanks{To whom correspondence should be addressed.\\ E-mail: \href{mailto:basvdm@science.ru.nl}{basvdm@science.ru.nl}}
\affiliation{Radboud University, Institute for Molecules and Materials, Heyendaalseweg 135, 6525 AJ Nijmegen, the Netherlands}

\begin{abstract}
	A scattering resonance is one of the most striking quantum effects in low-temperature molecular collisions. Predicted decades ago theoretically, they have only been resolved experimentally for systems involving at most four atoms. Extension to more complex systems is essential to probe the true quantum nature of chemically more relevant processes, but is thus far hampered by major obstacles. Here, we present a joint experimental and theoretical study of scattering resonances in state-to-state inelastic collisions for the six-atom ND$_3$-H$_2$/HD systems across the collision energy range \num{0.5}-\SI{25}{\per\cm}, bringing this type of experiment into the realm of polyatomic symmetric top molecules. Strong resonances are resolved in the integral cross sections, whereas differential cross sections are measured with high resolution using a laser ionization scheme involving VUV light. The experimental data could only be reproduced using theoretical predictions based on a potential energy surface at the CCSD(T) level of theory with corrections at the CCSDT(Q) level. 
\end{abstract}

\maketitle

The recent observation of scattering resonances in low-energy molecular collisions arguably has been one of the most exciting breakthroughs in modern atomic and molecular physics. Scattering resonances can only be understood from the framework of quantum mechanics, and they are testimony to the wavelike quantum nature of matter. In a simplified picture, these resonances may be regarded as the orbiting of the two collision partners around each other that only occurs at low energies, typically in the  $10^{-3}$ to \SI{10}{K} range, as here the de Broglie wavelength is sufficiently large for quantum effects to dominate. Even though classically forbidden, they can occur either by tunneling through a potential barrier (an orbiting or shape resonance), or by the transient excitation of the molecule to a state of higher energy (a Feshbach resonance). 

Scattering resonances manifest themselves by a dramatic increase in the integral cross sections (ICSs) at the resonance energies, accompanied by rapidly changing differential cross sections (DCSs). They are extremely sensitive to the details of the potential energy surface (PES) describing the interaction between the colliding molecules; a change in PES of typically less than a percent can already induce major differences in the resonance structures. Not surprisingly, there has been a long term quest to observe (and fully resolve!) these scattering resonances experimentally, ideally in collision experiments retrieving both state-to-state integral and differential cross sections~\cite{Chandler:JCP132:110901}. Using a crossed-beam approach, resonances were first observed in 1972 for H atoms scattering with Hg atoms~\cite{Schutte:PRL29:979} and later other collision partners~\cite{Toennies:JCP71:614}. Total integral cross sections were recorded by scanning the collision energy with mechanical velocity selectors while monitoring the hydrogen beam depletion. Over the next few decades, resonance states were mostly probed through IR-spectroscopy of weakly bound dimers~\cite{Lovejoy1990}. In 1993, evidence of a resonance in reactive scattering was observed for F + H$_2$ $\rightarrow$ HF + H collisions~\cite{Wang2018}. By applying merged-beam approaches, resonances in ICSs were observed since 2012 in Penning ionization reactions,  at collision energies down to \SI{0.01}{\per\cm} ~\cite{Henson:Science338:234, Lavert-Ofir:NatChem6:332, Jankunas:JCP142:164305, Shagam2015, Klein:NatPhys13:35, Bibelnik2019, Margulis2022}. 

Resonances in state-to-state inelastic ICSs were first recorded in 2012 for CO-H$_2$ at collision energies down to \SI{4}{\per\cm}~\cite{Chefdeville:PRL109:023201, Chefdeville:AJL799:L9}, using crossed molecular beams with a small and variable intersection angle in combination with state-selective Resonance Enhanced Multi Photon Ionization (REMPI) detection. This approach has since been used to study scattering resonances in state-to-state ICSs for O$_2$-H$_2$, CO-He, C-He/H$_2$/D$_2$ and D$_2$O-H$_2$ collisions~\cite{Chefdeville:Science341:06092013, Bergeat:NatChem7:349, Klos2018, Bergeat:NatChem10:519, Bergeat2019, Bergeat2020,Bergeat2022}, although pronounced resonance structures could not always be resolved in these studies. The highest experimental resolution thus far is obtained using the Stark deceleration and velocity map imaging (VMI) techniques, that in addition to measurements of ICSs enabled the probing of the energy dependence of DCSs in the resonance region for the NO-He and NO-H$_2$ systems at collision energies down to \SI{0.2}{\per\cm}~\cite{Vogels:Science350:787,Vogels:NatChem10:435,Jongh2020,Shuai2020,Jongh2022}. The unprecedented high resolution obtained in these experiments required quantum chemistry calculations beyond the CCSD(T) gold standard level of theory. For the benchmark NO-He system, only theory with corrections at the CCSDT(Q) level could reproduce the experimental obervations, epitomizing the extreme sensitivity of resonance features to details of the PES \cite{Jongh2022}. VMI has recently also been applied to record ICSs and DCSs for elastic scattering of He* with D$_2$ down to \SI{0.7}{\per\cm}~\cite{Paliwal2021}, as well as to probe resonance effects in inelastic collisions between Zeeman decelerated C atoms and H$_2$ molecules at energies down to \SI{0.5}{\per\cm}~\cite{Plomp2024}.

Despite these breakthroughs, many open questions still remain. Can we, for instance, extend the unprecedented experimental precision and exquisit agreement with state-of-the-art quantum theory to more complex systems beyond benchmark systems like NO-He? Stepping up the complexity ladder is essential to test new multi-electron quantum chemistry methods and validate the approximations inevitably needed to describe larger systems, and would help bridge the gap between ab initio methods typically used for small weakly-interacting systems and density-functional or semi-empirical methods that are more relevant for heavier and strongly interacting systems. It is also essential to test quantum scattering methods, since for larger molecules there is an increasing number of effects that need to be taken into account to properly describe how the system evolves over the PES. Change of system may also facilitate the manipulation of resonance structures using external electric or magnetic fields. The idea is that at energies below $\sim$ 1 Kelvin, the interaction energy of a polar molecule with external electric and magnetic fields is on the order of the collision energy itself, offering the distinctive opportunity to engineer interaction Hamiltonians and control the collision outcome. 
  
The systems used thus far to probe scattering resonances are often predominantly chosen for reasons of experimental feasibility, but they are unfortunately not very favorable to further break new grounds. The NO radical, for instance, is particularly easy to produce and detect, but its modest dipole moment of \SI{0.16}{D} makes scattering resonances involving NO rather immune to electric fields. This low dipole moment also prohibits reaching lower energies by beam merging, that requires a sufficiently strong electric field induced force to bend the beam's trajectory. In these respects, molecules like OH and ND$_3$ are much more appealing, and have been prime candidates in cold molecular research ever since the field started in the 1990's \cite{Meerakker:CR112:4828}. However, they are either difficult to produce in large quantities required for controlled scattering experiments, or lack sensitive detection schemes that allow for high-resolution VMI detection. For the latter, one of the strongest bottlenecks is the requirement for a state-selective REMPI detection scheme that ionizes the molecules near threshold, thus imparting negligible recoil energy to the detected ions. Such schemes are generally lacking for most molecules of interest, or are impedingly insensitive. Being a tour-de-force experimentally, measurements of state-to-state resolved ICSs and DCSs of quantum scattering resonances involving molecules thus remain restricted to systems involving the NO radical, and it is until now unclear how and if the major hurdles can be overcome to extend these studies to other systems.

Cold collision studies involving ND$_3$ are particularly relevant, as ND$_3$ has been the system of choice in many seminal experiments on the manipulation of neutral polar molecules. The molecule was used in the first demonstration of electrostatic trapping ~\cite{Bethlem:Nature406:491,Bethlem:PRA65:053416}, AC trapping ~\cite{Veldhoven:PRL94:083001,Schnell:JPCA111:7411}, a buncher ~\cite{Crompvoets:PRL89:093004}, mirror~\cite{Schulz:PRL93:020406}, storage ring~\cite{Crompvoets:Nat411:174}, synchrotron~\cite{Heiner:NatPhys3:115,Zieger2010}, beamsplitter~\cite{Deng2011,Gordon2017:3D}, fountain~\cite{Cheng:PRL117:253201}, cryofuge \cite{Wu:Science358:645}, co-trapping with laser-cooled atoms~\cite{Parazzoli:PRL106:193201}, as well as in the first demonstration of the increased spectral resolution by the elongated interaction time afforded by decelerated molecules ~\cite{Veldhoven:PRA66:032501}. Since the discovery of NH$_3$ in the interstellar medium in 1968~\cite{Cheung1968}, and of ND$_3$ in 2002~\cite{Lis2002}, rotationally inelastic collisions involving ammonia have attracted considerable interest \cite{Meyer1995,Schleipen1991,Sanden1996,Tkac2014,Tkac2015b,Meyer1986,Das1986,Daly1970,Oka1968,Oka1968a,Broquier1987,Gao2019,Gubbels:JCP136:074301,Danby1987,Loreau2015,Bouhafs2017,Demes2023,Loreau2023}. Since then, observed inversion transitions are used to probe the temperature of molecular clouds~\cite{Walmsley1983,Maret2009}. 
	
Here, we report measurements of scattering resonances in the ICS and DCS for inelastic inversion-deexcitation collisions between ND$_3$ ($1_1^-\rightarrow1_1^+$) and H$_2 $ or HD. The collision energy is varied between \num{0.5}-\SI{25}{\per\cm}, scanning over several groups of resonances for both systems. DCSs are probed with high resolution using VMI in combination with a near-recoil free 1+1$'$ REMPI scheme using Vacuum Ultra Violet (VUV) light \cite{Kuijpers:JPCA128:10993}, solving a long-standing bottleneck in using ND$_3$ in high-resolution imaging experiments. We find that a new PES at the CCSD(T)/AVTZ+MB level of theory, with a correction based on CCSDT(Q)/AVDZ calculations, is required to find quantitative agreement between the predicted and measured resonance positions.

\section*{Results}
The experiments were performed by crossing a state-selected and velocity-controlled ND$_3$ packet emerging from a Stark decelerator with a beam of para-H$_2$ or HD at an angle of \SI{5.2}{\degree}. The H$_2$ (HD) beam traveled at a fixed velocity of $v_2=\SI{865}{m/s}$ (\SI{815}{m/s}) after expanding from a cryogenically cooled pulsed valve at \SI{35}{K} (\SI{40}{K}). Using the Stark decelerator, the velocity of the ND$_3$ packets was tuned between \num{350} and \num{980} {m/s}, resulting in collision energies $E_\text{col}$ between \num{0.5} and \SI{25}{\per\cm}. The collision energy resolution ranged from \SI{0.1}{\per\cm} at the lowest energies to \SI{2.5}{\per\cm} at the highest energies. The $j_k=1_1$ rotational ground state of para-ammonia is split into two inversion components with opposite parity, of which the Stark decelerator only transmits molecules in the $1_1^-$ upper inversion component. 

\begin{figure}[b!]
	\includegraphics{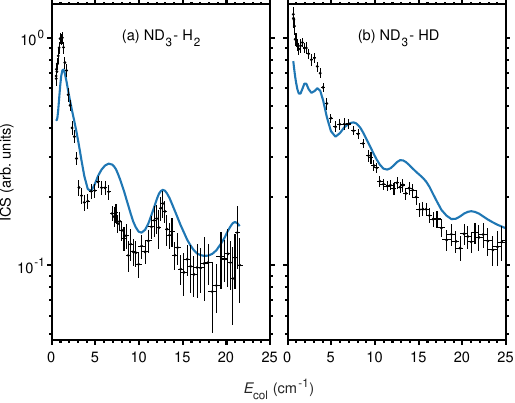}
	\caption{
		Integral Cross Section (ICS) as a function of the collision energy, $E_\text{col}$, for ND$_3$ ($1_1^-\rightarrow1_1^+$) collisions with (a) H$_2$ or (b) HD, both with $j_2=0$. Each panel shows a comparison between experimental (black) and predicted cross sections based on the CCSD(T)+$E^\text{T(Q)}$ PES (blue). Horizontal error bars reflect the energy calibration uncertainty, computed by propagating the uncertainty in $v_2$ (see SI). Vertical error bars represent statistical uncertainties, calculated as the standard deviation of the mean over hundreds of samples (see Methods section). All error bars represent a \SI{95}{\percent} confidence interval. The smoothing of the predicted cross sections due to the experimental resolution was taken into account based on simulations.
	}
	\label{fig:ND3_exp_ICS}
\end{figure}

We first measured ICSs for $1_1^-\rightarrow1_1^+$ inelastic inversion-deexcitation collisions by state-selectively ionizing scattered ND$_3$ ($1_1^+$) molecules using a convenient 2+1 REMPI scheme at \SI{321}{nm}, see Fig. \ref{fig:ND3_exp_ICS}. For ND$_3$-H$_2$, three distinct resonance features were observed at energies around $1$, $6$ and \SI{13}{\per\cm}. For ND$_3$-HD, the resonances are less pronounced but three resonance features causing a series of inflection points could clearly be discerned.

\begin{figure*}[t!]
	\includegraphics{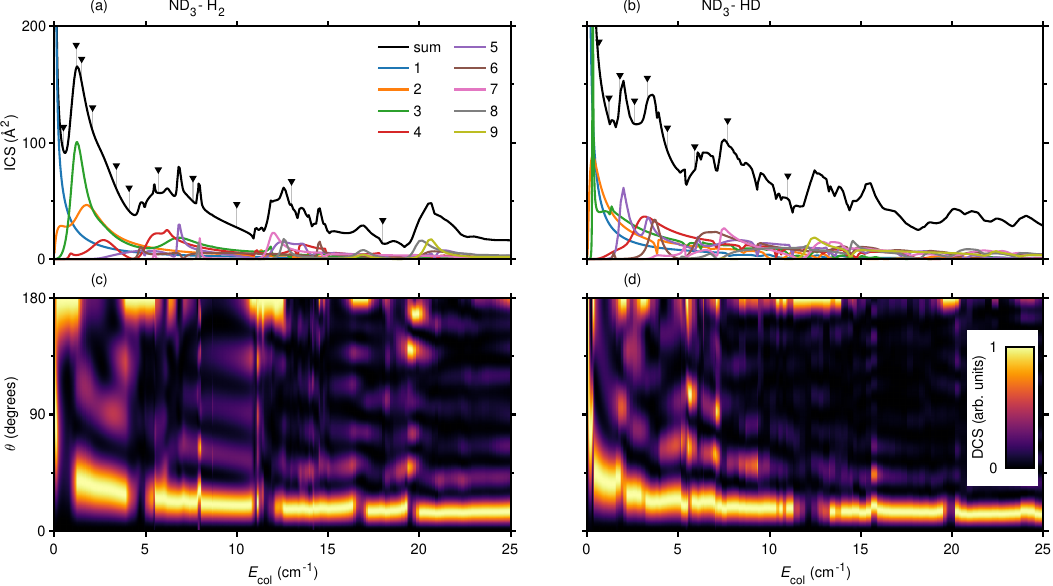}
	\caption{
		Calculated cross sections based on the CCSD(T)+$E^\text{T(Q)}$ PES as a function of the collision energy for ND$_3$ ($1_1^-\rightarrow1_1^+$) collisions with H$_2$ (a,c) or HD (b,d). (a,b) Calculated ICS (black), with the contribution of angular momentum states $\mathcal{J}$ (see legend). Black triangles mark the energies at which the DCS was probed experimentally. (c,d) 2D image plot of the calculated Differential Cross Section (DCS) as function of the collision energy and scattering angle. The DCS is normalized for each energy separately to emphasize the angular structure.
	}
	\label{fig:ND3_theory_ICS}
\end{figure*}
	
We compared the experimentally observed ICSs with the calculated ICSs based on the available PES computed by Maret et al.~\cite{Maret2009,Ma2015}. This PES was calculated at the CCSD(T)/AVDZ level of theory, after which the correlation part of the interaction energy was scaled using CCSD(T)/AVTZ calculations by performing a two-point CBS extrapolation (see SI). In the scattering calculations, a rotational basis up to $j=6$ for ND$_3$, $j_2=2$ for para-H$_2$ and $j_2=4$ for HD was used. The resulting ICSs, convoluted with the experimental resolution, gave unsatisfactory agreement with the experimentally observed ICSs, as the resonance structures appeared shifted to higher collision energies (see SI). 

We therefore attempted to calculate PESs at a higher level of theory. We found that CBS extrapolation to AVTZ and AVQZ gave incorrect results. Unfortunately, a complete basis set extrapolation, as performed for NO-He (using AVnZ for $\text{n}=4,5,6$~\cite{Jongh2020}), is currently unrealistic. Compared to NO-He, ND$_3$-H$_2$ is much more computationally demanding even though it has fewer electrons (12 instead of 17) and no open-shell character. The increase in computational cost is caused by the increased number of degrees of freedom (5 instead of 2) in the rigid rotor approximation. Where the NO-He potential could be defined by a grid of \num{912} geometries, our ND$_3$-H$_2$ potential required \num{29569} points. We computed this PES at the CCSD(T)/\{AVTZ+MB\} level of theory and applied an additional correction $E^\text{T(Q)}$ that depended on the radial coordinate only (see SI). This correction function was determined by comparing calculations at the CCSD(T)/AVDZ and CCSDT(Q)/AVDZ level of theory, similar to recent work for NO-He~\cite{Jongh2020}. Instead of evaluating $E^\text{T(Q)}$ at every point of the PES as was done previously for NO-He, we averaged the radial dependence found for a few fixed angular coordinates. For the N-D bond distance a value of \SI{1.946}{\bohr} was used, which is the vibrational average of NH$_3$~\cite{Huang2011}. The inclusion of midbond functions was found to yield more accurate energies at a lower computational cost compared to using the AVQZ basis set, based on a handful of test geometries for which calculations up to CCSD(T)/AV6Z were performed (see SI). We found that the resonances at energies near 1~cm$^{-1}$ and below responded extremely sensitively to the level of theory we used (see SI). 

Our new CCSD(T)+$E^\text{T(Q)}$ PES was found to be deeper than the Maret PES by about \SI{2}{\percent}, which caused the resonances to shift to lower energies in better agreement with the experiments, although an intensity mismatch across the sampled collision energies remained, see Fig. \ref{fig:ND3_exp_ICS}. We tested several further modifications to the potential at less computationally expensive levels of theory. Most notably, we explicitly included the umbrella coordinate of ND$_3$, as vibrational motion may impact the low-energy scattering behavior \cite{Yang2015a,Faure2016}. Although full dimensional calculations are currently not feasible for ND$_3$-H$_2$, we evaluated every point of the 5D PES at ten different umbrella angles to yield a new 6D PES beyond the rigid rotor approximation. This modification was found to have a too small effect to explain the discrepancy between experiment and theory, however, consistent with previous results for collisions of NH$_3$ with rare gas atoms \cite{Loreau2015,Loreau2024}. Furthermore, we changed the N-D bond length and used a global scaling factor, but these efforts did not result in a better agreement between experiment and theory (see SI).
	
\begin{figure*}[t!]
	\includegraphics{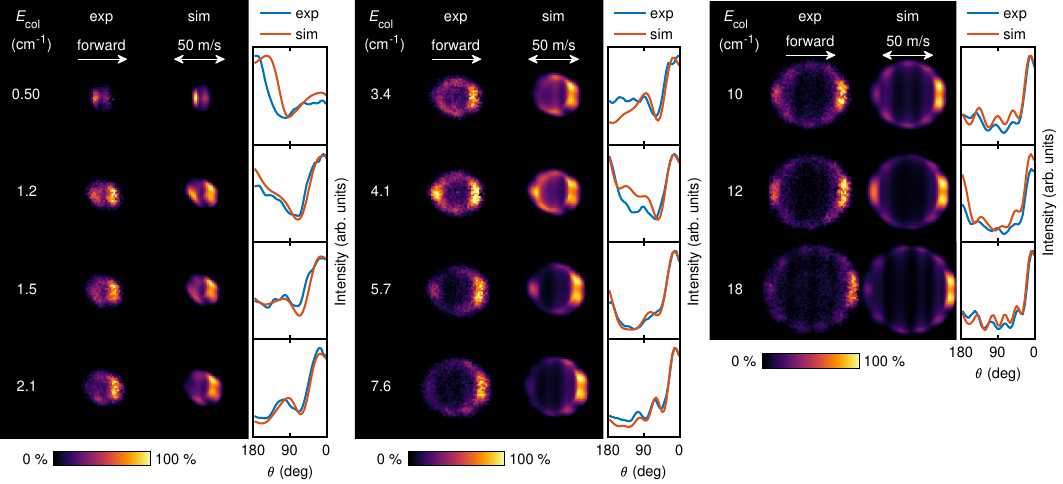}
	\caption{
		Experimental (exp) and simulated (sim) velocity mapped images for ND$_3$($1_1^-\rightarrow1_1^+$)-H$_2$ scattering at several collision energies. The simulated images were generated using the calculated DCSs based on the CCSD(T)+$E^\text{T(Q)}$ PES, as shown in Figure \ref{fig:ND3_theory_ICS}(b). They are oriented such that the forward direction faces right. Each image pair was normalized by their integrated intensity. The angular distributions extracted from the images are shown for every collision energy to the right of every image pair.
	}
	\label{fig:H2_DCS}
\end{figure*}

Characterization of the resonances was achieved by performing a full partial wave analysis and by calculating the scattering wave functions at the resonance energies. While scattering, the total parity $\mathcal{P}$ as well as the total angular momentum with quantum number $\mathcal{J}$ are conserved (see SI). We found that both parities contribute near-equally to the scattering cross sections, such that a detailed analysis for one value of $\mathcal{P}$ sufficed. The total angular momentum is obtained by coupling the partial wave with quantum number $\ell$ with the rotational angular momenta $j$ and $j_2$ of the ND$_3$ and H$_2$ molecules, respectively. We calculated for each value of $\mathcal{J}$ the individual contribution to the scattering cross section (see Fig. \ref{fig:ND3_theory_ICS}), and found that groups of overlapping resonances cause the observed resonance features in the ICSs. The resonance observed at a collision energy near 1.2 cm$^{-1}$ in ND$_3$-H$_2$ appeared relatively pure, with a dominant contribution of $\mathcal{J}=3$.  

For each resonance, we could derive the values of $\ell_\text{in}$ and $\ell_\text{out}$ that represent the relevant partial wave of the entrance and exit channels, respectively, as well as the resonant partial wave $\ell_\text{res}$ that characterizes the quasi-bound state from which the resonance originates. Since we exclusively studied $1_1^- \rightarrow 1_1^+$ inversion changing collisions, the value for $\ell$ can only change from $\ell_\text{in}$ even to $\ell_\text{out}$ odd or vice versa. The partial waves are further constrained by the conservation of $\mathcal{J}$, which for the $1_1^- \rightarrow 1_1^+$ transition implies that $\ell_\text{in/out}=\left\{ \mathcal{J}-1,\mathcal{J},\mathcal{J}+1 \right\}$. Together with the calculated scattering wavefunction, we could infer that during the collision the partial waves evolve from $\ell_\text{in}=\left\{ 2,4 \right\}$ via a resonance state with $\ell_\text{res}=4$ to $\ell_\text{out}=3$ (see SI). The resonance state could be associated with the $2_1^-$ rotational level of ND$_3$, and could hence be characterized as a Feshbach resonance (the $2_1^-$ state is asymptotically closed at a collision energy of 1.2 cm$^{-1}$). The resonance appears as a relatively broad feature in the ICS, indicating that the corresponding quasi-bound state is short-lived. 
	
Using similar reasoning, we could fully characterize the ten most prominent resonances for ND$_3$-H$_2$. We found that nearly all resonances are of Feshbach character, except for the two resonances at $E_{col}=\num{7.87}$ and \SI{7.97}{\per\cm} that could be characterized as shape resonances, and the resonance at $E_{col}=\SI{14.47}{\per\cm}$ that could be best described as a combined Feshbach-shape resonance (see SI). 

We further investigated the resonances by calculating the DCSs as a function of collision energy, that directly reflect the partial wave composition of a resonance. The theoretical DCSs computed from the CCSD(T)+$E^\text{T(Q)}$ PES, see Fig. \ref{fig:ND3_theory_ICS}, showed a pattern of diffraction oscillations whose spacing scales with $1/\sqrt{E_\text{col}}$ ~\cite{Faubel1984,Buck:JCP68:5654,Zastrow:NatChem6:216}. The diffraction pattern was interrupted at energies that coincided with a resonance, reflecting the vastly different scattering behavior in which only selected partial waves dominated when a resonance was accessed. The angular distributions changed rapidly as the collision energy was tuned over the resonances, with the appearance and disappearance of pronounced scattering flux in particular angular regions, such as the strong backward scattering in energy windows between adjacent groups of resonances. Such rapid variation was also observed in previous work on NO-He \cite{Jongh2020}, and is the result of the rapidly changing partial wave composition underlying each resonance. The interference between individual partial waves can strongly enhance or reduce the flux in specific angles, and the observation of such fast evolution of DCSs is testimony of the quantum nature of scattering resonances in low energy collisions. 

\begin{figure*}[t!]
	\includegraphics{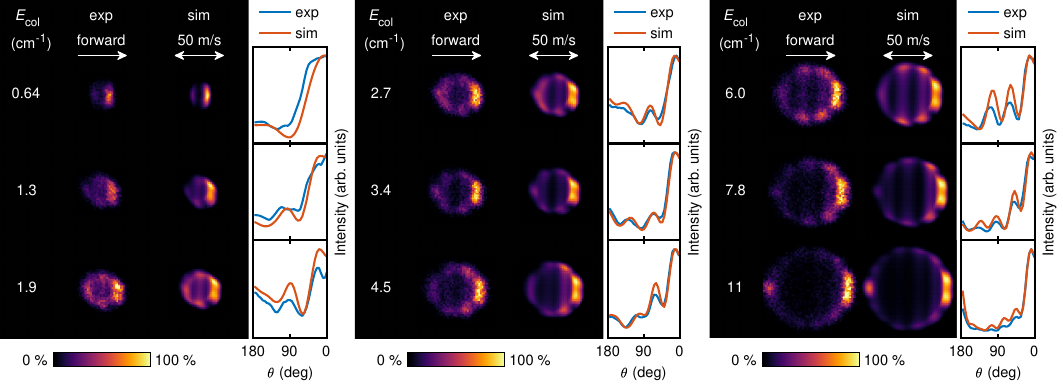}
	\caption{
		Experimental (exp) and simulated (sim) velocity mapped images for ND$_3$($1_1^-\rightarrow1_1^+$)-HD scattering at several collision energies. The simulated images were generated using the calculated DCSs based on the CCSD(T)+$E^\text{T(Q)}$ PES, as shown in Figure \ref{fig:ND3_theory_ICS}(d). They are oriented such that the forward direction faces right. Each image pair was normalized by their integrated intensity. The angular distributions extracted from the images are shown for every collision energy to the right of every image pair.
	}
	\label{fig:HD_DCS}
\end{figure*}

We probed the DCSs experimentally using VMI in combination with a new recoil-free 1+1$'$ REMPI scheme involving VUV \cite{Kuijpers:JPCA128:10993}. The 2+1 REMPI scheme used for ICS measurements was not suitable for this purpose, as this scheme imparts \SI{17}{m/s} recoil velocity to the ND$_3$ ions, blurring the images. The lack of a suitable REMPI scheme hampered measurements of DCSs in low-energy collisions before, but was essential in our experiment to record angular distributions with sufficient resolution to infer the energy dependence of DCSs in the resonance region. We recorded a total of 20 high-resolution scattering images at different collision energies accross the resonance regions, see Figs \ref{fig:H2_DCS} and \ref{fig:HD_DCS}. We observed clear diffraction oscillations with additional structures featuring strong backscattering at selected energies, consistent with the rapidly changing DCSs around the resonances we found theoretically. We quantitatively compared the experimental images with simulated images based on the theoretically predicted DCSs and the kinematics of the experiment. Angular scattering distributions were extracted from all images, and in general excellent agreement was found between the experimental and simulated distributions. For ND$_3$-H$_2$, a deviation was found in the backscattered region at $E_\text{col}=3.4$ and \SI{4.1}{\per\cm}, which we attributed to the small shift between the experimentally observed and theoretically predicted resonance position in the ICS, in combination with the sudden appearance of backscattering close to these collision energies. We found that only the CCSD(T)+$E^\text{T(Q)}$ PES gave good agreement between experiment and theory; as resonant features in the DCS are shifted along with those in the ICS, we found that simulated images based on the PES by Maret et al. did not capture the observed energy dependence of the DCSs well.   
	
\section*{Discussion}
Our joint experimental and theoretical study of partial wave scattering resonances in ND$_3$-H$_2$ and ND$_3$-HD at collision energies down to \SI{0.5}{\per\cm} underlines the level of detail that can now be obtained in scattering experiments involving polyatomic systems. As only the second molecular system for which we were able to probe resonances in both state-to-state ICS and DCS with high resolution, the success attained here unlocks possiblities to probe fully controlled quantum dynamics studies beyond the benchmark NO-He system. Since ND$_3$ has a strong, near-linear Stark effect, collision systems involving ND$_3$ appear the most attractive forum to study the effects of external electric fields on resonance structures and partial wave dynamics. Unlike NO, Stark energies of \SI{\sim 1}{\per\cm} are readily obtained for ND$_3$ in experimentally attainable electric fields of \SI{\sim 75}{kV\per\cm}. The distinct resonance features as observed here at collision energies near \SI{1}{\per\cm} are expected to sensitively respond to external fields, paving the way to modify the collision dynamics and control the scattering outcome. Moreover, in the initial $1_1^-$ and final $1_1^+$ state, the ND$_3$ molecules possesses two distinctive angular momentum projection states belonging to $m_j=0$ and $|m_j|=1$. Molecules in either of these states can be prepared before the collision using the decelerator \cite{Herbers2024}, whereas transitions to either of the two final projection states would appear as distinctive rings in the images separared by the field induced Stark shift. The recoil-free detection scheme for ND$_3$ as demonstrated here enables sufficient resolution to fully separate these rings, offering the unprecendented opportunity to steer and control low-energy resonances, and to simultaneously probe how stereodynamics influences partial wave dynamics.

\section*{Methods}
The experiments were performed using a crossed molecular beam setup described before \cite{Jongh2020}. A supersonic beam of ND$_3$ was created by expanding 2\% ND$_3$ seeded in a carrier gas into a vacuum chamber through a Nijmegen Pulsed Valve \cite{Yan2013}. The mean velocity of the molecular beam could be coarsely tuned between 450~m/s and 900~m/s by using different carrier gas mixtures. For a given carrier gas, precise tuning of the velocity was achieved by passing the beam through a 2.6~m long Stark decelerator. The velocity controlled packets of ND$_3$ emerging from the decelerator exclusively resided in the $j_k^p=1_1^-$ upper inversion component of the rotational ground state of $E$-symmetry ND$_3$, often called para-ammonia by analogy to NH$_3$ \cite{Bethlem:PRA65:053416,Veldhoven:PRA66:032501}. Population in the $1_1^+$ lower inversion component was effectively eliminated from the beam. Molecules initially in the $1_1^+, |m_j|=1$ component are high-field seeking and deflected from the beam axis inside the decelerator, whereas molecules initially in the $1_1^+, m_j=0$ component are immune to electric fields and travel though the decelerator in free flight greatly reducing particle densities. 

The ND$_3$ packets traveled in free flight for 529.5 mm towards the interaction region, where they were intercepted by a cryogenic beam of H$_2$ or HD at an angle of \SI{5.2}{\degree}. These beams were produced by supersonically expanding neat beams of H$_2$ or HD through a temperature-stabilized Even-Lavie valve that was mounted on the second stage of a cold head. A pure sample of para-H$_2$ ($j_2=0$) could be produced by first condensing normal H$_2$ over a NiSO$_4$ catalyst. The beams were collimated 47.5~mm before the beam crossing point by 3~mm diameter pinholes. 

Inelastic $1_1^-\rightarrow1_1^+$ inversion-deexcitation collisions were probed by state-selectively ionizing scattered ND$_3$ ($1_1^+$) molecules. As this is the only open channel at the lowest collision energies, studying cold inelastic collisions involving other rotational levels requires preparing the system in a rotationally excited state \cite{Jongh2022,Herbers2024}. The ions were then mapped onto a microchannel plate using an advanced high-resolution velocity map imaging spectrometer \cite{Plomp2020}. The extraction field was \SI{20}{V/cm}, too low to cause a significant Stark shift or associated change in the collision dynamics. For ICS measurements, a 2+1 REMPI scheme was employed at 320~nm by inducing the $B \leftarrow X$ transition in ND$_3$ using a single dye laser. For DCS measurements, we employed a 1+1$'$ REMPI scheme that ionized ND$_3$ at threshold, thereby minimizing blurring effects in the images due to ion recoil. In this scheme, the $B \leftarrow X$ transition was induced using a single photon near 160~nm, after which a photon of 448~nm excited the ND$_3$ molecule to a Rydberg state above the ionization potential that autoionizes with near-zero recoil \cite{Kuijpers:JPCA128:10993}. The 160~nm photons were generated using difference frequency mixing in xenon gas using two dye lasers, while a third dye laser was used to generate the ionizing photons at 448~nm.        

To cover a range of collision energies between \num{0.5} and \SI{25}{\per\cm}, seven seed-gas mixtures were used to prepare the ND$_3$ beam. For a given seed gas, ICSs were recorded by switching to a different velocity (energy) every four seconds using the decelerator, continuously cycling back-and-forth over the collision energy range allowed by the seed gas. Each data point except the outer edges was covered by at least two different seed gas mixtures, such that ICSs measured during different experimental runs could be stitched together. Every four seconds, a background measurement was performed by detuning the secondary beam in time such that no collisions were probed. Every two hours, the measurement was interrupted to probe the density of the incoming $1_1^-$ beam, confirming that long term drifts in the experiment did not deteriorate the beam intensity. Each velocity range was repeated during at least two days, with a typical amount of $\sim$200 recorded cycles per day. The normalized signal was then computed as the difference between the accumulated scattering and background measurement, divided by the initial ND$_3$ beam density. Calibration of the collision energy was performed using three independent methods by recording ICSs and scattering images while scanning the ND$_3$ velocity accross the region of minimal collision energy (see SI). The experimentally obtained normalized signals were corrected for flux to density effects using extensive Monte Carlo simulations of the experiment (see SI). Scattering images were recorded one energy at a time, over a period ranging between several hours to a few days. For these measurements, the background signal was sampled by toggling the overlap of the secondary beam on/off every 30 seconds. For each image the laser powers were reduced to yield only a few events per shot, such that event counting and centroiding could be applied to accumulate a high resolution image.

The scattering calculations were performed by means of the close-coupling method in the body-fixed frame, as already described elsewhere for collisions of NH$_3$ and ND$_3$ with H$_2$ and D$_2$ \cite{Ma2015,Tkac2015b,Gao2019}. 

For this purpose, various PESs were employed. The five-dimensional PES of Maret et al. \cite{Maret2009}, in which ammonia and hydrogen are considered as rigid rotors, was first used. Additional five-dimensional PESs were generated at various levels of theory (CCSD(T), CCSD(T)-F12a, CCSDT, CCSDT(Q)) using the {\small MOLPRO} quantum chemistry package \cite{Molpro2018} and various basis sets (see SI) on a grid of 29,568 unique geometries. The PES was expanded in angular functions and the radial coefficients were interpolated with the reproducing kernel Hilbert space (RKHS) method. The long-range part of the PES was constructed by developing the radial coefficients in inverse powers of the distance. 
A six-dimensional PES with non-rigid NH$_3$ was also constructed by repeating this process for several values of the umbrella angle that describes the inversion motion of ammonia. The impact of isotopic substitution (in the case of ND$_3$-H$_2$ and ND$_3$-HD) was taken into account by performing a coordinate transformation on the fitted PESs to reflect the change in the position of the centers of mass, after which a new angular expansion of the PES was carried out.

Integral and differential cross sections were computed by solving the coupled channel equations in the body-fixed frame over a grid of 400 values of the kinetic energy in the range \num{0.01}-\SI{25}{\per\cm}. The rotational basis included all levels with $j\leq 6$ (for ND$_3$) and $j_2\leq 4$ (for H$_2$/HD). 

The inversion of ND$_3$ was treated using two different models. For PESs with rigid ND$_3$, a two-state model was adopted in which the inversion-tunnelling states are taken as linear combinations of the two rigid equilibrium structures. When the PES included an explicit dependence on the inversion angle, the umbrella motion of ammonia was treated using an Hamiltonian that comprises a specific kinetic and potential term dependent on the inversion angle. This follows the approach tested for low-energy collisions of ammonia with rare gas atoms (see e.g., \cite{Gubbels:JCP136:074301, Loreau2024}).

To examine the character of scattering resonances, a partial wave analysis was performed for all values of the total angular momentum $\mathcal{J}$ and for each parity $\mathcal{P}$ by computing the scattering wavefunctions.

\section*{Data availability}
All data are available online at DANS:

\noindent\href{https://doi.org/10.17026/PT/IZNJNM}{https://doi.org/10.17026/PT/IZNJNM}

\section*{Acknowledgements}
This work is part of the research program of the Dutch Research Council (NWO). S.Y.T.v.d.M. acknowledges support from the European Research Council (ERC) under the European Union's Horizon 2020 Research and Innovation Program (Grant Agreement No. 817947 FICOMOL and No. 101141163 QUCUMBER). We thank N. Janssen, E. Sweers and A. van Roij for expert technical support.

\section*{Author Contributions}
The project was conceived by S.Y.T.v.d.M. The experiments were carried out by S.E.J.K. and D.H.P.. Data analysis and simulations were performed by S.E.J.K.. Theoretical calculations were performed by J.L. and A.v.d.A.. The manuscript was written by S.E.J.K, J.L. and S.Y.T.v.d.M. with contributions from all authors. All authors were involved in the interpretation of the data and the preparation of the manuscript.

\section*{Competing interests}
The authors declare no competing interests.

\bibliographystyle{apsrev4-1.bst}
\bibliography{bibliography-ND3H2}

\begin{thebibliography}{10}
\expandafter\ifx\csname url\endcsname\relax
  \def\url#1{\texttt{#1}}\fi
\expandafter\ifx\csname urlprefix\endcsname\relax\def\urlprefix{URL }\fi
\providecommand{\bibinfo}[2]{#2}
\providecommand{\eprint}[2][]{\url{#2}}

\bibitem{Onvlee:PCCP16:15768}
\bibinfo{author}{Onvlee, J.}, \bibinfo{author}{Vogels, S.~N.},
  \bibinfo{author}{von Zastrow, A.}, \bibinfo{author}{Parker, D.~H.} \&
  \bibinfo{author}{van~de Meerakker, S. Y.~T.}
\newblock \bibinfo{title}{Molecular collisions coming into focus}.
\newblock \emph{\bibinfo{journal}{Phys. Chem. Chem. Phys.}}
  \textbf{\bibinfo{volume}{16}}, \bibinfo{pages}{15768--15779}
  (\bibinfo{year}{2014}).

\bibitem{Zastrow:NatChem6:216}
\bibinfo{author}{von Zastrow, A.} \emph{et~al.}
\newblock \bibinfo{title}{State-resolved diffraction oscillations imaged for
  inelastic collisions of {NO} radicals with {H}e, {N}e and {A}r}.
\newblock \emph{\bibinfo{journal}{Nat. Chem.}} \textbf{\bibinfo{volume}{6}},
  \bibinfo{pages}{216--221} (\bibinfo{year}{2014}).

\bibitem{Jongh2020}
\bibinfo{author}{de~Jongh, T.} \emph{et~al.}
\newblock \bibinfo{title}{Imaging the onset of the resonance regime in
  low-energy {NO}-{H}e collisions}.
\newblock \emph{\bibinfo{journal}{Science}} \textbf{\bibinfo{volume}{368}},
  \bibinfo{pages}{626--630} (\bibinfo{year}{2020}).

\bibitem{Blender}
\bibinfo{author}{{Blender Online Community}}.
\newblock \emph{\bibinfo{title}{Blender - a 3D modelling and rendering
  package}}.
\newblock \bibinfo{organization}{Blender Foundation},
  \bibinfo{address}{Stichting Blender Foundation, Amsterdam}
  (\bibinfo{year}{2018}).
\newblock \urlprefix\url{http://www.blender.org}.

\bibitem{Yan:RSI84:023102}
\bibinfo{author}{Yan, B.} \emph{et~al.}
\newblock \bibinfo{title}{A new high intensity and short-pulse molecular beam
  valve}.
\newblock \emph{\bibinfo{journal}{Rev. Sci. Inst.}}
  \textbf{\bibinfo{volume}{84}}, \bibinfo{pages}{023102}
  (\bibinfo{year}{2013}).

\bibitem{_Scoles:MolBeam:1}
\bibinfo{editor}{Scoles, G.} (ed.) \emph{\bibinfo{title}{Atomic and molecular
  beam methods}}, vol.~\bibinfo{volume}{1} (\bibinfo{address}{Oxford, UK},
  \bibinfo{year}{1988}).

\bibitem{Even:AiC2014:636042}
\bibinfo{author}{Even, U.}
\newblock \bibinfo{title}{Pulsed supersonic beams from high pressure source:
  Simulation results and experimental measurements}.
\newblock \emph{\bibinfo{journal}{Adv. Chem.}} \textbf{\bibinfo{volume}{2014}},
  \bibinfo{pages}{636042} (\bibinfo{year}{2014}).

\bibitem{Hoge1951}
\bibinfo{author}{Hoge, H.~J.} \& \bibinfo{author}{Arnold, R.~D.}
\newblock \bibinfo{title}{Vapor pressures of hydrogen, deuterium, and hydrogen
  deuteride and dew-point pressures of their mixtures}.
\newblock \emph{\bibinfo{journal}{J. Res. Natl. Bur. Stand.}}
  \textbf{\bibinfo{volume}{47}}, \bibinfo{pages}{63--74}
  (\bibinfo{year}{1951}).

\bibitem{Vogels:NatChem10:435}
\bibinfo{author}{Vogels, S.~N.} \emph{et~al.}
\newblock \bibinfo{title}{Scattering resonances in bimolecular collisions
  between {NO} radicals and {H}$_2$ challenge the theoretical gold standard}.
\newblock \emph{\bibinfo{journal}{Nat. Chem.}} \textbf{\bibinfo{volume}{10}},
  \bibinfo{pages}{435} (\bibinfo{year}{2018}).

\bibitem{Rinnen1991}
\bibinfo{author}{Rinnen, K.-D.}, \bibinfo{author}{Buntine, M.~A.},
  \bibinfo{author}{Kliner, D. A.~V.}, \bibinfo{author}{Zare, R.~N.} \&
  \bibinfo{author}{Huo, W.~M.}
\newblock \bibinfo{title}{Quantitative determination of {H$_2$}, {HD}, and
  {D$_2$} internal-state distributions by (2+1) resonance-enhanced multiphoton
  ionization}.
\newblock \emph{\bibinfo{journal}{J. Comp. Phys.}}
  \textbf{\bibinfo{volume}{95}}, \bibinfo{pages}{214--225}
  (\bibinfo{year}{1991}).

\bibitem{Ashfold:CPL138:201}
\bibinfo{author}{Ashfold, M. N.~R.}, \bibinfo{author}{Dixon, R.~N.},
  \bibinfo{author}{Stickland, R.~J.} \& \bibinfo{author}{Western, C.~M.}
\newblock \bibinfo{title}{2+1 {MPI} spectroscopy of $\widetilde{B} {}^{1}{E}''$
  state {NH$_3$} and {ND$_3$}: rotational analysis of the origin bands}
  \textbf{\bibinfo{volume}{138}}, \bibinfo{pages}{201 -- 208}
  (\bibinfo{year}{1987}).

\bibitem{Ashfold:JCP89:1754}
\bibinfo{author}{Ashfold, M. N.~R.}, \bibinfo{author}{Dixon, R.~N.},
  \bibinfo{author}{Little, N.}, \bibinfo{author}{Stickland, R.~J.} \&
  \bibinfo{author}{Western, C.~M.}
\newblock \bibinfo{title}{The $\widetilde{B} {}^{1}{E}''$ state of ammonia:
  sub-{D}oppler spectroscopy at vacuum ultraviolet energies}.
\newblock \emph{\bibinfo{journal}{J. Comp. Phys.}}
  \textbf{\bibinfo{volume}{89}}, \bibinfo{pages}{1754--1761}
  (\bibinfo{year}{1988}).

\bibitem{Tkac2014}
\bibinfo{author}{Tk{\'a}{\v{c}}, O.} \emph{et~al.}
\newblock \bibinfo{title}{State-to-state resolved differential cross sections
  for rotationally inelastic scattering of {ND$_3$} with {H}e}.
\newblock \emph{\bibinfo{journal}{Phys. Chem. Chem. Phys.}}
  \textbf{\bibinfo{volume}{16}}, \bibinfo{pages}{477--488}
  (\bibinfo{year}{2014}).

\bibitem{Tkac2015b}
\bibinfo{author}{Tk{\'a}{\v{c}}, O.} \emph{et~al.}
\newblock \bibinfo{title}{Rotationally inelastic scattering of {ND$_3$} with
  {H$_2$} as a probe of the intermolecular potential energy surface}.
\newblock \emph{\bibinfo{journal}{Mol. Phys.}} \textbf{\bibinfo{volume}{113}},
  \bibinfo{pages}{3925--3933} (\bibinfo{year}{2015}).

\bibitem{Gao2019}
\bibinfo{author}{Gao, Z.}, \bibinfo{author}{Loreau, J.}, \bibinfo{author}{Van
  Der~Avoird, A.} \& \bibinfo{author}{Van De~Meerakker, S. Y.~T.}
\newblock \bibinfo{title}{Direct observation of product-pair correlations in
  rotationally inelastic collisions of {ND$_3$} with {D$_2$}}.
\newblock \emph{\bibinfo{journal}{Phys. Chem. Chem. Phys.}}
  \textbf{\bibinfo{volume}{21}}, \bibinfo{pages}{14033--14041}
  (\bibinfo{year}{2019}).

\bibitem{Kuijpers:JPCA128:10993}
\bibinfo{author}{Kuijpers, S. E.~J.} \emph{et~al.}
\newblock \bibinfo{title}{Sensitive low-recoil {VUV} 1+1' {REMPI} detection of
  {ND}$_3$}.
\newblock \emph{\bibinfo{journal}{J. Phys. Chem. A}}
  \textbf{\bibinfo{volume}{128}}, \bibinfo{pages}{10993--11004}
  (\bibinfo{year}{2024}).

\bibitem{Hilbig1983}
\bibinfo{author}{Hilbig, R.} \& \bibinfo{author}{Wallenstein, R.}
\newblock \bibinfo{title}{Tunable {VUV} radiation generated by two-photon
  resonant frequency mixing in xenon}.
\newblock \emph{\bibinfo{journal}{IEEE J. Quantum Electron.}}
  \textbf{\bibinfo{volume}{19}}, \bibinfo{pages}{194--201}
  (\bibinfo{year}{1983}).

\bibitem{Miyazaki1989}
\bibinfo{author}{Miyazaki, K.}, \bibinfo{author}{Sakai, H.} \&
  \bibinfo{author}{Sato, T.}
\newblock \bibinfo{title}{Two-photon resonances in {X}e and {K}r for the
  generation of tunable coherent extreme {UV} radiation}.
\newblock \emph{\bibinfo{journal}{Appl. Optics}} \textbf{\bibinfo{volume}{28}},
  \bibinfo{pages}{699} (\bibinfo{year}{1989}).

\bibitem{Eppink:RSI68:3477}
\bibinfo{author}{Eppink, A. T. J.~B.} \& \bibinfo{author}{Parker, D.~H.}
\newblock \bibinfo{title}{Velocity map imaging of ions and electrons using
  electrostatic lenses: Application in photoelectron and photofragment ion
  imaging of molecular oxygen}.
\newblock \emph{\bibinfo{journal}{Rev. Sci. Inst.}}
  \textbf{\bibinfo{volume}{68}}, \bibinfo{pages}{3477--3484}
  (\bibinfo{year}{1997}).

\bibitem{Plomp2020}
\bibinfo{author}{Plomp, V.}, \bibinfo{author}{Gao, Z.} \&
  \bibinfo{author}{van~de Meerakker, S. Y.~T.}
\newblock \bibinfo{title}{A velocity map imaging apparatus optimised for
  high-resolution crossed molecular beam experiments}.
\newblock \emph{\bibinfo{journal}{Mol. Phys.}} \textbf{\bibinfo{volume}{119}},
  \bibinfo{pages}{e1814437} (\bibinfo{year}{2020}).

\bibitem{Townsend:RSI74:2530}
\bibinfo{author}{Townsend, D.}, \bibinfo{author}{Minitti, M.~P.} \&
  \bibinfo{author}{Suits, A.~G.}
\newblock \bibinfo{title}{Direct current slice imaging}.
\newblock \emph{\bibinfo{journal}{Rev. Sci. Inst.}}
  \textbf{\bibinfo{volume}{74}}, \bibinfo{pages}{2530--2539}
  (\bibinfo{year}{2003}).

\bibitem{Lin:RSI74:2495}
\bibinfo{author}{Lin, J.~J.}, \bibinfo{author}{Zhou, J.},
  \bibinfo{author}{Shiu, W.} \& \bibinfo{author}{Liu, K.}
\newblock \bibinfo{title}{Application of time-sliced ion velocity imaging to
  crossed molecular beam experiments}.
\newblock \emph{\bibinfo{journal}{Rev. Sci. Inst.}}
  \textbf{\bibinfo{volume}{74}}, \bibinfo{pages}{2495--2500}
  (\bibinfo{year}{2003}).

\bibitem{Cremers2019}
\bibinfo{author}{Cremers, T.~L.}
\newblock \emph{\bibinfo{title}{A Multistage Zeeman Decelerator for
  Molecular-Beam Scattering Experiments}}.
\newblock Ph.D. thesis, \bibinfo{school}{Radboud University},
  \bibinfo{address}{Nijmegen, The Netherlands} (\bibinfo{year}{2019}).
\newblock \urlprefix\url{https://hdl.handle.net/2066/204153}.

\bibitem{Shuai2020}
\bibinfo{author}{Shuai, Q.} \emph{et~al.}
\newblock \bibinfo{title}{Experimental and theoretical investigation of
  resonances in low-energy {NO-H$_2$} collisions}.
\newblock \emph{\bibinfo{journal}{J. Chem. Phys.}}
  \textbf{\bibinfo{volume}{153}}, \bibinfo{pages}{244302}
  (\bibinfo{year}{2020}).

\bibitem{Western:pgopher}
\bibinfo{author}{Western, C.~M.}
\newblock \bibinfo{title}{Pgopher version 10.1} (\bibinfo{year}{2018}).

\bibitem{Western2017}
\bibinfo{author}{Western, C.~M.}
\newblock \bibinfo{title}{Pgopher: A program for simulating rotational,
  vibrational and electronic spectra}.
\newblock \emph{\bibinfo{journal}{J. Quant. Spectrosc. Radiat. Transfer}}
  \textbf{\bibinfo{volume}{186}}, \bibinfo{pages}{221--242}
  (\bibinfo{year}{2017}).

\bibitem{Veldhoven:thesis:2006}
\bibinfo{author}{van Veldhoven, J.}
\newblock \emph{\bibinfo{title}{AC trapping and high-resolution spectroscopy of
  ammonia molecules}}.
\newblock \bibinfo{type}{{Ph.D.} thesis}, \bibinfo{school}{Radboud Universiteit
  Nijmegen} (\bibinfo{year}{2006}).
\newblock \urlprefix\url{https://hdl.handle.net/2066/29874}.

\bibitem{Zastrow2015}
\bibinfo{author}{von Zastrow, A.}, \bibinfo{author}{Onvlee, J.},
  \bibinfo{author}{Parker, D.~H.} \& \bibinfo{author}{van~de Meerakker, S.
  Y.~T.}
\newblock \bibinfo{title}{Analysis of velocity-mapped ion images from
  high-resolution crossed-beam scattering experiments: a tutorial review}.
\newblock \emph{\bibinfo{journal}{EPJ Tech. Instrum.}}
  \textbf{\bibinfo{volume}{2}}, \bibinfo{pages}{11} (\bibinfo{year}{2015}).

\bibitem{Tang2023}
\bibinfo{author}{Tang, G.} \emph{et~al.}
\newblock \bibinfo{title}{Quantum state-resolved molecular dipolar collisions
  over four decades of energy}.
\newblock \emph{\bibinfo{journal}{Science}} \textbf{\bibinfo{volume}{379}},
  \bibinfo{pages}{1031--1036} (\bibinfo{year}{2023}).

\bibitem{Simion8.1:2012}
\bibinfo{author}{Dahl, D.~A.}
\newblock \bibinfo{title}{{\it SIMION Version 8.1}} (\bibinfo{year}{2012}).
\newblock \urlprefix\url{https://simion.com/}.

\bibitem{Dahl2000}
\bibinfo{author}{Dahl, D.~A.}
\newblock \bibinfo{title}{{SIMION} for the personal computer in reflection}.
\newblock \emph{\bibinfo{journal}{Int. J. Mass Spectrom.}}
  \textbf{\bibinfo{volume}{200}}, \bibinfo{pages}{3--25}
  (\bibinfo{year}{2000}).

\bibitem{Fehlberg1970}
\bibinfo{author}{Fehlberg, E.}
\newblock \bibinfo{title}{Klassische runge-kutta-formeln vierter und
  niedrigerer ordnung mit schrittweiten-kontrolle und ihre anwendung auf
  warmeleitungs-probleme}.
\newblock \emph{\bibinfo{journal}{Computing}} \textbf{\bibinfo{volume}{6}},
  \bibinfo{pages}{61--71} (\bibinfo{year}{1970}).

\bibitem{Meerakker:PRA71:053409}
\bibinfo{author}{van~de Meerakker, S. Y.~T.}, \bibinfo{author}{Vanhaecke, N.},
  \bibinfo{author}{Bethlem, H.~L.} \& \bibinfo{author}{Meijer, G.}
\newblock \bibinfo{title}{Higher-order resonances in a {S}tark decelerator}.
\newblock \emph{\bibinfo{journal}{Phys. Rev. A}} \textbf{\bibinfo{volume}{71}},
  \bibinfo{pages}{053409} (\bibinfo{year}{2005}).

\bibitem{Meerakker:PRA73:023401}
\bibinfo{author}{van~de Meerakker, S. Y.~T.}, \bibinfo{author}{Vanhaecke, N.},
  \bibinfo{author}{Bethlem, H.~L.} \& \bibinfo{author}{Meijer, G.}
\newblock \bibinfo{title}{Transverse stability in a {S}tark decelerator}.
\newblock \emph{\bibinfo{journal}{Phys. Rev. A}} \textbf{\bibinfo{volume}{73}},
  \bibinfo{pages}{023401} (\bibinfo{year}{2006}).

\bibitem{Scharfenberg:PRA79:023410}
\bibinfo{author}{Scharfenberg, L.}, \bibinfo{author}{Haak, H.},
  \bibinfo{author}{Meijer, G.} \& \bibinfo{author}{van~de Meerakker, S. Y.~T.}
\newblock \bibinfo{title}{Operation of a {S}tark decelerator with optimum
  acceptance}.
\newblock \emph{\bibinfo{journal}{Phys. Rev. A}} \textbf{\bibinfo{volume}{79}},
  \bibinfo{pages}{023410} (\bibinfo{year}{2009}).

\bibitem{Maret2009}
\bibinfo{author}{Maret, S.}, \bibinfo{author}{Faure, A.},
  \bibinfo{author}{Scifoni, E.} \& \bibinfo{author}{Wiesenfeld, L.}
\newblock \bibinfo{title}{On the robustness of the ammonia thermometer}.
\newblock \emph{\bibinfo{journal}{Mon. Not. R. Astron. Soc.}}
  \textbf{\bibinfo{volume}{399}}, \bibinfo{pages}{425--431}
  (\bibinfo{year}{2009}).

\bibitem{Daniel2014}
\bibinfo{author}{Daniel, F.} \emph{et~al.}
\newblock \bibinfo{title}{Collisional excitation of singly deuterated ammonia
  {NH$_2$D} by {H$_2$}}.
\newblock \emph{\bibinfo{journal}{Mon. Not. R. Astron Soc.}}
  \textbf{\bibinfo{volume}{444}}, \bibinfo{pages}{2544--2554}
  (\bibinfo{year}{2014}).

\bibitem{Daniel2016}
\bibinfo{author}{Daniel, F.} \emph{et~al.}
\newblock \bibinfo{title}{Collisional excitation of doubly and triply
  deuterated ammonia {ND$_2$H} and {ND$_3$} by {H$_2$}}.
\newblock \emph{\bibinfo{journal}{Mon. Not. R. Astron Soc.}}
  \textbf{\bibinfo{volume}{457}}, \bibinfo{pages}{1535--1549}
  (\bibinfo{year}{2016}).

\bibitem{Ma2015}
\bibinfo{author}{Ma, Q.} \emph{et~al.}
\newblock \bibinfo{title}{Resonances in rotationally inelastic scattering of
  {NH$_3$} and {ND$_3$} with {H$_2$}}.
\newblock \emph{\bibinfo{journal}{J. Chem. Phys.}}
  \textbf{\bibinfo{volume}{143}}, \bibinfo{pages}{044312}
  (\bibinfo{year}{2015}).

\bibitem{Bouhafs2017}
\bibinfo{author}{Bouhafs, N.} \emph{et~al.}
\newblock \bibinfo{title}{Collisional excitation of {NH$_3$} by atomic and
  molecular hydrogen}.
\newblock \emph{\bibinfo{journal}{Mon. Not. R. Astron Soc.}}
  \textbf{\bibinfo{volume}{470}}, \bibinfo{pages}{2204--2211}
  (\bibinfo{year}{2017}).

\bibitem{Demes2023}
\bibinfo{author}{Demes, S.}, \bibinfo{author}{Lique, F.},
  \bibinfo{author}{Loreau, J.} \& \bibinfo{author}{Faure, A.}
\newblock \bibinfo{title}{Collision-induced excitation of ammonia in warm
  interstellar and circumstellar environments}.
\newblock \emph{\bibinfo{journal}{Mon. Not. R. Astron Soc.}}
  \textbf{\bibinfo{volume}{524}}, \bibinfo{pages}{2368--2378}
  (\bibinfo{year}{2023}).

\bibitem{Loreau2023}
\bibinfo{author}{Loreau, J.}, \bibinfo{author}{Faure, A.},
  \bibinfo{author}{Lique, F.}, \bibinfo{author}{Demes, S.} \&
  \bibinfo{author}{Dagdigian, P.~J.}
\newblock \bibinfo{title}{Hyperfine collisional excitation of ammonia by
  molecular hydrogen}.
\newblock \emph{\bibinfo{journal}{Mon. Not. R. Astron. Soc.}}
  \textbf{\bibinfo{volume}{526}}, \bibinfo{pages}{3213--3218}
  (\bibinfo{year}{2023}).

\bibitem{Hanwell2012}
\bibinfo{author}{Hanwell, M.~D.} \emph{et~al.}
\newblock \bibinfo{title}{Avogadro: an advanced semantic chemical editor,
  visualization, and analysis platform}.
\newblock \emph{\bibinfo{journal}{J. Cheminf.}} \textbf{\bibinfo{volume}{4}}
  (\bibinfo{year}{2012}).

\bibitem{Phillips1994}
\bibinfo{author}{Phillips, T.~R.}, \bibinfo{author}{Maluendes, S.},
  \bibinfo{author}{McLean, A.~D.} \& \bibinfo{author}{Green, S.}
\newblock \bibinfo{title}{Anisotropic rigid rotor potential energy function for
  {H$_2$O–H$_2$}}.
\newblock \emph{\bibinfo{journal}{J. Chem. Phys.}}
  \textbf{\bibinfo{volume}{101}}, \bibinfo{pages}{5824--5830}
  (\bibinfo{year}{1994}).

\bibitem{Gubbels:JCP136:074301}
\bibinfo{author}{Gubbels, K.~B.}, \bibinfo{author}{van~de Meerakker, S. Y.~T.},
  \bibinfo{author}{Groenenboom, G.~C.}, \bibinfo{author}{Meijer, G.} \&
  \bibinfo{author}{van~der Avoird, A.}
\newblock \bibinfo{title}{Scattering resonances in slow {NH}$_3$-{H}e
  collisions}.
\newblock \emph{\bibinfo{journal}{J. Comp. Phys.}}
  \textbf{\bibinfo{volume}{136}}, \bibinfo{pages}{074301}
  (\bibinfo{year}{2012}).

\bibitem{Loreau2015}
\bibinfo{author}{Loreau, J.} \& \bibinfo{author}{Van~der Avoird, A.}
\newblock \bibinfo{title}{Scattering of {NH$_3$} and {ND$_3$} with rare gas
  atoms at low collision energy}.
\newblock \emph{\bibinfo{journal}{J. Chem. Phys.}}
  \textbf{\bibinfo{volume}{143}}, \bibinfo{pages}{184303}
  (\bibinfo{year}{2015}).

\bibitem{Loreau2024}
\bibinfo{author}{Loreau, J.} \& \bibinfo{author}{van~der Avoird, A.}
\newblock \bibinfo{title}{Vibrational energy transfer in ammonia–helium
  collisions}.
\newblock \emph{\bibinfo{journal}{Faraday Discuss.}}
  \textbf{\bibinfo{volume}{251}}, \bibinfo{pages}{249--261}
  (\bibinfo{year}{2024}).

\bibitem{Huang2008}
\bibinfo{author}{Huang, X.}, \bibinfo{author}{Schwenke, D.~W.} \&
  \bibinfo{author}{Lee, T.~J.}
\newblock \bibinfo{title}{An accurate global potential energy surface, dipole
  moment surface, and rovibrational frequencies for {NH$_3$}}.
\newblock \emph{\bibinfo{journal}{J. Chem. Phys.}}
  \textbf{\bibinfo{volume}{129}}, \bibinfo{pages}{214304}
  (\bibinfo{year}{2008}).

\bibitem{Lee_priv_comm}
\bibinfo{author}{Lee, T.}
\newblock \bibinfo{title}{private communication.}

\bibitem{Tkac2015a}
\bibinfo{author}{Tkáč, O.} \emph{et~al.}
\newblock \bibinfo{title}{Rotationally inelastic scattering of
  quantum-state-selected {ND$_3$} with {Ar}}.
\newblock \emph{\bibinfo{journal}{J. Phys. Chem. A}}
  \textbf{\bibinfo{volume}{119}}, \bibinfo{pages}{5979--5987}
  (\bibinfo{year}{2015}).

\bibitem{Avoird2011}
\bibinfo{author}{van~der Avoird, A.} \& \bibinfo{author}{Nesbitt, D.~J.}
\newblock \bibinfo{title}{Rovibrational states of the {H$_2$O}–{H$_2$}
  complex: An ab initio calculation}.
\newblock \emph{\bibinfo{journal}{J. Chem. Phys.}}
  \textbf{\bibinfo{volume}{134}}, \bibinfo{pages}{044314}
  (\bibinfo{year}{2011}).

\bibitem{rkhs96}
\bibinfo{author}{Ho, T.-S.} \& \bibinfo{author}{Rabitz, H.}
\newblock \bibinfo{title}{A general method for constructing multidimensional
  molecular potential energy surfaces from ab initio calculations}.
\newblock \emph{\bibinfo{journal}{J. Chem. Phys.}}
  \textbf{\bibinfo{volume}{104}}, \bibinfo{pages}{2584--2597}
  (\bibinfo{year}{1996}).

\bibitem{Cybulski1999}
\bibinfo{author}{Cybulski, S.~M.} \& \bibinfo{author}{Toczyłowski, R.~R.}
\newblock \bibinfo{title}{Ground state potential energy curves for {He$_2$},
  {Ne$_2$}, {Ar$_2$}, {He-Ne}, {He-Ar}, and {Ne-Ar}: A coupled-cluster study}.
\newblock \emph{\bibinfo{journal}{J. Chem. Phys.}}
  \textbf{\bibinfo{volume}{111}}, \bibinfo{pages}{10520--10528}
  (\bibinfo{year}{1999}).

\bibitem{Loreau2014b}
\bibinfo{author}{Loreau, J.}, \bibinfo{author}{Liévin, J.},
  \bibinfo{author}{Scribano, Y.} \& \bibinfo{author}{van~der Avoird, A.}
\newblock \bibinfo{title}{Potential energy surface and bound states of the
  {NH$_3$–Ar} and {ND$_3$–Ar} complexes}.
\newblock \emph{\bibinfo{journal}{J. Chem. Phys.}}
  \textbf{\bibinfo{volume}{141}}, \bibinfo{pages}{224303}
  (\bibinfo{year}{2014}).

\bibitem{MRCC}
\bibinfo{author}{Kállay, M.} \emph{et~al.}
\newblock \bibinfo{title}{The {MRCC} program system: Accurate quantum chemistry
  from water to proteins}.
\newblock \emph{\bibinfo{journal}{J. Chem. Phys.}}
  \textbf{\bibinfo{volume}{152}}, \bibinfo{pages}{074107}
  (\bibinfo{year}{2020}).

\bibitem{Green1980}
\bibinfo{author}{Green, S.}
\newblock \bibinfo{title}{Energy transfer in {NH$_3$}–{H}e collisions}.
\newblock \emph{\bibinfo{journal}{J. Chem. Phys.}}
  \textbf{\bibinfo{volume}{73}}, \bibinfo{pages}{2740--2750}
  (\bibinfo{year}{1980}).

\end{thebibliography}


\begin{thebibliography}{79}%
\makeatletter
\providecommand \@ifxundefined [1]{%
 \@ifx{#1\undefined}
}%
\providecommand \@ifnum [1]{%
 \ifnum #1\expandafter \@firstoftwo
 \else \expandafter \@secondoftwo
 \fi
}%
\providecommand \@ifx [1]{%
 \ifx #1\expandafter \@firstoftwo
 \else \expandafter \@secondoftwo
 \fi
}%
\providecommand \natexlab [1]{#1}%
\providecommand \enquote  [1]{``#1''}%
\providecommand \bibnamefont  [1]{#1}%
\providecommand \bibfnamefont [1]{#1}%
\providecommand \citenamefont [1]{#1}%
\providecommand \href@noop [0]{\@secondoftwo}%
\providecommand \href [0]{\begingroup \@sanitize@url \@href}%
\providecommand \@href[1]{\@@startlink{#1}\@@href}%
\providecommand \@@href[1]{\endgroup#1\@@endlink}%
\providecommand \@sanitize@url [0]{\catcode `\\12\catcode `\$12\catcode
  `\&12\catcode `\#12\catcode `\^12\catcode `\_12\catcode `\%12\relax}%
\providecommand \@@startlink[1]{}%
\providecommand \@@endlink[0]{}%
\providecommand \url  [0]{\begingroup\@sanitize@url \@url }%
\providecommand \@url [1]{\endgroup\@href {#1}{\urlprefix }}%
\providecommand \urlprefix  [0]{URL }%
\providecommand \Eprint [0]{\href }%
\providecommand \doibase [0]{http://dx.doi.org/}%
\providecommand \selectlanguage [0]{\@gobble}%
\providecommand \bibinfo  [0]{\@secondoftwo}%
\providecommand \bibfield  [0]{\@secondoftwo}%
\providecommand \translation [1]{[#1]}%
\providecommand \BibitemOpen [0]{}%
\providecommand \bibitemStop [0]{}%
\providecommand \bibitemNoStop [0]{.\EOS\space}%
\providecommand \EOS [0]{\spacefactor3000\relax}%
\providecommand \BibitemShut  [1]{\csname bibitem#1\endcsname}%
\let\auto@bib@innerbib\@empty
\bibitem [{\citenamefont {Chandler}(2010)}]{Chandler:JCP132:110901}%
  \BibitemOpen
  \bibfield  {author} {\bibinfo {author} {\bibfnamefont {D.~W.}\ \bibnamefont
  {Chandler}},\ }\href {\doibase https://doi.org/10.1063/1.3357286} {\bibfield
  {journal} {\bibinfo  {journal} {J. Comp. Phys.}\ }\textbf {\bibinfo {volume}
  {132}},\ \bibinfo {pages} {110901} (\bibinfo {year} {2010})}\BibitemShut
  {NoStop}%
\bibitem [{\citenamefont {Schutte}\ \emph {et~al.}(1972)\citenamefont
  {Schutte}, \citenamefont {Bassi}, \citenamefont {Tommasini},\ and\
  \citenamefont {Scoles}}]{Schutte:PRL29:979}%
  \BibitemOpen
  \bibfield  {author} {\bibinfo {author} {\bibfnamefont {A.}~\bibnamefont
  {Schutte}}, \bibinfo {author} {\bibfnamefont {D.}~\bibnamefont {Bassi}},
  \bibinfo {author} {\bibfnamefont {F.}~\bibnamefont {Tommasini}}, \ and\
  \bibinfo {author} {\bibfnamefont {G.}~\bibnamefont {Scoles}},\ }\href
  {\doibase https://doi.org/10.1103/PhysRevLett.29.979} {\bibfield  {journal}
  {\bibinfo  {journal} {Phys. Rev. Lett.}\ }\textbf {\bibinfo {volume} {29}},\
  \bibinfo {pages} {979} (\bibinfo {year} {1972})}\BibitemShut {NoStop}%
\bibitem [{\citenamefont {Toennies}\ \emph {et~al.}(1979)\citenamefont
  {Toennies}, \citenamefont {Welz},\ and\ \citenamefont
  {Wolf}}]{Toennies:JCP71:614}%
  \BibitemOpen
  \bibfield  {author} {\bibinfo {author} {\bibfnamefont {J.~P.}\ \bibnamefont
  {Toennies}}, \bibinfo {author} {\bibfnamefont {W.}~\bibnamefont {Welz}}, \
  and\ \bibinfo {author} {\bibfnamefont {G.}~\bibnamefont {Wolf}},\ }\href
  {\doibase https://doi.org/10.1063/1.438414} {\bibfield  {journal} {\bibinfo
  {journal} {J. Comp. Phys.}\ }\textbf {\bibinfo {volume} {71}},\ \bibinfo
  {pages} {614} (\bibinfo {year} {1979})}\BibitemShut {NoStop}%
\bibitem [{\citenamefont {Lovejoy}\ and\ \citenamefont
  {Nesbitt}(1990)}]{Lovejoy1990}%
  \BibitemOpen
  \bibfield  {author} {\bibinfo {author} {\bibfnamefont {C.~M.}\ \bibnamefont
  {Lovejoy}}\ and\ \bibinfo {author} {\bibfnamefont {D.~J.}\ \bibnamefont
  {Nesbitt}},\ }\href {\doibase 10.1063/1.459663} {\bibfield  {journal}
  {\bibinfo  {journal} {J. Comp. Phys.}\ }\textbf {\bibinfo {volume} {93}},\
  \bibinfo {pages} {5387} (\bibinfo {year} {1990})}\BibitemShut {NoStop}%
\bibitem [{\citenamefont {Wang}\ \emph {et~al.}(2018)\citenamefont {Wang},
  \citenamefont {Yang}, \citenamefont {Xiao}, \citenamefont {Sun},
  \citenamefont {Zhang}, \citenamefont {Yang}, \citenamefont {Weichman},\ and\
  \citenamefont {Neumark}}]{Wang2018}%
  \BibitemOpen
  \bibfield  {author} {\bibinfo {author} {\bibfnamefont {T.}~\bibnamefont
  {Wang}}, \bibinfo {author} {\bibfnamefont {T.}~\bibnamefont {Yang}}, \bibinfo
  {author} {\bibfnamefont {C.}~\bibnamefont {Xiao}}, \bibinfo {author}
  {\bibfnamefont {Z.}~\bibnamefont {Sun}}, \bibinfo {author} {\bibfnamefont
  {D.}~\bibnamefont {Zhang}}, \bibinfo {author} {\bibfnamefont
  {X.}~\bibnamefont {Yang}}, \bibinfo {author} {\bibfnamefont {M.~L.}\
  \bibnamefont {Weichman}}, \ and\ \bibinfo {author} {\bibfnamefont {D.~M.}\
  \bibnamefont {Neumark}},\ }\href {\doibase 10.1039/c8cs00041g} {\bibfield
  {journal} {\bibinfo  {journal} {Chem. Soc. Rev.}\ }\textbf {\bibinfo {volume}
  {47}},\ \bibinfo {pages} {6744} (\bibinfo {year} {2018})}\BibitemShut
  {NoStop}%
\bibitem [{\citenamefont {Henson}\ \emph {et~al.}(2012)\citenamefont {Henson},
  \citenamefont {Gersten}, \citenamefont {Shagam}, \citenamefont {Narevicius},\
  and\ \citenamefont {Narevicius}}]{Henson:Science338:234}%
  \BibitemOpen
  \bibfield  {author} {\bibinfo {author} {\bibfnamefont {A.~B.}\ \bibnamefont
  {Henson}}, \bibinfo {author} {\bibfnamefont {S.}~\bibnamefont {Gersten}},
  \bibinfo {author} {\bibfnamefont {Y.}~\bibnamefont {Shagam}}, \bibinfo
  {author} {\bibfnamefont {J.}~\bibnamefont {Narevicius}}, \ and\ \bibinfo
  {author} {\bibfnamefont {E.}~\bibnamefont {Narevicius}},\ }\href {\doibase
  https://doi.org/10.1126/science.1229141} {\bibfield  {journal} {\bibinfo
  {journal} {Science}\ }\textbf {\bibinfo {volume} {338}},\ \bibinfo {pages}
  {234} (\bibinfo {year} {2012})}\BibitemShut {NoStop}%
\bibitem [{\citenamefont {Lavert-Ofir}\ \emph {et~al.}(2014)\citenamefont
  {Lavert-Ofir}, \citenamefont {Shagam}, \citenamefont {Henson}, \citenamefont
  {Gersten}, \citenamefont {K{\l}os}, \citenamefont {{\.Z}uchowski},
  \citenamefont {Narevicius},\ and\ \citenamefont
  {Narevicius}}]{Lavert-Ofir:NatChem6:332}%
  \BibitemOpen
  \bibfield  {author} {\bibinfo {author} {\bibfnamefont {E.}~\bibnamefont
  {Lavert-Ofir}}, \bibinfo {author} {\bibfnamefont {Y.}~\bibnamefont {Shagam}},
  \bibinfo {author} {\bibfnamefont {A.~B.}\ \bibnamefont {Henson}}, \bibinfo
  {author} {\bibfnamefont {S.}~\bibnamefont {Gersten}}, \bibinfo {author}
  {\bibfnamefont {J.}~\bibnamefont {K{\l}os}}, \bibinfo {author} {\bibfnamefont
  {P.~S.}\ \bibnamefont {{\.Z}uchowski}}, \bibinfo {author} {\bibfnamefont
  {J.}~\bibnamefont {Narevicius}}, \ and\ \bibinfo {author} {\bibfnamefont
  {E.}~\bibnamefont {Narevicius}},\ }\href {\doibase
  https://doi.org/10.1038/nchem.1857} {\bibfield  {journal} {\bibinfo
  {journal} {Nat. Chem.}\ }\textbf {\bibinfo {volume} {6}},\ \bibinfo {pages}
  {332} (\bibinfo {year} {2014})}\BibitemShut {NoStop}%
\bibitem [{\citenamefont {Jankunas}\ \emph {et~al.}(2015)\citenamefont
  {Jankunas}, \citenamefont {Jachymski}, \citenamefont {Hapka},\ and\
  \citenamefont {Osterwalder}}]{Jankunas:JCP142:164305}%
  \BibitemOpen
  \bibfield  {author} {\bibinfo {author} {\bibfnamefont {J.}~\bibnamefont
  {Jankunas}}, \bibinfo {author} {\bibfnamefont {K.}~\bibnamefont {Jachymski}},
  \bibinfo {author} {\bibfnamefont {M.}~\bibnamefont {Hapka}}, \ and\ \bibinfo
  {author} {\bibfnamefont {A.}~\bibnamefont {Osterwalder}},\ }\href {\doibase
  10.1063/1.4919369} {\bibfield  {journal} {\bibinfo  {journal} {J. Comp.
  Phys.}\ }\textbf {\bibinfo {volume} {142}},\ \bibinfo {pages} {164305}
  (\bibinfo {year} {2015})}\BibitemShut {NoStop}%
\bibitem [{\citenamefont {Shagam}\ \emph {et~al.}(2015)\citenamefont {Shagam},
  \citenamefont {Klein}, \citenamefont {Skomorowski}, \citenamefont {Yun},
  \citenamefont {Averbukh}, \citenamefont {Koch},\ and\ \citenamefont
  {Narevicius}}]{Shagam2015}%
  \BibitemOpen
  \bibfield  {author} {\bibinfo {author} {\bibfnamefont {Y.}~\bibnamefont
  {Shagam}}, \bibinfo {author} {\bibfnamefont {A.}~\bibnamefont {Klein}},
  \bibinfo {author} {\bibfnamefont {W.}~\bibnamefont {Skomorowski}}, \bibinfo
  {author} {\bibfnamefont {R.}~\bibnamefont {Yun}}, \bibinfo {author}
  {\bibfnamefont {V.}~\bibnamefont {Averbukh}}, \bibinfo {author}
  {\bibfnamefont {C.~P.}\ \bibnamefont {Koch}}, \ and\ \bibinfo {author}
  {\bibfnamefont {E.}~\bibnamefont {Narevicius}},\ }\href {\doibase
  http://dx.doi.org/10.1038/nchem.2359} {\bibfield  {journal} {\bibinfo
  {journal} {Nat. Chem.}\ }\textbf {\bibinfo {volume} {7}},\ \bibinfo {pages}
  {921} (\bibinfo {year} {2015})}\BibitemShut {NoStop}%
\bibitem [{\citenamefont {Klein}\ \emph {et~al.}(2017)\citenamefont {Klein},
  \citenamefont {Shagam}, \citenamefont {Skomorowski}, \citenamefont
  {{\.Z}uchowski}, \citenamefont {Pawlak}, \citenamefont {Janssen},
  \citenamefont {Moiseyev}, \citenamefont {van~de Meerakker}, \citenamefont
  {van~der Avoird}, \citenamefont {Koch},\ and\ \citenamefont
  {Narevicius}}]{Klein:NatPhys13:35}%
  \BibitemOpen
  \bibfield  {author} {\bibinfo {author} {\bibfnamefont {A.}~\bibnamefont
  {Klein}}, \bibinfo {author} {\bibfnamefont {Y.}~\bibnamefont {Shagam}},
  \bibinfo {author} {\bibfnamefont {W.}~\bibnamefont {Skomorowski}}, \bibinfo
  {author} {\bibfnamefont {P.~S.}\ \bibnamefont {{\.Z}uchowski}}, \bibinfo
  {author} {\bibfnamefont {M.}~\bibnamefont {Pawlak}}, \bibinfo {author}
  {\bibfnamefont {L.~M.}\ \bibnamefont {Janssen}}, \bibinfo {author}
  {\bibfnamefont {N.}~\bibnamefont {Moiseyev}}, \bibinfo {author}
  {\bibfnamefont {S.~Y.}\ \bibnamefont {van~de Meerakker}}, \bibinfo {author}
  {\bibfnamefont {A.}~\bibnamefont {van~der Avoird}}, \bibinfo {author}
  {\bibfnamefont {C.~P.}\ \bibnamefont {Koch}}, \ and\ \bibinfo {author}
  {\bibfnamefont {E.}~\bibnamefont {Narevicius}},\ }\href {\doibase
  https://doi.org/10.1038/nphys3904} {\bibfield  {journal} {\bibinfo  {journal}
  {Nat. Phys.}\ }\textbf {\bibinfo {volume} {13}},\ \bibinfo {pages} {35}
  (\bibinfo {year} {2017})}\BibitemShut {NoStop}%
\bibitem [{\citenamefont {Bibelnik}\ \emph {et~al.}(2019)\citenamefont
  {Bibelnik}, \citenamefont {Gersten}, \citenamefont {Henson}, \citenamefont
  {Lavert-Ofir}, \citenamefont {Shagam}, \citenamefont {Skomorowski},
  \citenamefont {Koch},\ and\ \citenamefont {Narevicius}}]{Bibelnik2019}%
  \BibitemOpen
  \bibfield  {author} {\bibinfo {author} {\bibfnamefont {N.}~\bibnamefont
  {Bibelnik}}, \bibinfo {author} {\bibfnamefont {S.}~\bibnamefont {Gersten}},
  \bibinfo {author} {\bibfnamefont {A.~B.}\ \bibnamefont {Henson}}, \bibinfo
  {author} {\bibfnamefont {E.}~\bibnamefont {Lavert-Ofir}}, \bibinfo {author}
  {\bibfnamefont {Y.}~\bibnamefont {Shagam}}, \bibinfo {author} {\bibfnamefont
  {W.}~\bibnamefont {Skomorowski}}, \bibinfo {author} {\bibfnamefont {C.~P.}\
  \bibnamefont {Koch}}, \ and\ \bibinfo {author} {\bibfnamefont
  {E.}~\bibnamefont {Narevicius}},\ }\href {\doibase
  https://doi.org/10.1080/00268976.2019.1594421} {\bibfield  {journal}
  {\bibinfo  {journal} {Mol. Phys.}\ }\textbf {\bibinfo {volume} {117}},\
  \bibinfo {pages} {2128} (\bibinfo {year} {2019})}\BibitemShut {NoStop}%
\bibitem [{\citenamefont {Margulis}\ \emph {et~al.}(2022)\citenamefont
  {Margulis}, \citenamefont {Paliwal}, \citenamefont {Skomorowski},
  \citenamefont {Pawlak}, \citenamefont {{\.Z}uchowski},\ and\ \citenamefont
  {Narevicius}}]{Margulis2022}%
  \BibitemOpen
  \bibfield  {author} {\bibinfo {author} {\bibfnamefont {B.}~\bibnamefont
  {Margulis}}, \bibinfo {author} {\bibfnamefont {P.}~\bibnamefont {Paliwal}},
  \bibinfo {author} {\bibfnamefont {W.}~\bibnamefont {Skomorowski}}, \bibinfo
  {author} {\bibfnamefont {M.}~\bibnamefont {Pawlak}}, \bibinfo {author}
  {\bibfnamefont {P.~S.}\ \bibnamefont {{\.Z}uchowski}}, \ and\ \bibinfo
  {author} {\bibfnamefont {E.}~\bibnamefont {Narevicius}},\ }\href {\doibase
  https://doi.org/10.1103/PhysRevResearch.4.043042} {\bibfield  {journal}
  {\bibinfo  {journal} {Phys. Rev. Research}\ }\textbf {\bibinfo {volume}
  {4}},\ \bibinfo {pages} {043042} (\bibinfo {year} {2022})}\BibitemShut
  {NoStop}%
\bibitem [{\citenamefont {Chefdeville}\ \emph {et~al.}(2012)\citenamefont
  {Chefdeville}, \citenamefont {Stoecklin}, \citenamefont {Bergeat},
  \citenamefont {Hickson}, \citenamefont {Naulin},\ and\ \citenamefont
  {Costes}}]{Chefdeville:PRL109:023201}%
  \BibitemOpen
  \bibfield  {author} {\bibinfo {author} {\bibfnamefont {S.}~\bibnamefont
  {Chefdeville}}, \bibinfo {author} {\bibfnamefont {T.}~\bibnamefont
  {Stoecklin}}, \bibinfo {author} {\bibfnamefont {A.}~\bibnamefont {Bergeat}},
  \bibinfo {author} {\bibfnamefont {K.~M.}\ \bibnamefont {Hickson}}, \bibinfo
  {author} {\bibfnamefont {C.}~\bibnamefont {Naulin}}, \ and\ \bibinfo {author}
  {\bibfnamefont {M.}~\bibnamefont {Costes}},\ }\href {\doibase
  https://doi.org/10.1103/PhysRevLett.109.023201} {\bibfield  {journal}
  {\bibinfo  {journal} {Phys. Rev. Lett.}\ }\textbf {\bibinfo {volume} {109}},\
  \bibinfo {pages} {023201} (\bibinfo {year} {2012})}\BibitemShut {NoStop}%
\bibitem [{\citenamefont {Chefdeville}\ \emph {et~al.}(2015)\citenamefont
  {Chefdeville}, \citenamefont {Stoecklin}, \citenamefont {Naulin},
  \citenamefont {Jankowski}, \citenamefont {Szalewicz}, \citenamefont {Faure},
  \citenamefont {Costes},\ and\ \citenamefont
  {Bergeat}}]{Chefdeville:AJL799:L9}%
  \BibitemOpen
  \bibfield  {author} {\bibinfo {author} {\bibfnamefont {S.}~\bibnamefont
  {Chefdeville}}, \bibinfo {author} {\bibfnamefont {T.}~\bibnamefont
  {Stoecklin}}, \bibinfo {author} {\bibfnamefont {C.}~\bibnamefont {Naulin}},
  \bibinfo {author} {\bibfnamefont {P.}~\bibnamefont {Jankowski}}, \bibinfo
  {author} {\bibfnamefont {K.}~\bibnamefont {Szalewicz}}, \bibinfo {author}
  {\bibfnamefont {A.}~\bibnamefont {Faure}}, \bibinfo {author} {\bibfnamefont
  {M.}~\bibnamefont {Costes}}, \ and\ \bibinfo {author} {\bibfnamefont
  {A.}~\bibnamefont {Bergeat}},\ }\href {\doibase
  http://dx.doi.org/10.1088/2041-8205/799/1/L9} {\bibfield  {journal} {\bibinfo
   {journal} {Astr. J. Lett.}\ }\textbf {\bibinfo {volume} {799}},\ \bibinfo
  {pages} {L9} (\bibinfo {year} {2015})}\BibitemShut {NoStop}%
\bibitem [{\citenamefont {Chefdeville}\ \emph {et~al.}(2013)\citenamefont
  {Chefdeville}, \citenamefont {Kalugina}, \citenamefont {van~de Meerakker},
  \citenamefont {Naulin}, \citenamefont {Lique},\ and\ \citenamefont
  {Costes}}]{Chefdeville:Science341:06092013}%
  \BibitemOpen
  \bibfield  {author} {\bibinfo {author} {\bibfnamefont {S.}~\bibnamefont
  {Chefdeville}}, \bibinfo {author} {\bibfnamefont {Y.}~\bibnamefont
  {Kalugina}}, \bibinfo {author} {\bibfnamefont {S.~Y.~T.}\ \bibnamefont
  {van~de Meerakker}}, \bibinfo {author} {\bibfnamefont {C.}~\bibnamefont
  {Naulin}}, \bibinfo {author} {\bibfnamefont {F.}~\bibnamefont {Lique}}, \
  and\ \bibinfo {author} {\bibfnamefont {M.}~\bibnamefont {Costes}},\ }\href
  {\doibase https://doi.org/10.1126/science.1241395} {\bibfield  {journal}
  {\bibinfo  {journal} {Science}\ }\textbf {\bibinfo {volume} {341}},\ \bibinfo
  {pages} {1094} (\bibinfo {year} {2013})}\BibitemShut {NoStop}%
\bibitem [{\citenamefont {Bergeat}\ \emph {et~al.}(2015)\citenamefont
  {Bergeat}, \citenamefont {Onvlee}, \citenamefont {Naulin}, \citenamefont
  {van~der Avoird},\ and\ \citenamefont {Costes}}]{Bergeat:NatChem7:349}%
  \BibitemOpen
  \bibfield  {author} {\bibinfo {author} {\bibfnamefont {A.}~\bibnamefont
  {Bergeat}}, \bibinfo {author} {\bibfnamefont {J.}~\bibnamefont {Onvlee}},
  \bibinfo {author} {\bibfnamefont {C.}~\bibnamefont {Naulin}}, \bibinfo
  {author} {\bibfnamefont {A.}~\bibnamefont {van~der Avoird}}, \ and\ \bibinfo
  {author} {\bibfnamefont {M.}~\bibnamefont {Costes}},\ }\href {\doibase
  https://doi.org/10.1038/nchem.2204} {\bibfield  {journal} {\bibinfo
  {journal} {Nat. Chem.}\ }\textbf {\bibinfo {volume} {7}},\ \bibinfo {pages}
  {349} (\bibinfo {year} {2015})}\BibitemShut {NoStop}%
\bibitem [{\citenamefont {Kłos}\ \emph {et~al.}(2018)\citenamefont {Kłos},
  \citenamefont {Bergeat}, \citenamefont {Vanuzzo}, \citenamefont {Morales},
  \citenamefont {Naulin},\ and\ \citenamefont {Lique}}]{Klos2018}%
  \BibitemOpen
  \bibfield  {author} {\bibinfo {author} {\bibfnamefont {J.}~\bibnamefont
  {Kłos}}, \bibinfo {author} {\bibfnamefont {A.}~\bibnamefont {Bergeat}},
  \bibinfo {author} {\bibfnamefont {G.}~\bibnamefont {Vanuzzo}}, \bibinfo
  {author} {\bibfnamefont {S.~B.}\ \bibnamefont {Morales}}, \bibinfo {author}
  {\bibfnamefont {C.}~\bibnamefont {Naulin}}, \ and\ \bibinfo {author}
  {\bibfnamefont {F.}~\bibnamefont {Lique}},\ }\href {\doibase
  http://dx.doi.org/10.1021/acs.jpclett.8b03025} {\bibfield  {journal}
  {\bibinfo  {journal} {J. Phys. Chem. Lett.}\ }\textbf {\bibinfo {volume}
  {9}},\ \bibinfo {pages} {6496} (\bibinfo {year} {2018})}\BibitemShut
  {NoStop}%
\bibitem [{\citenamefont {Bergeat}\ \emph {et~al.}(2018)\citenamefont
  {Bergeat}, \citenamefont {Chefdeville}, \citenamefont {Costes}, \citenamefont
  {Morales}, \citenamefont {Naulin}, \citenamefont {Even}, \citenamefont
  {K{\l}os},\ and\ \citenamefont {Lique}}]{Bergeat:NatChem10:519}%
  \BibitemOpen
  \bibfield  {author} {\bibinfo {author} {\bibfnamefont {A.}~\bibnamefont
  {Bergeat}}, \bibinfo {author} {\bibfnamefont {S.}~\bibnamefont
  {Chefdeville}}, \bibinfo {author} {\bibfnamefont {M.}~\bibnamefont {Costes}},
  \bibinfo {author} {\bibfnamefont {S.~B.}\ \bibnamefont {Morales}}, \bibinfo
  {author} {\bibfnamefont {C.}~\bibnamefont {Naulin}}, \bibinfo {author}
  {\bibfnamefont {U.}~\bibnamefont {Even}}, \bibinfo {author} {\bibfnamefont
  {J.}~\bibnamefont {K{\l}os}}, \ and\ \bibinfo {author} {\bibfnamefont
  {F.}~\bibnamefont {Lique}},\ }\href {\doibase
  https://doi.org/10.1038/s41557-018-0030-y} {\bibfield  {journal} {\bibinfo
  {journal} {Nat. Chem.}\ }\textbf {\bibinfo {volume} {10}},\ \bibinfo {pages}
  {519} (\bibinfo {year} {2018})}\BibitemShut {NoStop}%
\bibitem [{\citenamefont {Bergeat}\ \emph {et~al.}(2019)\citenamefont
  {Bergeat}, \citenamefont {Morales}, \citenamefont {Naulin}, \citenamefont
  {Kłos},\ and\ \citenamefont {Lique}}]{Bergeat2019}%
  \BibitemOpen
  \bibfield  {author} {\bibinfo {author} {\bibfnamefont {A.}~\bibnamefont
  {Bergeat}}, \bibinfo {author} {\bibfnamefont {S.~B.}\ \bibnamefont
  {Morales}}, \bibinfo {author} {\bibfnamefont {C.}~\bibnamefont {Naulin}},
  \bibinfo {author} {\bibfnamefont {J.}~\bibnamefont {Kłos}}, \ and\ \bibinfo
  {author} {\bibfnamefont {F.}~\bibnamefont {Lique}},\ }\href {\doibase
  https://doi.org/10.3389/fchem.2019.00164} {\bibfield  {journal} {\bibinfo
  {journal} {Front. Chem.}\ }\textbf {\bibinfo {volume} {7}},\ \bibinfo {pages}
  {164} (\bibinfo {year} {2019})}\BibitemShut {NoStop}%
\bibitem [{\citenamefont {Bergeat}\ \emph {et~al.}(2020)\citenamefont
  {Bergeat}, \citenamefont {Morales}, \citenamefont {Naulin}, \citenamefont
  {Wiesenfeld},\ and\ \citenamefont {Faure}}]{Bergeat2020}%
  \BibitemOpen
  \bibfield  {author} {\bibinfo {author} {\bibfnamefont {A.}~\bibnamefont
  {Bergeat}}, \bibinfo {author} {\bibfnamefont {S.~B.}\ \bibnamefont
  {Morales}}, \bibinfo {author} {\bibfnamefont {C.}~\bibnamefont {Naulin}},
  \bibinfo {author} {\bibfnamefont {L.}~\bibnamefont {Wiesenfeld}}, \ and\
  \bibinfo {author} {\bibfnamefont {A.}~\bibnamefont {Faure}},\ }\href
  {\doibase https://doi.org/10.1103/PhysRevLett.125.143402} {\bibfield
  {journal} {\bibinfo  {journal} {Phys. Rev. Lett.}\ }\textbf {\bibinfo
  {volume} {125}},\ \bibinfo {pages} {143402} (\bibinfo {year}
  {2020})}\BibitemShut {NoStop}%
\bibitem [{\citenamefont {Bergeat}\ \emph {et~al.}(2022)\citenamefont
  {Bergeat}, \citenamefont {Faure}, \citenamefont {Wiesenfeld}, \citenamefont
  {Miossec}, \citenamefont {Morales},\ and\ \citenamefont
  {Naulin}}]{Bergeat2022}%
  \BibitemOpen
  \bibfield  {author} {\bibinfo {author} {\bibfnamefont {A.}~\bibnamefont
  {Bergeat}}, \bibinfo {author} {\bibfnamefont {A.}~\bibnamefont {Faure}},
  \bibinfo {author} {\bibfnamefont {L.}~\bibnamefont {Wiesenfeld}}, \bibinfo
  {author} {\bibfnamefont {C.}~\bibnamefont {Miossec}}, \bibinfo {author}
  {\bibfnamefont {S.~B.}\ \bibnamefont {Morales}}, \ and\ \bibinfo {author}
  {\bibfnamefont {C.}~\bibnamefont {Naulin}},\ }\href {\doibase
  https://doi.org/10.3390/molecules27217535} {\bibfield  {journal} {\bibinfo
  {journal} {Molecules}\ }\textbf {\bibinfo {volume} {27}},\ \bibinfo {pages}
  {7535} (\bibinfo {year} {2022})}\BibitemShut {NoStop}%
\bibitem [{\citenamefont {Vogels}\ \emph {et~al.}(2015)\citenamefont {Vogels},
  \citenamefont {Onvlee}, \citenamefont {Chefdeville}, \citenamefont {van~der
  Avoird}, \citenamefont {Groenenboom},\ and\ \citenamefont {van~de
  Meerakker}}]{Vogels:Science350:787}%
  \BibitemOpen
  \bibfield  {author} {\bibinfo {author} {\bibfnamefont {S.~N.}\ \bibnamefont
  {Vogels}}, \bibinfo {author} {\bibfnamefont {J.}~\bibnamefont {Onvlee}},
  \bibinfo {author} {\bibfnamefont {S.}~\bibnamefont {Chefdeville}}, \bibinfo
  {author} {\bibfnamefont {A.}~\bibnamefont {van~der Avoird}}, \bibinfo
  {author} {\bibfnamefont {G.~C.}\ \bibnamefont {Groenenboom}}, \ and\ \bibinfo
  {author} {\bibfnamefont {S.~Y.~T.}\ \bibnamefont {van~de Meerakker}},\ }\href
  {\doibase https://doi.org/10.1126/science.aad2356} {\bibfield  {journal}
  {\bibinfo  {journal} {Science}\ }\textbf {\bibinfo {volume} {350}},\ \bibinfo
  {pages} {787} (\bibinfo {year} {2015})}\BibitemShut {NoStop}%
\bibitem [{\citenamefont {Vogels}\ \emph {et~al.}(2018)\citenamefont {Vogels},
  \citenamefont {Karman}, \citenamefont {K{\l}os}, \citenamefont {Besemer},
  \citenamefont {Onvlee}, \citenamefont {van~der Avoird}, \citenamefont
  {Groenenboom},\ and\ \citenamefont {van~de
  Meerakker}}]{Vogels:NatChem10:435}%
  \BibitemOpen
  \bibfield  {author} {\bibinfo {author} {\bibfnamefont {S.~N.}\ \bibnamefont
  {Vogels}}, \bibinfo {author} {\bibfnamefont {T.}~\bibnamefont {Karman}},
  \bibinfo {author} {\bibfnamefont {J.}~\bibnamefont {K{\l}os}}, \bibinfo
  {author} {\bibfnamefont {M.}~\bibnamefont {Besemer}}, \bibinfo {author}
  {\bibfnamefont {J.}~\bibnamefont {Onvlee}}, \bibinfo {author} {\bibfnamefont
  {A.}~\bibnamefont {van~der Avoird}}, \bibinfo {author} {\bibfnamefont
  {G.~C.}\ \bibnamefont {Groenenboom}}, \ and\ \bibinfo {author} {\bibfnamefont
  {S.~Y.~T.}\ \bibnamefont {van~de Meerakker}},\ }\href {\doibase
  https://doi.org/10.1038/s41557-018-0001-3} {\bibfield  {journal} {\bibinfo
  {journal} {Nat. Chem.}\ }\textbf {\bibinfo {volume} {10}},\ \bibinfo {pages}
  {435} (\bibinfo {year} {2018})}\BibitemShut {NoStop}%
\bibitem [{\citenamefont {de~Jongh}\ \emph {et~al.}(2020)\citenamefont
  {de~Jongh}, \citenamefont {Besemer}, \citenamefont {Shuai}, \citenamefont
  {Karman}, \citenamefont {van~der Avoird}, \citenamefont {Groenenboom},\ and\
  \citenamefont {van~de Meerakker}}]{Jongh2020}%
  \BibitemOpen
  \bibfield  {author} {\bibinfo {author} {\bibfnamefont {T.}~\bibnamefont
  {de~Jongh}}, \bibinfo {author} {\bibfnamefont {M.}~\bibnamefont {Besemer}},
  \bibinfo {author} {\bibfnamefont {Q.}~\bibnamefont {Shuai}}, \bibinfo
  {author} {\bibfnamefont {T.}~\bibnamefont {Karman}}, \bibinfo {author}
  {\bibfnamefont {A.}~\bibnamefont {van~der Avoird}}, \bibinfo {author}
  {\bibfnamefont {G.~C.}\ \bibnamefont {Groenenboom}}, \ and\ \bibinfo {author}
  {\bibfnamefont {S.~Y.~T.}\ \bibnamefont {van~de Meerakker}},\ }\href
  {\doibase https://doi.org/10.1126/science.aba3990} {\bibfield  {journal}
  {\bibinfo  {journal} {Science}\ }\textbf {\bibinfo {volume} {368}},\ \bibinfo
  {pages} {626} (\bibinfo {year} {2020})}\BibitemShut {NoStop}%
\bibitem [{\citenamefont {Shuai}\ \emph {et~al.}(2020)\citenamefont {Shuai},
  \citenamefont {de~Jongh}, \citenamefont {Besemer}, \citenamefont {van~der
  Avoird}, \citenamefont {Groenenboom},\ and\ \citenamefont {van~de
  Meerakker}}]{Shuai2020}%
  \BibitemOpen
  \bibfield  {author} {\bibinfo {author} {\bibfnamefont {Q.}~\bibnamefont
  {Shuai}}, \bibinfo {author} {\bibfnamefont {T.}~\bibnamefont {de~Jongh}},
  \bibinfo {author} {\bibfnamefont {M.}~\bibnamefont {Besemer}}, \bibinfo
  {author} {\bibfnamefont {A.}~\bibnamefont {van~der Avoird}}, \bibinfo
  {author} {\bibfnamefont {G.~C.}\ \bibnamefont {Groenenboom}}, \ and\ \bibinfo
  {author} {\bibfnamefont {S.~Y.~T.}\ \bibnamefont {van~de Meerakker}},\ }\href
  {\doibase https://doi.org/10.1063/5.0033488} {\bibfield  {journal} {\bibinfo
  {journal} {J. Chem. Phys.}\ }\textbf {\bibinfo {volume} {153}},\ \bibinfo
  {pages} {244302} (\bibinfo {year} {2020})}\BibitemShut {NoStop}%
\bibitem [{\citenamefont {de~Jongh}\ \emph {et~al.}(2022)\citenamefont
  {de~Jongh}, \citenamefont {Shuai}, \citenamefont {Abma}, \citenamefont
  {Kuijpers}, \citenamefont {Besemer}, \citenamefont {van~der Avoird},
  \citenamefont {Groenenboom},\ and\ \citenamefont {van~de
  Meerakker}}]{Jongh2022}%
  \BibitemOpen
  \bibfield  {author} {\bibinfo {author} {\bibfnamefont {T.}~\bibnamefont
  {de~Jongh}}, \bibinfo {author} {\bibfnamefont {Q.}~\bibnamefont {Shuai}},
  \bibinfo {author} {\bibfnamefont {G.~L.}\ \bibnamefont {Abma}}, \bibinfo
  {author} {\bibfnamefont {S.}~\bibnamefont {Kuijpers}}, \bibinfo {author}
  {\bibfnamefont {M.}~\bibnamefont {Besemer}}, \bibinfo {author} {\bibfnamefont
  {A.}~\bibnamefont {van~der Avoird}}, \bibinfo {author} {\bibfnamefont
  {G.~C.}\ \bibnamefont {Groenenboom}}, \ and\ \bibinfo {author} {\bibfnamefont
  {S.~Y.~T.}\ \bibnamefont {van~de Meerakker}},\ }\href {\doibase
  https://doi.org/10.17026/dans-x8q-pcuk} {\bibfield  {journal} {\bibinfo
  {journal} {Nat. Chem.}\ }\textbf {\bibinfo {volume} {14}},\ \bibinfo {pages}
  {538} (\bibinfo {year} {2022})}\BibitemShut {NoStop}%
\bibitem [{\citenamefont {Paliwal}\ \emph {et~al.}(2021)\citenamefont
  {Paliwal}, \citenamefont {Deb}, \citenamefont {Reich}, \citenamefont
  {Avoird}, \citenamefont {Koch},\ and\ \citenamefont
  {Narevicius}}]{Paliwal2021}%
  \BibitemOpen
  \bibfield  {author} {\bibinfo {author} {\bibfnamefont {P.}~\bibnamefont
  {Paliwal}}, \bibinfo {author} {\bibfnamefont {N.}~\bibnamefont {Deb}},
  \bibinfo {author} {\bibfnamefont {D.~M.}\ \bibnamefont {Reich}}, \bibinfo
  {author} {\bibfnamefont {A.~v.~d.}\ \bibnamefont {Avoird}}, \bibinfo {author}
  {\bibfnamefont {C.~P.}\ \bibnamefont {Koch}}, \ and\ \bibinfo {author}
  {\bibfnamefont {E.}~\bibnamefont {Narevicius}},\ }\href {\doibase
  https://doi.org/10.1038/s41557-020-00578-x} {\bibfield  {journal} {\bibinfo
  {journal} {Nat. Chem.}\ }\textbf {\bibinfo {volume} {13}},\ \bibinfo {pages}
  {94} (\bibinfo {year} {2021})}\BibitemShut {NoStop}%
\bibitem [{\citenamefont {Plomp}\ \emph {et~al.}(2024)\citenamefont {Plomp},
  \citenamefont {Wang}, \citenamefont {Kłos}, \citenamefont {Dagdigian},
  \citenamefont {Lique}, \citenamefont {Onvlee},\ and\ \citenamefont {van~de
  Meerakker}}]{Plomp2024}%
  \BibitemOpen
  \bibfield  {author} {\bibinfo {author} {\bibfnamefont {V.}~\bibnamefont
  {Plomp}}, \bibinfo {author} {\bibfnamefont {X.-D.}\ \bibnamefont {Wang}},
  \bibinfo {author} {\bibfnamefont {J.}~\bibnamefont {Kłos}}, \bibinfo
  {author} {\bibfnamefont {P.~J.}\ \bibnamefont {Dagdigian}}, \bibinfo {author}
  {\bibfnamefont {F.}~\bibnamefont {Lique}}, \bibinfo {author} {\bibfnamefont
  {J.}~\bibnamefont {Onvlee}}, \ and\ \bibinfo {author} {\bibfnamefont {S.~Y.}\
  \bibnamefont {van~de Meerakker}},\ }\href {\doibase
  https://doi.org/10.1021/acs.jpclett.3c03379} {\bibfield  {journal} {\bibinfo
  {journal} {J. Phys. Chem. Lett.}\ }\textbf {\bibinfo {volume} {15}},\
  \bibinfo {pages} {4602} (\bibinfo {year} {2024})}\BibitemShut {NoStop}%
\bibitem [{\citenamefont {van~de Meerakker}\ \emph {et~al.}(2012)\citenamefont
  {van~de Meerakker}, \citenamefont {Bethlem}, \citenamefont {Vanhaecke},\ and\
  \citenamefont {Meijer}}]{Meerakker:CR112:4828}%
  \BibitemOpen
  \bibfield  {author} {\bibinfo {author} {\bibfnamefont {S.~Y.~T.}\
  \bibnamefont {van~de Meerakker}}, \bibinfo {author} {\bibfnamefont {H.~L.}\
  \bibnamefont {Bethlem}}, \bibinfo {author} {\bibfnamefont {N.}~\bibnamefont
  {Vanhaecke}}, \ and\ \bibinfo {author} {\bibfnamefont {G.}~\bibnamefont
  {Meijer}},\ }\href {\doibase https://doi.org/10.1021/cr200349r} {\bibfield
  {journal} {\bibinfo  {journal} {Chem. Rev.}\ }\textbf {\bibinfo {volume}
  {112}},\ \bibinfo {pages} {4828} (\bibinfo {year} {2012})}\BibitemShut
  {NoStop}%
\bibitem [{\citenamefont {Bethlem}\ \emph {et~al.}(2000)\citenamefont
  {Bethlem}, \citenamefont {Berden}, \citenamefont {Crompvoets}, \citenamefont
  {Jongma}, \citenamefont {van Roij},\ and\ \citenamefont
  {Meijer}}]{Bethlem:Nature406:491}%
  \BibitemOpen
  \bibfield  {author} {\bibinfo {author} {\bibfnamefont {H.~L.}\ \bibnamefont
  {Bethlem}}, \bibinfo {author} {\bibfnamefont {G.}~\bibnamefont {Berden}},
  \bibinfo {author} {\bibfnamefont {F.~M.~H.}\ \bibnamefont {Crompvoets}},
  \bibinfo {author} {\bibfnamefont {R.~T.}\ \bibnamefont {Jongma}}, \bibinfo
  {author} {\bibfnamefont {A.~J.~A.}\ \bibnamefont {van Roij}}, \ and\ \bibinfo
  {author} {\bibfnamefont {G.}~\bibnamefont {Meijer}},\ }\href {\doibase
  https://doi.org/10.1038/35020030} {\bibfield  {journal} {\bibinfo  {journal}
  {Nature}\ }\textbf {\bibinfo {volume} {406}},\ \bibinfo {pages} {491}
  (\bibinfo {year} {2000})}\BibitemShut {NoStop}%
\bibitem [{\citenamefont {Bethlem}\ \emph {et~al.}(2002)\citenamefont
  {Bethlem}, \citenamefont {Crompvoets}, \citenamefont {Jongma}, \citenamefont
  {van~de Meerakker},\ and\ \citenamefont {Meijer}}]{Bethlem:PRA65:053416}%
  \BibitemOpen
  \bibfield  {author} {\bibinfo {author} {\bibfnamefont {H.~L.}\ \bibnamefont
  {Bethlem}}, \bibinfo {author} {\bibfnamefont {F.~M.~H.}\ \bibnamefont
  {Crompvoets}}, \bibinfo {author} {\bibfnamefont {R.~T.}\ \bibnamefont
  {Jongma}}, \bibinfo {author} {\bibfnamefont {S.~Y.~T.}\ \bibnamefont {van~de
  Meerakker}}, \ and\ \bibinfo {author} {\bibfnamefont {G.}~\bibnamefont
  {Meijer}},\ }\href {\doibase https://doi.org/10.1103/PhysRevA.65.053416}
  {\bibfield  {journal} {\bibinfo  {journal} {Phys. Rev. A}\ }\textbf {\bibinfo
  {volume} {65}},\ \bibinfo {pages} {053416} (\bibinfo {year}
  {2002})}\BibitemShut {NoStop}%
\bibitem [{\citenamefont {van Veldhoven}\ \emph {et~al.}(2005)\citenamefont
  {van Veldhoven}, \citenamefont {Bethlem},\ and\ \citenamefont
  {Meijer}}]{Veldhoven:PRL94:083001}%
  \BibitemOpen
  \bibfield  {author} {\bibinfo {author} {\bibfnamefont {J.}~\bibnamefont {van
  Veldhoven}}, \bibinfo {author} {\bibfnamefont {H.~L.}\ \bibnamefont
  {Bethlem}}, \ and\ \bibinfo {author} {\bibfnamefont {G.}~\bibnamefont
  {Meijer}},\ }\href {\doibase https://doi.org/10.1103/PhysRevLett.94.083001}
  {\bibfield  {journal} {\bibinfo  {journal} {Phys. Rev. Lett.}\ }\textbf
  {\bibinfo {volume} {94}},\ \bibinfo {pages} {083001} (\bibinfo {year}
  {2005})}\BibitemShut {NoStop}%
\bibitem [{\citenamefont {Schnell}\ \emph {et~al.}(2007)\citenamefont
  {Schnell}, \citenamefont {L\"utzow}, \citenamefont {van Veldhoven},
  \citenamefont {Bethlem}, \citenamefont {K\"upper}, \citenamefont {Friedrich},
  \citenamefont {Schleier-Smith}, \citenamefont {Haak},\ and\ \citenamefont
  {Meijer}}]{Schnell:JPCA111:7411}%
  \BibitemOpen
  \bibfield  {author} {\bibinfo {author} {\bibfnamefont {M.}~\bibnamefont
  {Schnell}}, \bibinfo {author} {\bibfnamefont {P.}~\bibnamefont {L\"utzow}},
  \bibinfo {author} {\bibfnamefont {J.}~\bibnamefont {van Veldhoven}}, \bibinfo
  {author} {\bibfnamefont {H.}~\bibnamefont {Bethlem}}, \bibinfo {author}
  {\bibfnamefont {J.}~\bibnamefont {K\"upper}}, \bibinfo {author}
  {\bibfnamefont {B.}~\bibnamefont {Friedrich}}, \bibinfo {author}
  {\bibfnamefont {M.}~\bibnamefont {Schleier-Smith}}, \bibinfo {author}
  {\bibfnamefont {H.}~\bibnamefont {Haak}}, \ and\ \bibinfo {author}
  {\bibfnamefont {G.}~\bibnamefont {Meijer}},\ }\href {\doibase
  https://doi.org/10.1021/jp070902n} {\bibfield  {journal} {\bibinfo  {journal}
  {J. Phys. Chem. A}\ }\textbf {\bibinfo {volume} {111}},\ \bibinfo {pages}
  {7411} (\bibinfo {year} {2007})}\BibitemShut {NoStop}%
\bibitem [{\citenamefont {Crompvoets}\ \emph {et~al.}(2002)\citenamefont
  {Crompvoets}, \citenamefont {Jongma}, \citenamefont {Bethlem}, \citenamefont
  {van Roij},\ and\ \citenamefont {Meijer}}]{Crompvoets:PRL89:093004}%
  \BibitemOpen
  \bibfield  {author} {\bibinfo {author} {\bibfnamefont {F.~M.~H.}\
  \bibnamefont {Crompvoets}}, \bibinfo {author} {\bibfnamefont {R.~T.}\
  \bibnamefont {Jongma}}, \bibinfo {author} {\bibfnamefont {H.~L.}\
  \bibnamefont {Bethlem}}, \bibinfo {author} {\bibfnamefont {A.~J.~A.}\
  \bibnamefont {van Roij}}, \ and\ \bibinfo {author} {\bibfnamefont
  {G.}~\bibnamefont {Meijer}},\ }\href {\doibase
  https://doi.org/10.1103/PhysRevLett.89.093004} {\bibfield  {journal}
  {\bibinfo  {journal} {Phys. Rev. Lett.}\ }\textbf {\bibinfo {volume} {89}},\
  \bibinfo {pages} {093004} (\bibinfo {year} {2002})}\BibitemShut {NoStop}%
\bibitem [{\citenamefont {Schulz}\ \emph {et~al.}(2004)\citenamefont {Schulz},
  \citenamefont {Bethlem}, \citenamefont {van Veldhoven}, \citenamefont
  {K\"upper}, \citenamefont {Conrad},\ and\ \citenamefont
  {Meijer}}]{Schulz:PRL93:020406}%
  \BibitemOpen
  \bibfield  {author} {\bibinfo {author} {\bibfnamefont {S.~A.}\ \bibnamefont
  {Schulz}}, \bibinfo {author} {\bibfnamefont {H.~L.}\ \bibnamefont {Bethlem}},
  \bibinfo {author} {\bibfnamefont {J.}~\bibnamefont {van Veldhoven}}, \bibinfo
  {author} {\bibfnamefont {J.}~\bibnamefont {K\"upper}}, \bibinfo {author}
  {\bibfnamefont {H.}~\bibnamefont {Conrad}}, \ and\ \bibinfo {author}
  {\bibfnamefont {G.}~\bibnamefont {Meijer}},\ }\href {\doibase
  https://doi.org/10.1103/PhysRevLett.93.020406} {\bibfield  {journal}
  {\bibinfo  {journal} {Phys. Rev. Lett.}\ }\textbf {\bibinfo {volume} {93}},\
  \bibinfo {pages} {020406} (\bibinfo {year} {2004})}\BibitemShut {NoStop}%
\bibitem [{\citenamefont {Crompvoets}\ \emph {et~al.}(2001)\citenamefont
  {Crompvoets}, \citenamefont {Bethlem}, \citenamefont {Jongma},\ and\
  \citenamefont {Meijer}}]{Crompvoets:Nat411:174}%
  \BibitemOpen
  \bibfield  {author} {\bibinfo {author} {\bibfnamefont {F.~M.~H.}\
  \bibnamefont {Crompvoets}}, \bibinfo {author} {\bibfnamefont {H.~L.}\
  \bibnamefont {Bethlem}}, \bibinfo {author} {\bibfnamefont {R.~T.}\
  \bibnamefont {Jongma}}, \ and\ \bibinfo {author} {\bibfnamefont
  {G.}~\bibnamefont {Meijer}},\ }\href {\doibase
  https://doi.org/10.1038/35075537} {\bibfield  {journal} {\bibinfo  {journal}
  {Nature}\ }\textbf {\bibinfo {volume} {411}},\ \bibinfo {pages} {174}
  (\bibinfo {year} {2001})}\BibitemShut {NoStop}%
\bibitem [{\citenamefont {Heiner}\ \emph {et~al.}(2007)\citenamefont {Heiner},
  \citenamefont {Carty}, \citenamefont {Meijer},\ and\ \citenamefont
  {Bethlem}}]{Heiner:NatPhys3:115}%
  \BibitemOpen
  \bibfield  {author} {\bibinfo {author} {\bibfnamefont {C.~E.}\ \bibnamefont
  {Heiner}}, \bibinfo {author} {\bibfnamefont {D.}~\bibnamefont {Carty}},
  \bibinfo {author} {\bibfnamefont {G.}~\bibnamefont {Meijer}}, \ and\ \bibinfo
  {author} {\bibfnamefont {H.~L.}\ \bibnamefont {Bethlem}},\ }\href {\doibase
  https://doi.org/10.1038/nphys513} {\bibfield  {journal} {\bibinfo  {journal}
  {Nat. Phys.}\ }\textbf {\bibinfo {volume} {3}},\ \bibinfo {pages} {115}
  (\bibinfo {year} {2007})}\BibitemShut {NoStop}%
\bibitem [{\citenamefont {Zieger}\ \emph {et~al.}(2010)\citenamefont {Zieger},
  \citenamefont {van~de Meerakker}, \citenamefont {Heiner}, \citenamefont
  {Bethlem}, \citenamefont {van Roij},\ and\ \citenamefont
  {Meijer}}]{Zieger2010}%
  \BibitemOpen
  \bibfield  {author} {\bibinfo {author} {\bibfnamefont {P.~C.}\ \bibnamefont
  {Zieger}}, \bibinfo {author} {\bibfnamefont {S.~Y.~T.}\ \bibnamefont {van~de
  Meerakker}}, \bibinfo {author} {\bibfnamefont {C.~E.}\ \bibnamefont
  {Heiner}}, \bibinfo {author} {\bibfnamefont {H.~L.}\ \bibnamefont {Bethlem}},
  \bibinfo {author} {\bibfnamefont {A.~J.~A.}\ \bibnamefont {van Roij}}, \ and\
  \bibinfo {author} {\bibfnamefont {G.}~\bibnamefont {Meijer}},\ }\href
  {\doibase http://dx.doi.org/10.1103/PhysRevLett.105.173001} {\bibfield
  {journal} {\bibinfo  {journal} {Phys. Rev. Lett.}\ }\textbf {\bibinfo
  {volume} {105}},\ \bibinfo {pages} {173001} (\bibinfo {year}
  {2010})}\BibitemShut {NoStop}%
\bibitem [{\citenamefont {Deng}\ \emph {et~al.}(2011)\citenamefont {Deng},
  \citenamefont {Liang}, \citenamefont {Gu}, \citenamefont {Hou}, \citenamefont
  {Li}, \citenamefont {Xia},\ and\ \citenamefont {Yin}}]{Deng2011}%
  \BibitemOpen
  \bibfield  {author} {\bibinfo {author} {\bibfnamefont {L.}~\bibnamefont
  {Deng}}, \bibinfo {author} {\bibfnamefont {Y.}~\bibnamefont {Liang}},
  \bibinfo {author} {\bibfnamefont {Z.}~\bibnamefont {Gu}}, \bibinfo {author}
  {\bibfnamefont {S.}~\bibnamefont {Hou}}, \bibinfo {author} {\bibfnamefont
  {S.}~\bibnamefont {Li}}, \bibinfo {author} {\bibfnamefont {Y.}~\bibnamefont
  {Xia}}, \ and\ \bibinfo {author} {\bibfnamefont {J.}~\bibnamefont {Yin}},\
  }\href {\doibase http://dx.doi.org/10.1103/PhysRevLett.106.140401} {\bibfield
   {journal} {\bibinfo  {journal} {Phys. Rev. Lett.}\ }\textbf {\bibinfo
  {volume} {106}},\ \bibinfo {pages} {140401} (\bibinfo {year}
  {2011})}\BibitemShut {NoStop}%
\bibitem [{\citenamefont {Gordon}\ and\ \citenamefont
  {Osterwalder}(2017)}]{Gordon2017:3D}%
  \BibitemOpen
  \bibfield  {author} {\bibinfo {author} {\bibfnamefont {S.~D.}\ \bibnamefont
  {Gordon}}\ and\ \bibinfo {author} {\bibfnamefont {A.}~\bibnamefont
  {Osterwalder}},\ }\href {\doibase
  https://doi.org/10.1103/PhysRevApplied.7.044022} {\bibfield  {journal}
  {\bibinfo  {journal} {Phys. Rev. Appl.}\ }\textbf {\bibinfo {volume} {7}},\
  \bibinfo {pages} {044022} (\bibinfo {year} {2017})}\BibitemShut {NoStop}%
\bibitem [{\citenamefont {Cheng}\ \emph {et~al.}(2016)\citenamefont {Cheng},
  \citenamefont {van~der Poel}, \citenamefont {Jansen}, \citenamefont
  {Quintero-P{\'{e}}rez}, \citenamefont {Wall}, \citenamefont {Ubachs},\ and\
  \citenamefont {Bethlem}}]{Cheng:PRL117:253201}%
  \BibitemOpen
  \bibfield  {author} {\bibinfo {author} {\bibfnamefont {C.}~\bibnamefont
  {Cheng}}, \bibinfo {author} {\bibfnamefont {A.~P.}\ \bibnamefont {van~der
  Poel}}, \bibinfo {author} {\bibfnamefont {P.}~\bibnamefont {Jansen}},
  \bibinfo {author} {\bibfnamefont {M.}~\bibnamefont {Quintero-P{\'{e}}rez}},
  \bibinfo {author} {\bibfnamefont {T.~E.}\ \bibnamefont {Wall}}, \bibinfo
  {author} {\bibfnamefont {W.}~\bibnamefont {Ubachs}}, \ and\ \bibinfo {author}
  {\bibfnamefont {H.~L.}\ \bibnamefont {Bethlem}},\ }\href {\doibase
  10.1103/physrevlett.117.253201} {\bibfield  {journal} {\bibinfo  {journal}
  {Phys. Rev. Lett.}\ }\textbf {\bibinfo {volume} {117}},\ \bibinfo {pages}
  {253201} (\bibinfo {year} {2016})}\BibitemShut {NoStop}%
\bibitem [{\citenamefont {Wu}\ \emph {et~al.}(2017)\citenamefont {Wu},
  \citenamefont {Gantner}, \citenamefont {Koller}, \citenamefont {Zeppenfeld},
  \citenamefont {Chervenkov},\ and\ \citenamefont {Rempe}}]{Wu:Science358:645}%
  \BibitemOpen
  \bibfield  {author} {\bibinfo {author} {\bibfnamefont {X.}~\bibnamefont
  {Wu}}, \bibinfo {author} {\bibfnamefont {T.}~\bibnamefont {Gantner}},
  \bibinfo {author} {\bibfnamefont {M.}~\bibnamefont {Koller}}, \bibinfo
  {author} {\bibfnamefont {M.}~\bibnamefont {Zeppenfeld}}, \bibinfo {author}
  {\bibfnamefont {S.}~\bibnamefont {Chervenkov}}, \ and\ \bibinfo {author}
  {\bibfnamefont {G.}~\bibnamefont {Rempe}},\ }\href {\doibase
  https://doi.org/10.1126/science.aan3029} {\bibfield  {journal} {\bibinfo
  {journal} {Science}\ }\textbf {\bibinfo {volume} {358}},\ \bibinfo {pages}
  {645} (\bibinfo {year} {2017})}\BibitemShut {NoStop}%
\bibitem [{\citenamefont {Parazzoli}\ \emph {et~al.}(2011)\citenamefont
  {Parazzoli}, \citenamefont {Fitch}, \citenamefont {\ifmmode~\dot{Z}\else
  \.{Z}\fi{}uchowski}, \citenamefont {Hutson},\ and\ \citenamefont
  {Lewandowski}}]{Parazzoli:PRL106:193201}%
  \BibitemOpen
  \bibfield  {author} {\bibinfo {author} {\bibfnamefont {L.~P.}\ \bibnamefont
  {Parazzoli}}, \bibinfo {author} {\bibfnamefont {N.~J.}\ \bibnamefont
  {Fitch}}, \bibinfo {author} {\bibfnamefont {P.~S.}\ \bibnamefont
  {\ifmmode~\dot{Z}\else \.{Z}\fi{}uchowski}}, \bibinfo {author} {\bibfnamefont
  {J.~M.}\ \bibnamefont {Hutson}}, \ and\ \bibinfo {author} {\bibfnamefont
  {H.~J.}\ \bibnamefont {Lewandowski}},\ }\href {\doibase
  http://dx.doi.org/10.1103/PhysRevLett.106.193201} {\bibfield  {journal}
  {\bibinfo  {journal} {Phys. Rev. Lett.}\ }\textbf {\bibinfo {volume} {106}},\
  \bibinfo {pages} {193201} (\bibinfo {year} {2011})}\BibitemShut {NoStop}%
\bibitem [{\citenamefont {van Veldhoven}\ \emph {et~al.}(2002)\citenamefont
  {van Veldhoven}, \citenamefont {Jongma}, \citenamefont {Sartakov},
  \citenamefont {Bongers},\ and\ \citenamefont
  {Meij\-er}}]{Veldhoven:PRA66:032501}%
  \BibitemOpen
  \bibfield  {author} {\bibinfo {author} {\bibfnamefont {J.}~\bibnamefont {van
  Veldhoven}}, \bibinfo {author} {\bibfnamefont {R.~T.}\ \bibnamefont
  {Jongma}}, \bibinfo {author} {\bibfnamefont {B.}~\bibnamefont {Sartakov}},
  \bibinfo {author} {\bibfnamefont {W.~A.}\ \bibnamefont {Bongers}}, \ and\
  \bibinfo {author} {\bibfnamefont {G.}~\bibnamefont {Meij\-er}},\ }\href
  {\doibase https://doi.org/10.1103/PhysRevA.66.032501} {\bibfield  {journal}
  {\bibinfo  {journal} {Phys. Rev. A}\ }\textbf {\bibinfo {volume} {66}},\
  \bibinfo {pages} {32501} (\bibinfo {year} {2002})}\BibitemShut {NoStop}%
\bibitem [{\citenamefont {Cheung}\ \emph {et~al.}(1968)\citenamefont {Cheung},
  \citenamefont {Rank}, \citenamefont {Townes}, \citenamefont {Thornton},\ and\
  \citenamefont {Welch}}]{Cheung1968}%
  \BibitemOpen
  \bibfield  {author} {\bibinfo {author} {\bibfnamefont {A.~C.}\ \bibnamefont
  {Cheung}}, \bibinfo {author} {\bibfnamefont {D.~M.}\ \bibnamefont {Rank}},
  \bibinfo {author} {\bibfnamefont {C.~H.}\ \bibnamefont {Townes}}, \bibinfo
  {author} {\bibfnamefont {D.~D.}\ \bibnamefont {Thornton}}, \ and\ \bibinfo
  {author} {\bibfnamefont {W.~J.}\ \bibnamefont {Welch}},\ }\href {\doibase
  10.1103/PhysRevLett.21.1701} {\bibfield  {journal} {\bibinfo  {journal}
  {Phys. Rev. Lett.}\ }\textbf {\bibinfo {volume} {21}},\ \bibinfo {pages}
  {1701} (\bibinfo {year} {1968})}\BibitemShut {NoStop}%
\bibitem [{\citenamefont {Lis}\ \emph {et~al.}(2002)\citenamefont {Lis},
  \citenamefont {Roueff}, \citenamefont {Gerin}, \citenamefont {Phillips},
  \citenamefont {Coudert}, \citenamefont {van~der Tak},\ and\ \citenamefont
  {Schilke}}]{Lis2002}%
  \BibitemOpen
  \bibfield  {author} {\bibinfo {author} {\bibfnamefont {D.~C.}\ \bibnamefont
  {Lis}}, \bibinfo {author} {\bibfnamefont {E.}~\bibnamefont {Roueff}},
  \bibinfo {author} {\bibfnamefont {M.}~\bibnamefont {Gerin}}, \bibinfo
  {author} {\bibfnamefont {T.~G.}\ \bibnamefont {Phillips}}, \bibinfo {author}
  {\bibfnamefont {L.~H.}\ \bibnamefont {Coudert}}, \bibinfo {author}
  {\bibfnamefont {F.~F.~S.}\ \bibnamefont {van~der Tak}}, \ and\ \bibinfo
  {author} {\bibfnamefont {P.}~\bibnamefont {Schilke}},\ }\href {\doibase
  https://doi.org/10.1086/341132} {\bibfield  {journal} {\bibinfo  {journal}
  {Astrophys. J.}\ }\textbf {\bibinfo {volume} {571}},\ \bibinfo {pages} {L55}
  (\bibinfo {year} {2002})}\BibitemShut {NoStop}%
\bibitem [{\citenamefont {Meyer}(1995)}]{Meyer1995}%
  \BibitemOpen
  \bibfield  {author} {\bibinfo {author} {\bibfnamefont {H.}~\bibnamefont
  {Meyer}},\ }\href {\doibase 10.1021/j100004a008} {\bibfield  {journal}
  {\bibinfo  {journal} {J. Phys. Chem.}\ }\textbf {\bibinfo {volume} {99}},\
  \bibinfo {pages} {1101} (\bibinfo {year} {1995})}\BibitemShut {NoStop}%
\bibitem [{\citenamefont {Schleipen}\ and\ \citenamefont {ter
  Meulen}(1991)}]{Schleipen1991}%
  \BibitemOpen
  \bibfield  {author} {\bibinfo {author} {\bibfnamefont {J.}~\bibnamefont
  {Schleipen}}\ and\ \bibinfo {author} {\bibfnamefont {J.}~\bibnamefont {ter
  Meulen}},\ }\href {\doibase 10.1016/0301-0104(91)89016-4} {\bibfield
  {journal} {\bibinfo  {journal} {Chem. Phys.}\ }\textbf {\bibinfo {volume}
  {156}},\ \bibinfo {pages} {479} (\bibinfo {year} {1991})}\BibitemShut
  {NoStop}%
\bibitem [{\citenamefont {van~der Sanden}\ \emph {et~al.}(1996)\citenamefont
  {van~der Sanden}, \citenamefont {Wormer},\ and\ \citenamefont {van~der
  Avoird}}]{Sanden1996}%
  \BibitemOpen
  \bibfield  {author} {\bibinfo {author} {\bibfnamefont {G.~C.~M.}\
  \bibnamefont {van~der Sanden}}, \bibinfo {author} {\bibfnamefont {P.~E.~S.}\
  \bibnamefont {Wormer}}, \ and\ \bibinfo {author} {\bibfnamefont
  {A.}~\bibnamefont {van~der Avoird}},\ }\href {\doibase 10.1063/1.472177}
  {\bibfield  {journal} {\bibinfo  {journal} {J. Chem. Phys.}\ }\textbf
  {\bibinfo {volume} {105}},\ \bibinfo {pages} {3079} (\bibinfo {year}
  {1996})}\BibitemShut {NoStop}%
\bibitem [{\citenamefont {Tk{\'a}{\v{c}}}\ \emph {et~al.}(2014)\citenamefont
  {Tk{\'a}{\v{c}}}, \citenamefont {Saha}, \citenamefont {Onvlee}, \citenamefont
  {Yang}, \citenamefont {Sarma}, \citenamefont {Bishwakarma}, \citenamefont
  {van~de Meerakker}, \citenamefont {van~der Avoird}, \citenamefont {Parker},\
  and\ \citenamefont {Orr-Ewing}}]{Tkac2014}%
  \BibitemOpen
  \bibfield  {author} {\bibinfo {author} {\bibfnamefont {O.}~\bibnamefont
  {Tk{\'a}{\v{c}}}}, \bibinfo {author} {\bibfnamefont {A.~K.}\ \bibnamefont
  {Saha}}, \bibinfo {author} {\bibfnamefont {J.}~\bibnamefont {Onvlee}},
  \bibinfo {author} {\bibfnamefont {C.-H.}\ \bibnamefont {Yang}}, \bibinfo
  {author} {\bibfnamefont {G.}~\bibnamefont {Sarma}}, \bibinfo {author}
  {\bibfnamefont {C.~K.}\ \bibnamefont {Bishwakarma}}, \bibinfo {author}
  {\bibfnamefont {S.~Y.~T.}\ \bibnamefont {van~de Meerakker}}, \bibinfo
  {author} {\bibfnamefont {A.}~\bibnamefont {van~der Avoird}}, \bibinfo
  {author} {\bibfnamefont {D.~H.}\ \bibnamefont {Parker}}, \ and\ \bibinfo
  {author} {\bibfnamefont {A.~J.}\ \bibnamefont {Orr-Ewing}},\ }\href {\doibase
  https://doi.org/10.1039/C3CP53550A} {\bibfield  {journal} {\bibinfo
  {journal} {Phys. Chem. Chem. Phys.}\ }\textbf {\bibinfo {volume} {16}},\
  \bibinfo {pages} {477} (\bibinfo {year} {2014})}\BibitemShut {NoStop}%
\bibitem [{\citenamefont {Tk{\'a}{\v{c}}}\ \emph {et~al.}(2015)\citenamefont
  {Tk{\'a}{\v{c}}}, \citenamefont {Saha}, \citenamefont {Loreau}, \citenamefont
  {Ma}, \citenamefont {Dagdigian}, \citenamefont {Parker}, \citenamefont
  {Van~der Avoird},\ and\ \citenamefont {Orr-Ewing}}]{Tkac2015b}%
  \BibitemOpen
  \bibfield  {author} {\bibinfo {author} {\bibfnamefont {O.}~\bibnamefont
  {Tk{\'a}{\v{c}}}}, \bibinfo {author} {\bibfnamefont {A.~K.}\ \bibnamefont
  {Saha}}, \bibinfo {author} {\bibfnamefont {J.}~\bibnamefont {Loreau}},
  \bibinfo {author} {\bibfnamefont {Q.}~\bibnamefont {Ma}}, \bibinfo {author}
  {\bibfnamefont {P.~J.}\ \bibnamefont {Dagdigian}}, \bibinfo {author}
  {\bibfnamefont {D.~H.}\ \bibnamefont {Parker}}, \bibinfo {author}
  {\bibfnamefont {A.}~\bibnamefont {Van~der Avoird}}, \ and\ \bibinfo {author}
  {\bibfnamefont {A.~J.}\ \bibnamefont {Orr-Ewing}},\ }\href {\doibase
  http://dx.doi.org/10.1080/00268976.2015.1059958} {\bibfield  {journal}
  {\bibinfo  {journal} {Mol. Phys.}\ }\textbf {\bibinfo {volume} {113}},\
  \bibinfo {pages} {3925} (\bibinfo {year} {2015})}\BibitemShut {NoStop}%
\bibitem [{\citenamefont {Meyer}\ \emph {et~al.}(1986)\citenamefont {Meyer},
  \citenamefont {Buck}, \citenamefont {Schinke},\ and\ \citenamefont
  {Diercksen}}]{Meyer1986}%
  \BibitemOpen
  \bibfield  {author} {\bibinfo {author} {\bibfnamefont {H.}~\bibnamefont
  {Meyer}}, \bibinfo {author} {\bibfnamefont {U.}~\bibnamefont {Buck}},
  \bibinfo {author} {\bibfnamefont {R.}~\bibnamefont {Schinke}}, \ and\
  \bibinfo {author} {\bibfnamefont {G.~H.~F.}\ \bibnamefont {Diercksen}},\
  }\href {\doibase 10.1063/1.450849} {\bibfield  {journal} {\bibinfo  {journal}
  {J. Chem. Phys.}\ }\textbf {\bibinfo {volume} {84}},\ \bibinfo {pages} {4976}
  (\bibinfo {year} {1986})}\BibitemShut {NoStop}%
\bibitem [{\citenamefont {Das}\ and\ \citenamefont {Townes}(1986)}]{Das1986}%
  \BibitemOpen
  \bibfield  {author} {\bibinfo {author} {\bibfnamefont {A.}~\bibnamefont
  {Das}}\ and\ \bibinfo {author} {\bibfnamefont {C.~H.}\ \bibnamefont
  {Townes}},\ }\href {\doibase 10.1063/1.451635} {\bibfield  {journal}
  {\bibinfo  {journal} {J. Chem. Phys.}\ }\textbf {\bibinfo {volume} {85}},\
  \bibinfo {pages} {179} (\bibinfo {year} {1986})}\BibitemShut {NoStop}%
\bibitem [{\citenamefont {Daly}\ and\ \citenamefont {Oka}(1970)}]{Daly1970}%
  \BibitemOpen
  \bibfield  {author} {\bibinfo {author} {\bibfnamefont {P.~W.}\ \bibnamefont
  {Daly}}\ and\ \bibinfo {author} {\bibfnamefont {T.}~\bibnamefont {Oka}},\
  }\href {\doibase 10.1063/1.1674477} {\bibfield  {journal} {\bibinfo
  {journal} {J. Chem. Phys.}\ }\textbf {\bibinfo {volume} {53}},\ \bibinfo
  {pages} {3272} (\bibinfo {year} {1970})}\BibitemShut {NoStop}%
\bibitem [{\citenamefont {Oka}(1968{\natexlab{a}})}]{Oka1968}%
  \BibitemOpen
  \bibfield  {author} {\bibinfo {author} {\bibfnamefont {T.}~\bibnamefont
  {Oka}},\ }\href {\doibase 10.1063/1.1668157} {\bibfield  {journal} {\bibinfo
  {journal} {J. Chem. Phys.}\ }\textbf {\bibinfo {volume} {48}},\ \bibinfo
  {pages} {4919} (\bibinfo {year} {1968}{\natexlab{a}})}\BibitemShut {NoStop}%
\bibitem [{\citenamefont {Oka}(1968{\natexlab{b}})}]{Oka1968a}%
  \BibitemOpen
  \bibfield  {author} {\bibinfo {author} {\bibfnamefont {T.}~\bibnamefont
  {Oka}},\ }\href {\doibase 10.1063/1.1670561} {\bibfield  {journal} {\bibinfo
  {journal} {J. Chem. Phys.}\ }\textbf {\bibinfo {volume} {49}},\ \bibinfo
  {pages} {3135} (\bibinfo {year} {1968}{\natexlab{b}})}\BibitemShut {NoStop}%
\bibitem [{\citenamefont {Broquier}\ \emph {et~al.}(1987)\citenamefont
  {Broquier}, \citenamefont {Picard-Bersellini},\ and\ \citenamefont
  {Hall}}]{Broquier1987}%
  \BibitemOpen
  \bibfield  {author} {\bibinfo {author} {\bibfnamefont {M.}~\bibnamefont
  {Broquier}}, \bibinfo {author} {\bibfnamefont {A.}~\bibnamefont
  {Picard-Bersellini}}, \ and\ \bibinfo {author} {\bibfnamefont
  {J.}~\bibnamefont {Hall}},\ }\href {\doibase 10.1016/0009-2614(87)80512-2}
  {\bibfield  {journal} {\bibinfo  {journal} {Chem. Phys. Lett.}\ }\textbf
  {\bibinfo {volume} {136}},\ \bibinfo {pages} {531} (\bibinfo {year}
  {1987})}\BibitemShut {NoStop}%
\bibitem [{\citenamefont {Gao}\ \emph {et~al.}(2019)\citenamefont {Gao},
  \citenamefont {Loreau}, \citenamefont {Van Der~Avoird},\ and\ \citenamefont
  {Van De~Meerakker}}]{Gao2019}%
  \BibitemOpen
  \bibfield  {author} {\bibinfo {author} {\bibfnamefont {Z.}~\bibnamefont
  {Gao}}, \bibinfo {author} {\bibfnamefont {J.}~\bibnamefont {Loreau}},
  \bibinfo {author} {\bibfnamefont {A.}~\bibnamefont {Van Der~Avoird}}, \ and\
  \bibinfo {author} {\bibfnamefont {S.~Y.~T.}\ \bibnamefont {Van
  De~Meerakker}},\ }\href {\doibase https://doi.org/10.1039/C8CP07109H}
  {\bibfield  {journal} {\bibinfo  {journal} {Phys. Chem. Chem. Phys.}\
  }\textbf {\bibinfo {volume} {21}},\ \bibinfo {pages} {14033} (\bibinfo {year}
  {2019})}\BibitemShut {NoStop}%
\bibitem [{\citenamefont {Gubbels}\ \emph {et~al.}(2012)\citenamefont
  {Gubbels}, \citenamefont {van~de Meerakker}, \citenamefont {Groenenboom},
  \citenamefont {Meijer},\ and\ \citenamefont {van~der
  Avoird}}]{Gubbels:JCP136:074301}%
  \BibitemOpen
  \bibfield  {author} {\bibinfo {author} {\bibfnamefont {K.~B.}\ \bibnamefont
  {Gubbels}}, \bibinfo {author} {\bibfnamefont {S.~Y.~T.}\ \bibnamefont {van~de
  Meerakker}}, \bibinfo {author} {\bibfnamefont {G.~C.}\ \bibnamefont
  {Groenenboom}}, \bibinfo {author} {\bibfnamefont {G.}~\bibnamefont {Meijer}},
  \ and\ \bibinfo {author} {\bibfnamefont {A.}~\bibnamefont {van~der Avoird}},\
  }\href {\doibase https://doi.org/10.1063/1.3683219} {\bibfield  {journal}
  {\bibinfo  {journal} {J. Comp. Phys.}\ }\textbf {\bibinfo {volume} {136}},\
  \bibinfo {eid} {074301} (\bibinfo {year} {2012})}\BibitemShut {NoStop}%
\bibitem [{\citenamefont {Danby}\ \emph {et~al.}(1987)\citenamefont {Danby},
  \citenamefont {Flower}, \citenamefont {Valiron}, \citenamefont {Kochanski},
  \citenamefont {Kurdi},\ and\ \citenamefont {Diercksen}}]{Danby1987}%
  \BibitemOpen
  \bibfield  {author} {\bibinfo {author} {\bibfnamefont {G.}~\bibnamefont
  {Danby}}, \bibinfo {author} {\bibfnamefont {D.~R.}\ \bibnamefont {Flower}},
  \bibinfo {author} {\bibfnamefont {P.}~\bibnamefont {Valiron}}, \bibinfo
  {author} {\bibfnamefont {E.}~\bibnamefont {Kochanski}}, \bibinfo {author}
  {\bibfnamefont {L.}~\bibnamefont {Kurdi}}, \ and\ \bibinfo {author}
  {\bibfnamefont {G.~H.~F.}\ \bibnamefont {Diercksen}},\ }\href@noop {}
  {\bibfield  {journal} {\bibinfo  {journal} {J. Phys. B: At. Mol. Phys.}\
  }\textbf {\bibinfo {volume} {20}},\ \bibinfo {pages} {1039} (\bibinfo {year}
  {1987})}\BibitemShut {NoStop}%
\bibitem [{\citenamefont {Loreau}\ and\ \citenamefont {Van~der
  Avoird}(2015)}]{Loreau2015}%
  \BibitemOpen
  \bibfield  {author} {\bibinfo {author} {\bibfnamefont {J.}~\bibnamefont
  {Loreau}}\ and\ \bibinfo {author} {\bibfnamefont {A.}~\bibnamefont {Van~der
  Avoird}},\ }\href {\doibase https://doi.org/10.1063/1.4935259} {\bibfield
  {journal} {\bibinfo  {journal} {J. Chem. Phys.}\ }\textbf {\bibinfo {volume}
  {143}},\ \bibinfo {pages} {184303} (\bibinfo {year} {2015})}\BibitemShut
  {NoStop}%
\bibitem [{\citenamefont {Bouhafs}\ \emph {et~al.}(2017)\citenamefont
  {Bouhafs}, \citenamefont {Rist}, \citenamefont {Daniel}, \citenamefont
  {Dumouchel}, \citenamefont {Lique}, \citenamefont {Wiesenfeld},\ and\
  \citenamefont {Faure}}]{Bouhafs2017}%
  \BibitemOpen
  \bibfield  {author} {\bibinfo {author} {\bibfnamefont {N.}~\bibnamefont
  {Bouhafs}}, \bibinfo {author} {\bibfnamefont {C.}~\bibnamefont {Rist}},
  \bibinfo {author} {\bibfnamefont {F.}~\bibnamefont {Daniel}}, \bibinfo
  {author} {\bibfnamefont {F.}~\bibnamefont {Dumouchel}}, \bibinfo {author}
  {\bibfnamefont {F.}~\bibnamefont {Lique}}, \bibinfo {author} {\bibfnamefont
  {L.}~\bibnamefont {Wiesenfeld}}, \ and\ \bibinfo {author} {\bibfnamefont
  {A.}~\bibnamefont {Faure}},\ }\href {\doibase
  https://doi.org/10.1093/mnras/stx1331} {\bibfield  {journal} {\bibinfo
  {journal} {Mon. Not. R. Astron Soc.}\ }\textbf {\bibinfo {volume} {470}},\
  \bibinfo {pages} {2204} (\bibinfo {year} {2017})}\BibitemShut {NoStop}%
\bibitem [{\citenamefont {Demes}\ \emph {et~al.}(2023)\citenamefont {Demes},
  \citenamefont {Lique}, \citenamefont {Loreau},\ and\ \citenamefont
  {Faure}}]{Demes2023}%
  \BibitemOpen
  \bibfield  {author} {\bibinfo {author} {\bibfnamefont {S.}~\bibnamefont
  {Demes}}, \bibinfo {author} {\bibfnamefont {F.}~\bibnamefont {Lique}},
  \bibinfo {author} {\bibfnamefont {J.}~\bibnamefont {Loreau}}, \ and\ \bibinfo
  {author} {\bibfnamefont {A.}~\bibnamefont {Faure}},\ }\href {\doibase
  https://doi.org/10.1093/mnras/stad1970} {\bibfield  {journal} {\bibinfo
  {journal} {Mon. Not. R. Astron Soc.}\ }\textbf {\bibinfo {volume} {524}},\
  \bibinfo {pages} {2368} (\bibinfo {year} {2023})}\BibitemShut {NoStop}%
\bibitem [{\citenamefont {Loreau}\ \emph {et~al.}(2023)\citenamefont {Loreau},
  \citenamefont {Faure}, \citenamefont {Lique}, \citenamefont {Demes},\ and\
  \citenamefont {Dagdigian}}]{Loreau2023}%
  \BibitemOpen
  \bibfield  {author} {\bibinfo {author} {\bibfnamefont {J.}~\bibnamefont
  {Loreau}}, \bibinfo {author} {\bibfnamefont {A.}~\bibnamefont {Faure}},
  \bibinfo {author} {\bibfnamefont {F.}~\bibnamefont {Lique}}, \bibinfo
  {author} {\bibfnamefont {S.}~\bibnamefont {Demes}}, \ and\ \bibinfo {author}
  {\bibfnamefont {P.~J.}\ \bibnamefont {Dagdigian}},\ }\href {\doibase
  https://doi.org/10.1093/mnras/stad2979} {\bibfield  {journal} {\bibinfo
  {journal} {Mon. Not. R. Astron. Soc.}\ }\textbf {\bibinfo {volume} {526}},\
  \bibinfo {pages} {3213} (\bibinfo {year} {2023})}\BibitemShut {NoStop}%
\bibitem [{\citenamefont {Walmsley}\ and\ \citenamefont
  {Ungerechts}(1983)}]{Walmsley1983}%
  \BibitemOpen
  \bibfield  {author} {\bibinfo {author} {\bibfnamefont {C.~M.}\ \bibnamefont
  {Walmsley}}\ and\ \bibinfo {author} {\bibfnamefont {H.}~\bibnamefont
  {Ungerechts}},\ }\href@noop {} {\bibfield  {journal} {\bibinfo  {journal}
  {Astron. Astrophys.}\ }\textbf {\bibinfo {volume} {122}},\ \bibinfo {pages}
  {164} (\bibinfo {year} {1983})}\BibitemShut {NoStop}%
\bibitem [{\citenamefont {Maret}\ \emph {et~al.}(2009)\citenamefont {Maret},
  \citenamefont {Faure}, \citenamefont {Scifoni},\ and\ \citenamefont
  {Wiesenfeld}}]{Maret2009}%
  \BibitemOpen
  \bibfield  {author} {\bibinfo {author} {\bibfnamefont {S.}~\bibnamefont
  {Maret}}, \bibinfo {author} {\bibfnamefont {A.}~\bibnamefont {Faure}},
  \bibinfo {author} {\bibfnamefont {E.}~\bibnamefont {Scifoni}}, \ and\
  \bibinfo {author} {\bibfnamefont {L.}~\bibnamefont {Wiesenfeld}},\ }\href
  {\doibase https://doi.org/10.1111/j.1365-2966.2009.15294.x} {\bibfield
  {journal} {\bibinfo  {journal} {Mon. Not. R. Astron. Soc.}\ }\textbf
  {\bibinfo {volume} {399}},\ \bibinfo {pages} {425} (\bibinfo {year}
  {2009})}\BibitemShut {NoStop}%
\bibitem [{\citenamefont {Kuijpers}\ \emph {et~al.}(2024)\citenamefont
  {Kuijpers}, \citenamefont {Kalaitzis}, \citenamefont {Sakkoula},
  \citenamefont {van~de Meerakker}, \citenamefont {Softley},\ and\
  \citenamefont {Parker}}]{Kuijpers:JPCA128:10993}%
  \BibitemOpen
  \bibfield  {author} {\bibinfo {author} {\bibfnamefont {S.~E.~J.}\
  \bibnamefont {Kuijpers}}, \bibinfo {author} {\bibfnamefont {P.}~\bibnamefont
  {Kalaitzis}}, \bibinfo {author} {\bibfnamefont {E.}~\bibnamefont {Sakkoula}},
  \bibinfo {author} {\bibfnamefont {S.~Y.~T.}\ \bibnamefont {van~de
  Meerakker}}, \bibinfo {author} {\bibfnamefont {T.~P.}\ \bibnamefont
  {Softley}}, \ and\ \bibinfo {author} {\bibfnamefont {D.~H.}\ \bibnamefont
  {Parker}},\ }\href {\doibase https://doi.org/10.1021/acs.jpca.4c06253}
  {\bibfield  {journal} {\bibinfo  {journal} {J. Phys. Chem. A}\ }\textbf
  {\bibinfo {volume} {128}},\ \bibinfo {pages} {10993} (\bibinfo {year}
  {2024})}\BibitemShut {NoStop}%
\bibitem [{\citenamefont {Ma}\ \emph {et~al.}(2015)\citenamefont {Ma},
  \citenamefont {Van~der Avoird}, \citenamefont {Loreau}, \citenamefont
  {Alexander}, \citenamefont {van~de Meerakker},\ and\ \citenamefont
  {Dagdigian}}]{Ma2015}%
  \BibitemOpen
  \bibfield  {author} {\bibinfo {author} {\bibfnamefont {Q.}~\bibnamefont
  {Ma}}, \bibinfo {author} {\bibfnamefont {A.}~\bibnamefont {Van~der Avoird}},
  \bibinfo {author} {\bibfnamefont {J.}~\bibnamefont {Loreau}}, \bibinfo
  {author} {\bibfnamefont {M.~H.}\ \bibnamefont {Alexander}}, \bibinfo {author}
  {\bibfnamefont {S.~Y.~T.}\ \bibnamefont {van~de Meerakker}}, \ and\ \bibinfo
  {author} {\bibfnamefont {P.~J.}\ \bibnamefont {Dagdigian}},\ }\href {\doibase
  https://doi.org/10.1063/1.4927074} {\bibfield  {journal} {\bibinfo  {journal}
  {J. Chem. Phys.}\ }\textbf {\bibinfo {volume} {143}},\ \bibinfo {pages}
  {044312} (\bibinfo {year} {2015})}\BibitemShut {NoStop}%
\bibitem [{\citenamefont {Huang}\ \emph {et~al.}(2011)\citenamefont {Huang},
  \citenamefont {Schwenke},\ and\ \citenamefont {Lee}}]{Huang2011}%
  \BibitemOpen
  \bibfield  {author} {\bibinfo {author} {\bibfnamefont {X.}~\bibnamefont
  {Huang}}, \bibinfo {author} {\bibfnamefont {D.~W.}\ \bibnamefont {Schwenke}},
  \ and\ \bibinfo {author} {\bibfnamefont {T.~J.}\ \bibnamefont {Lee}},\ }\href
  {\doibase http://dx.doi.org/10.1063/1.3541351} {\bibfield  {journal}
  {\bibinfo  {journal} {J. Chem. Phys.}\ }\textbf {\bibinfo {volume} {134}},\
  \bibinfo {pages} {044320} (\bibinfo {year} {2011})}\BibitemShut {NoStop}%
\bibitem [{\citenamefont {Yang}\ \emph {et~al.}(2015)\citenamefont {Yang},
  \citenamefont {Zhang}, \citenamefont {Wang}, \citenamefont {Stancil},
  \citenamefont {Bowman}, \citenamefont {Balakrishnan},\ and\ \citenamefont
  {Forrey}}]{Yang2015a}%
  \BibitemOpen
  \bibfield  {author} {\bibinfo {author} {\bibfnamefont {B.}~\bibnamefont
  {Yang}}, \bibinfo {author} {\bibfnamefont {P.}~\bibnamefont {Zhang}},
  \bibinfo {author} {\bibfnamefont {X.}~\bibnamefont {Wang}}, \bibinfo {author}
  {\bibfnamefont {P.}~\bibnamefont {Stancil}}, \bibinfo {author} {\bibfnamefont
  {J.}~\bibnamefont {Bowman}}, \bibinfo {author} {\bibfnamefont
  {N.}~\bibnamefont {Balakrishnan}}, \ and\ \bibinfo {author} {\bibfnamefont
  {R.}~\bibnamefont {Forrey}},\ }\href {\doibase
  https://doi.org/10.1038/ncomms7629} {\bibfield  {journal} {\bibinfo
  {journal} {Nat. Commun.}\ }\textbf {\bibinfo {volume} {6}},\ \bibinfo {pages}
  {6629} (\bibinfo {year} {2015})}\BibitemShut {NoStop}%
\bibitem [{\citenamefont {Faure}\ \emph {et~al.}(2016)\citenamefont {Faure},
  \citenamefont {Jankowski}, \citenamefont {Stoecklin},\ and\ \citenamefont
  {Szalewicz}}]{Faure2016}%
  \BibitemOpen
  \bibfield  {author} {\bibinfo {author} {\bibfnamefont {A.}~\bibnamefont
  {Faure}}, \bibinfo {author} {\bibfnamefont {P.}~\bibnamefont {Jankowski}},
  \bibinfo {author} {\bibfnamefont {T.}~\bibnamefont {Stoecklin}}, \ and\
  \bibinfo {author} {\bibfnamefont {K.}~\bibnamefont {Szalewicz}},\ }\href
  {\doibase https://doi.org/10.1038/srep28449} {\bibfield  {journal} {\bibinfo
  {journal} {Sci. Rep.}\ }\textbf {\bibinfo {volume} {6}},\ \bibinfo {pages}
  {28449} (\bibinfo {year} {2016})}\BibitemShut {NoStop}%
\bibitem [{\citenamefont {Loreau}\ and\ \citenamefont {van~der
  Avoird}(2024)}]{Loreau2024}%
  \BibitemOpen
  \bibfield  {author} {\bibinfo {author} {\bibfnamefont {J.}~\bibnamefont
  {Loreau}}\ and\ \bibinfo {author} {\bibfnamefont {A.}~\bibnamefont {van~der
  Avoird}},\ }\href {\doibase https://doi.org/10.1039/D3FD00180F} {\bibfield
  {journal} {\bibinfo  {journal} {Faraday Discuss.}\ }\textbf {\bibinfo
  {volume} {251}},\ \bibinfo {pages} {249} (\bibinfo {year}
  {2024})}\BibitemShut {NoStop}%
\bibitem [{\citenamefont {Faubel}(1984)}]{Faubel1984}%
  \BibitemOpen
  \bibfield  {author} {\bibinfo {author} {\bibfnamefont {M.}~\bibnamefont
  {Faubel}},\ }\href {\doibase https://doi.org/10.1063/1.447658} {\bibfield
  {journal} {\bibinfo  {journal} {J. Chem. Phys.}\ }\textbf {\bibinfo {volume}
  {81}},\ \bibinfo {pages} {5559} (\bibinfo {year} {1984})}\BibitemShut
  {NoStop}%
\bibitem [{\citenamefont {Buck}\ \emph {et~al.}(1978)\citenamefont {Buck},
  \citenamefont {Huisken},\ and\ \citenamefont
  {Schleusener}}]{Buck:JCP68:5654}%
  \BibitemOpen
  \bibfield  {author} {\bibinfo {author} {\bibfnamefont {U.}~\bibnamefont
  {Buck}}, \bibinfo {author} {\bibfnamefont {F.}~\bibnamefont {Huisken}}, \
  and\ \bibinfo {author} {\bibfnamefont {J.}~\bibnamefont {Schleusener}},\
  }\href {\doibase https://doi.org/10.1063/1.435699} {\bibfield  {journal}
  {\bibinfo  {journal} {J. Comp. Phys.}\ }\textbf {\bibinfo {volume} {68}},\
  \bibinfo {pages} {5654} (\bibinfo {year} {1978})}\BibitemShut {NoStop}%
\bibitem [{\citenamefont {von Zastrow}\ \emph {et~al.}(2014)\citenamefont {von
  Zastrow}, \citenamefont {Onvlee}, \citenamefont {Vogels}, \citenamefont
  {Groenenboom}, \citenamefont {van~der Avoird},\ and\ \citenamefont {van~de
  Meerakker}}]{Zastrow:NatChem6:216}%
  \BibitemOpen
  \bibfield  {author} {\bibinfo {author} {\bibfnamefont {A.}~\bibnamefont {von
  Zastrow}}, \bibinfo {author} {\bibfnamefont {J.}~\bibnamefont {Onvlee}},
  \bibinfo {author} {\bibfnamefont {S.~N.}\ \bibnamefont {Vogels}}, \bibinfo
  {author} {\bibfnamefont {G.~C.}\ \bibnamefont {Groenenboom}}, \bibinfo
  {author} {\bibfnamefont {A.}~\bibnamefont {van~der Avoird}}, \ and\ \bibinfo
  {author} {\bibfnamefont {S.~Y.~T.}\ \bibnamefont {van~de Meerakker}},\ }\href
  {\doibase https://doi.org/10.1038/nchem.1860} {\bibfield  {journal} {\bibinfo
   {journal} {Nat. Chem.}\ }\textbf {\bibinfo {volume} {6}},\ \bibinfo {pages}
  {216} (\bibinfo {year} {2014})}\BibitemShut {NoStop}%
\bibitem [{\citenamefont {Herbers}\ \emph {et~al.}(2024)\citenamefont
  {Herbers}, \citenamefont {Sustar}, \citenamefont {Kuijpers}, \citenamefont
  {Sparling},\ and\ \citenamefont {van~de Meerakker}}]{Herbers2024}%
  \BibitemOpen
  \bibfield  {author} {\bibinfo {author} {\bibfnamefont {S.}~\bibnamefont
  {Herbers}}, \bibinfo {author} {\bibfnamefont {P.~A.}\ \bibnamefont {Sustar}},
  \bibinfo {author} {\bibfnamefont {S.~E.~J.}\ \bibnamefont {Kuijpers}},
  \bibinfo {author} {\bibfnamefont {C.}~\bibnamefont {Sparling}}, \ and\
  \bibinfo {author} {\bibfnamefont {S.~Y.~T.}\ \bibnamefont {van~de
  Meerakker}},\ }\href {\doibase https://doi.org/10.1080/00268976.2024.2353333}
  {\bibfield  {journal} {\bibinfo  {journal} {Mol. Phys.}\ }\textbf {\bibinfo
  {volume} {123}},\ \bibinfo {pages} {e2353333} (\bibinfo {year}
  {2024})}\BibitemShut {NoStop}%
\bibitem [{\citenamefont {Yan}\ \emph {et~al.}(2013)\citenamefont {Yan},
  \citenamefont {Claus}, \citenamefont {van Oorschot}, \citenamefont
  {Gerritsen}, \citenamefont {Eppink}, \citenamefont {van~de Meerakker},\ and\
  \citenamefont {Parker}}]{Yan2013}%
  \BibitemOpen
  \bibfield  {author} {\bibinfo {author} {\bibfnamefont {B.}~\bibnamefont
  {Yan}}, \bibinfo {author} {\bibfnamefont {P.~F.~H.}\ \bibnamefont {Claus}},
  \bibinfo {author} {\bibfnamefont {B.~G.~M.}\ \bibnamefont {van Oorschot}},
  \bibinfo {author} {\bibfnamefont {L.}~\bibnamefont {Gerritsen}}, \bibinfo
  {author} {\bibfnamefont {A.~T. J.~B.}\ \bibnamefont {Eppink}}, \bibinfo
  {author} {\bibfnamefont {S.~Y.~T.}\ \bibnamefont {van~de Meerakker}}, \ and\
  \bibinfo {author} {\bibfnamefont {D.~H.}\ \bibnamefont {Parker}},\ }\href
  {\doibase 10.1063/1.4790176} {\bibfield  {journal} {\bibinfo  {journal} {Rev.
  Sci. Instrum.}\ }\textbf {\bibinfo {volume} {84}},\ \bibinfo {pages} {023102}
  (\bibinfo {year} {2013})}\BibitemShut {NoStop}%
\bibitem [{\citenamefont {Plomp}\ \emph {et~al.}(2020)\citenamefont {Plomp},
  \citenamefont {Gao},\ and\ \citenamefont {van~de Meerakker}}]{Plomp2020}%
  \BibitemOpen
  \bibfield  {author} {\bibinfo {author} {\bibfnamefont {V.}~\bibnamefont
  {Plomp}}, \bibinfo {author} {\bibfnamefont {Z.}~\bibnamefont {Gao}}, \ and\
  \bibinfo {author} {\bibfnamefont {S.~Y.~T.}\ \bibnamefont {van~de
  Meerakker}},\ }\href {\doibase https://doi.org/10.1080/00268976.2020.1814437}
  {\bibfield  {journal} {\bibinfo  {journal} {Mol. Phys.}\ }\textbf {\bibinfo
  {volume} {119}},\ \bibinfo {pages} {e1814437} (\bibinfo {year}
  {2020})}\BibitemShut {NoStop}%
\bibitem [{\citenamefont {Werner}\ \emph {et~al.}(2020)\citenamefont {Werner},
  \citenamefont {Knowles}, \citenamefont {Manby}, \citenamefont {Black},
  \citenamefont {Doll}, \citenamefont {He{\ss}elmann}, \citenamefont {Kats},
  \citenamefont {K{\"o}hn}, \citenamefont {Korona}, \citenamefont {Kreplin},
  \citenamefont {Ma}, \citenamefont {Miller}, \citenamefont {Mitrushchenkov},
  \citenamefont {Peterson}, \citenamefont {Polyak}, \citenamefont {Rauhut},\
  and\ \citenamefont {Sibaev}}]{Molpro2018}%
  \BibitemOpen
  \bibfield  {author} {\bibinfo {author} {\bibfnamefont {H.-J.}\ \bibnamefont
  {Werner}}, \bibinfo {author} {\bibfnamefont {P.~J.}\ \bibnamefont {Knowles}},
  \bibinfo {author} {\bibfnamefont {F.~R.}\ \bibnamefont {Manby}}, \bibinfo
  {author} {\bibfnamefont {J.~A.}\ \bibnamefont {Black}}, \bibinfo {author}
  {\bibfnamefont {K.}~\bibnamefont {Doll}}, \bibinfo {author} {\bibfnamefont
  {A.}~\bibnamefont {He{\ss}elmann}}, \bibinfo {author} {\bibfnamefont
  {D.}~\bibnamefont {Kats}}, \bibinfo {author} {\bibfnamefont {A.}~\bibnamefont
  {K{\"o}hn}}, \bibinfo {author} {\bibfnamefont {T.}~\bibnamefont {Korona}},
  \bibinfo {author} {\bibfnamefont {D.~A.}\ \bibnamefont {Kreplin}}, \bibinfo
  {author} {\bibfnamefont {Q.}~\bibnamefont {Ma}}, \bibinfo {author}
  {\bibfnamefont {T.~F.}\ \bibnamefont {Miller}}, \bibinfo {author}
  {\bibfnamefont {A.}~\bibnamefont {Mitrushchenkov}}, \bibinfo {author}
  {\bibfnamefont {K.~A.}\ \bibnamefont {Peterson}}, \bibinfo {author}
  {\bibfnamefont {I.}~\bibnamefont {Polyak}}, \bibinfo {author} {\bibfnamefont
  {G.}~\bibnamefont {Rauhut}}, \ and\ \bibinfo {author} {\bibfnamefont
  {M.}~\bibnamefont {Sibaev}},\ }\href {\doibase
  https://doi.org/10.1063/5.0005081} {\bibfield  {journal} {\bibinfo  {journal}
  {J. Chem. Phys.}\ }\textbf {\bibinfo {volume} {152}},\ \bibinfo {pages}
  {144107} (\bibinfo {year} {2020})}\BibitemShut {NoStop}%
\end{thebibliography}%
\cleardoublepage

\end{document}


\date{\today}

\maketitle

\newpage

\section{Experimental methods}
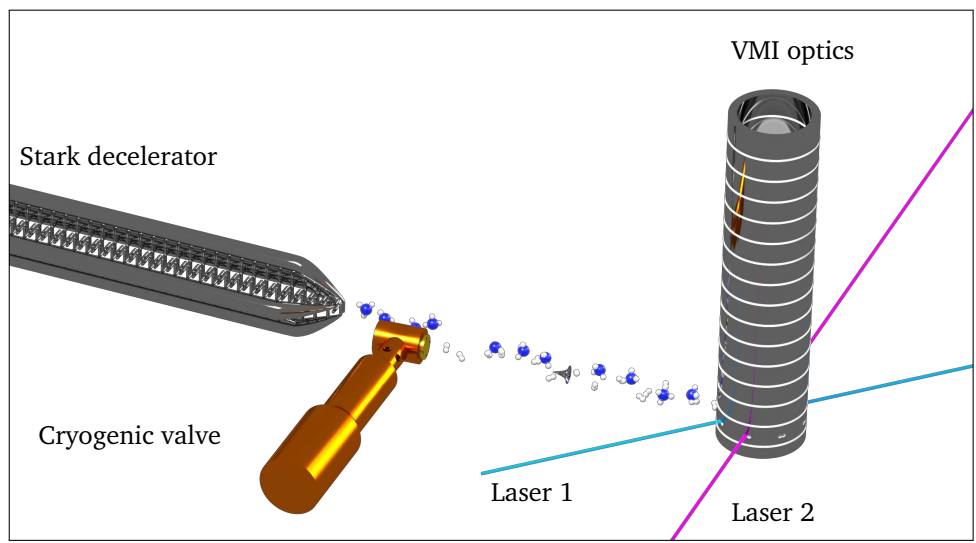
\begin{figure}[bt!]
	\begin{center}
		\setlength{\fboxsep}{0pt}
		\fbox{\begin{overpic}[width=0.8\textwidth]{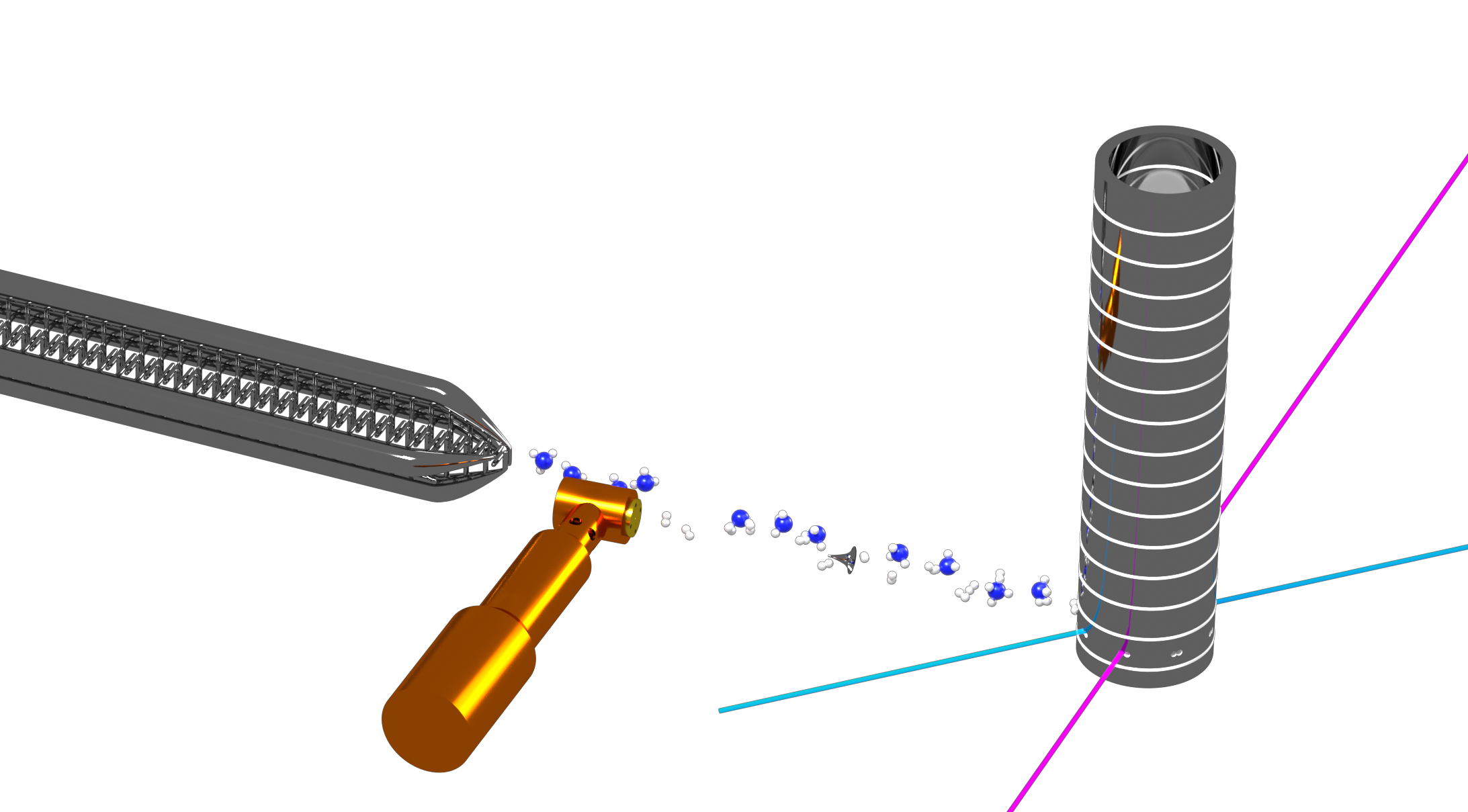}
				\put( 1,39) {Stark decelerator}
				\put( 3,10) {Cryogenic valve}
				\put(50, 4) {Laser 1}
				\put(75, 2) {Laser 2}
				\put(75,50) {VMI optics}
		\end{overpic}}
		\caption{Schematic representation of the setup. A packet of ammonia molecules is state- and velocity-selected by a 2.6-meter long Stark decelerator. A secondary beam of hydrogen molecules is created by a cryogenic source. The two beams cross at an angle of \SI{5.2}{\degree}. Inelastically scattered ammonia molecules are detected state-selectively by a REMPI scheme together with a VMI spectrometer. Only part of the Stark decelerator is shown. The figure was rendered in Blender~\cite{Blender}.}
		\label{fig:exp_setup}
	\end{center}
\end{figure}

The experiments presented were performed using a crossed beam scattering apparatus, which is depicted in \cref{fig:exp_setup}. This setup has been described in detail before \cite{Onvlee:PCCP16:15768,Zastrow:NatChem6:216,Jongh2020}, but an overview is discussed here. In this crossed-beam experiment two molecular beams were created, each by supersonically expanding a gas mixture through a pulsed valve into a differentially pumped source chamber. Both beams passed through a skimmer such that cold, collimated molecular packets entered the main vacuum chamber. The primary beam, consisting of ammonia molecules seeded in a carrier gas, was subsequently manipulated by a Stark decelerator to yield an even colder packet, i.e. one with high state-purity and a narrow velocity spread. The secondary beam consisted of hydrogen molecules. Both beams were crossed at an angle of \SI{5.2}{\degree} to ensure a low relative velocity and collision energy. Typically, only a small fraction of the incident molecules underwent a collision in this interaction region, but some ammonia molecules inelastically scattered into a different quantum state. Molecules in this product-state were state-selectively ionized through Resonance Enhanced Multi-Photon Ionization (REMPI) and mapped onto a position sensitive detector through Velocity Map Imaging (VMI).

\subsection{Primary beam}
A beam of ND$_3$ molecules was created by supersonically expanding \SI{2}{\percent} ND$_3$ seeded in a mixture of noble carrier gasses into a vacuum chamber through a Nijmegen Pulsed Valve (NPV)~\cite{Yan:RSI84:023102}. With a backing pressure of \SI{1}{\bar}, valve opening time of approximately \SI{20}{\us} and repetition rate of \SI{10}{\Hz}, the pressure of this source chamber typically was \SI{2e-6}{\milli\bar}. This chamber was pumped by a \SI{1380}{L/s} turbomolecular pump and reached a pressure of \SI{5e-8}{\milli\bar} with the valve closed. \SI{75}{mm} downstream from the nozzle the molecules passed through a \SI{3}{mm} diameter skimmer, which collimated the beam and allowed for differential pumping. During the supersonic expansion the ND$_3$ molecules were cooled by collisions with the carrier gas. This resulted in a beam with a velocity spread of around \SI{10}{\percent} (Full Width at Half Maximum (FWHM)) and a typical rotational temperature of \SI{5}{\K}. At this temperature, the ND$_3$ molecules predominantly resided in the rovibrational ground state ($\nu=0$, $j_k=0_0$ for ND$_3$ with $A_{1,2}$ symmetry, $j_k=1_1$ for ND$_3$ with $E$ symmetry).

\begin{figure}
	\centering
	\includegraphics[width=0.8\textwidth]{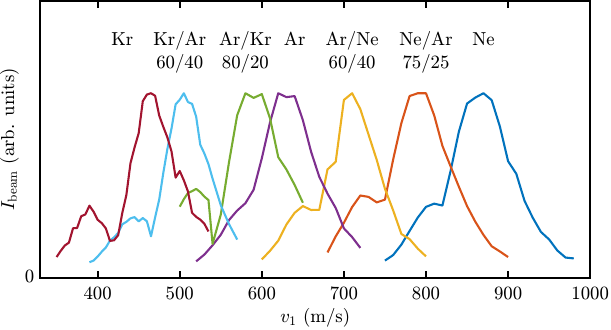}
	\caption{Normalized velocity profiles after preparing the ND$_3$ beam with different seeding gas mixtures. Mixing ratios are displayed above each profile. Velocities ranging from \num{350} to \SI{980}{m/s} could be prepared.}
	\label{fig:beam_densities}
\end{figure}

The velocity of the molecular beam could be coarsely tuned by using different carrier gas mixtures, since the final velocity of an ideal supersonic beam scales as \cite{_Scoles:MolBeam:1}

\begin{equation}
	\label{eq:molecular_beam}
	v_0 \propto \sqrt{\frac{T_0}{\langle m \rangle}},
\end{equation}

\noindent where $T_0$ is the temperature of the gas before the expansion and $\langle m \rangle$ the average mass of the particles in the gas mixture. With the carrier gasses Kr, Ar and Ne, an ND$_3$ beam containing velocities ranging from \num{350} to \SI{980}{m/s} could be prepared. The velocity profiles of different carrier gas mixtures are shown in \cref{fig:beam_densities}. These profiles could be sampled using the Stark decelerator, which will be discussed below. All profiles show a shoulder towards lower velocities, which could indicate either bouncing of the valve plunger or skimmer clogging. Since the valve settings described above yielded the largest amount of signal, and the Stark decelerator selects a narrow part of the initial velocity profiles, it was not necessary to minimize the low velocity shoulder.

\subsection{Secondary beam}
A beam of hydrogen molecules (H$_2$ or HD) was created by supersonically expanding the gas into a vacuum chamber through an Even-Lavie Valve (ELV) \cite{Even:AiC2014:636042}. The mean velocity of the resulting beam could be controlled by cooling the valve, following \cref{eq:molecular_beam}. This was achieved by mounting the valve on the second stage of a cold head (Oerlikon Coolpower 7/25). Temperature stabilization was provided by a temperature sensor and heating resistor mounted to the valve body, and connected to a PID-controller (Lake Shore 331 Cryogenic Temperature Controller). To absorb heat from black-body radiation emitted by the walls of the surrounding vacuum chamber, an extra layer of stainless steel shielding was applied around the valve. With the temperature stabilized to \SI{35}{K}, a backing pressure of \SI{2}{\bar}, valve opening time of \SI{15}{\us} and repetition rate of \SI{10}{\Hz}, the pressure inside this source chamber typically was \SI{1e-6}{\milli\bar}. This chamber was pumped by a \SI{255}{L/s} turbomolecular pump connected to a \SI{70}{L/s} booster pump, and reached a pressure of \SI{5e-8}{\milli\bar} with the valve closed. This booster pump was additionally connected to the exhaust of the turbomolecular pump evacuating the collision chamber. \SI{162}{mm} downstream from the nozzle the molecules passed through a \SI{50}{mm} long, $\diameter$\SI{3}{mm} skimmer, which collimated the beam and separated the source chamber from the collision chamber.

To prevent hydrogen from freezing and clogging or damaging the ELV, its backing pressure was kept below the vapor pressure at a given temperature~\cite{Hoge1951}. Cooling down near the critical point was not necessary as a temperature of \SI{35}{K} was sufficient to achieve an H$_2$ velocity of \SI{865}{m/s}. This velocity could be matched by the Stark decelerator to achieve collision energies below \SI{1}{\per\cm}. The collision energy could be continuously increased from here by decreasing the velocity produced by the Stark decelerator. With this single setting for the H$_2$ beam, a broad range of collision energies could be reached. At this temperature, the backing pressure can be increased to the critical pressure of \SI{12.8}{\bar}, but no signal gain was observed above a backing pressure of \SI{2}{\bar}. The HD beam was cooled to \SI{40}{K} with \SI{2}{\bar} backing pressure, resulting in a beam velocity of \SI{815}{m/s}.

For H$_2$, both \emph{para}- (even $j_2$) and \emph{ortho}-H$_2$ (odd $j_2$), or \emph{p}-H$_2$ and \emph{o}-H$_2$ in short, exist. At room temperature, \emph{p}- and \emph{o}-H$_2$ occur at a ratio of 1:3. Prior to the collision experiment, the H$_2$ gas was purified by converting nearly all hydrogen to the \emph{p}-H$_2$ ground state. This could be achieved by condensing the H$_2$ over NiSO$_4$ powder and storing it for a few hours~\cite{Vogels:NatChem10:435}. The magnetic center of this complex induced nuclear spin flips. Over time, the nuclear spins thermalized to the low temperature at which the H$_2$ was stored. The storage vessel containing the NiSO$_4$ catalyst was mounted on and cooled by the second stage of another cold head (Oerlikon Coolpower 7/25). Typically, the H$_2$ gas was stored for an hour, reheated and cooled down again for 5 times to achieve a good conversion. The reheating was found to speed up the conversion, presumably because different molecules absorb to the surface of the catalyst. After conversion, the \emph{p}-H$_2$ was transported through non-magnetic copper lines as much as possible to limit back-conversion. After collisional cooling during the supersonic expansion, the \emph{p}-H$_2$ molecules in the molecular beam now predominantly occupied the $j_2=0$ ground state.

This beam was crossed at an angle of \SI{5.2}{\degree} with the ND$_3$ packet. The low crossing angle enables low collision energies, even though both beams have a significant lab frame velocity. It also causes the $\sim\diameter\SI{3}{mm}$ molecular beams to spatially overlap over a distance of roughly \SI{50}{mm}. This overlap region is where collisions take place, and is centered around the detection volume. 

\subsection{Laser ionization}
\label{subsec:methods_rempi}
\begin{figure}[bt!]
	\centering
	\begin{circuitikz}[very thick]
		\tikzstyle{every node}=[font=\small]
		\node [font=\normalsize] at (2,27) {A};
		\draw [](1.25,19.25) to[short] (2.75,19.25);
		\draw [](1.25,24.25) to[short] (2.75,24.25);
		\draw [](1.25,25.5) to[short] (2.75,25.5);
		\draw [ color={rgb,255:red,142; green,0; blue,194}, ->, >=Stealth] (2,19.25) .. controls (2,20.5) and (2,20.5) .. (2,21.75);
		\draw [ color={rgb,255:red,142; green,0; blue,194}, ->, >=Stealth] (2,21.75) .. controls (2,23) and (2,23) .. (2,24.25);
		\draw [ color={rgb,255:red,142; green,0; blue,194}, ->, >=Stealth] (2,24.25) .. controls (2,25.5) and (2,25.5) .. (2,26.75);
		\node [] at (2,18.75) {H$_2$/HD};
		\node [] at (3.1,24.25) {$E,F$};
		\node [] at (3,19.25) {$X$};
		\node [] at (3,25.5) {$X^+$};
		\node [text width=2cm,align=right] at (0.75,22.8) {$(2+1)\times$ \SI{201}{nm}};
		\node [font=\normalsize] at (4.5,27) {B};
		\node [font=\normalsize] at (5.5,27) {C};
		\draw [](3.75,23) to[short] (6.25,23);
		\draw [](3.75,19.25) to[short] (6.25,19.25);
		\draw [](3.75,24) to[short] (6.25,24);
		\draw [ color={rgb,255:red,0; green,16; blue,245}, ->, >=Stealth] (4.5,19.25) .. controls (4.5,20.25) and (4.5,20.25) .. (4.5,21.25);
		\draw [ color={rgb,255:red,0; green,16; blue,245}, ->, >=Stealth] (4.5,21.25) .. controls (4.5,22) and (4.5,22) .. (4.5,23);
		\draw [ color={rgb,255:red,0; green,16; blue,245}, ->, >=Stealth] (4.5,23) .. controls (4.5,24) and (4.5,24) .. (4.5,25);
		\draw [ color={rgb,255:red,240; green,0; blue,176}, ->, >=Stealth] (5.5,19.25) .. controls (5.5,21) and (5.5,21) .. (5.5,23);
		\draw [ color={rgb,255:red,0; green,163; blue,245}, ->, >=Stealth] (5.5,23) .. controls (5.5,23.5) and (5.5,23.5) .. (5.5,24);
		\node [] at (5,18.75) {ND$_3$};
		\node [] at (6.5,19.25) {$X$};
		\node [] at (6.5,23) {$B$};
		\node [] at (6.5,24) {$X^+$};
		\node [text width=2cm,align=right] at (3.25,22.1) {$(2+1)\times$ \SI{320}{nm}};
		\node [] at (6.25,21.25) {\SI{160}{nm}};
		\node [] at (6.25,23.5) {\SI{448}{nm}};
		\node [font=\normalsize] at (8.75,27) {DFM};
		\draw [](7.5,19.25) to[short] (10,19.25);
		\draw [](7.5,23.75) to[short] (10,23.75);
		\draw [dashed] (6.75,23) .. controls (8.25,23) and (8.25,23) .. (10,23);
		\draw [ color={rgb,255:red,86; green,0; blue,245}, ->, >=Stealth] (8.25,19.25) .. controls (8.25,20.25) and (8.25,20.25) .. (8.25,21.5);
		\draw [ color={rgb,255:red,86; green,0; blue,245}, ->, >=Stealth] (8.25,21.5) .. controls (8.25,22.5) and (8.25,22.5) .. (8.25,23.75);
		\draw [ color={rgb,255:red,255; green,0; blue,0}, ->, >=Stealth] (9.25,23.75) .. controls (9.25,23.25) and (9.25,23.25) .. (9.25,23);
		\draw [ color={rgb,255:red,240; green,0; blue,176}, ->, >=Stealth] (9.25,23) .. controls (9.25,21) and (9.25,21) .. (9.25,19.25);
		\node [] at (8.75,18.75) {Xe};
		\node [text width=1.5cm,align=right] at (7.25,22) {$2\times$ \\ \SI{250}{nm}};
		\node [] at (10,23.4) {\SI{615}{nm}};
		\node [] at (10,21.25) {\SI{160}{nm}};
	\end{circuitikz}	
	\caption{ Overview of the REMPI and DFM schemes used for detection. (A) 2+1 REMPI scheme for H$_2$. (B) 2+1 REMPI scheme for ND$_3$. (C) low recoil 1+1$'$ REMPI scheme for ND$_3$. (DFM) Scheme for four wave difference frequency mixing in Xe to generate VUV radiation at \SI{160}{nm}.
	}
	\label{fig:exp_REMPI_schemes}
\end{figure}
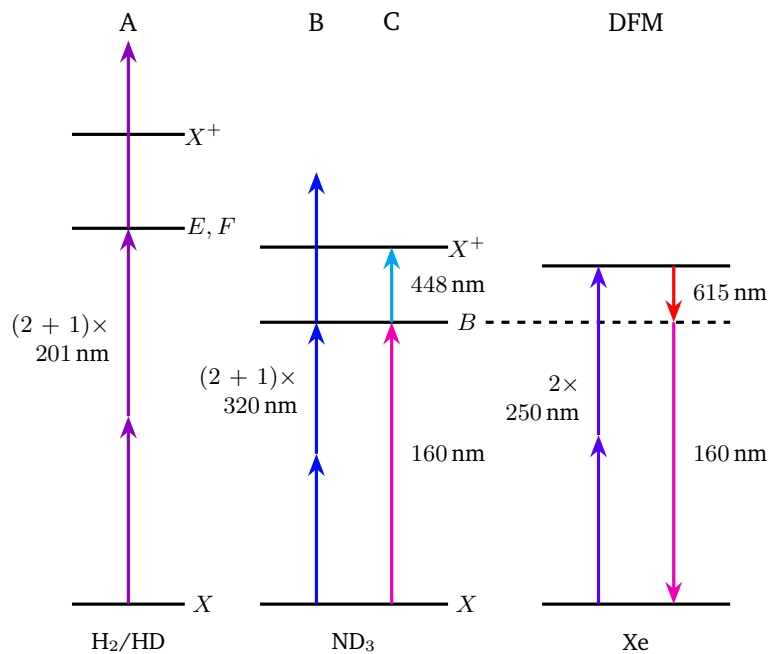

In all experiments, the molecules are detected using REMPI. Three different REMPI schemes were used, which are depicted in \cref{fig:exp_REMPI_schemes}. We used (A) to detect hydrogen molecules by inducing the $E,F \; {}^1 \! \Sigma_g^+ \leftarrow X \; {}^1 \! \Sigma_g^+$ transition in a two-photon step at \SI{201}{nm}~\cite{Rinnen1991}. A third photon of the same wavelength is absorbed to cross the Ionization Energy (IE). (B) was used to detect ND$_3$ through a 2+1 REMPI scheme at \SI{320}{nm} by inducing the $B \; {}^1 \! E'' \leftarrow X \; {}^1 \! A_1'$ transition~\cite{Ashfold:CPL138:201,Ashfold:JCP89:1754}. This is the conventional way to detect ND$_3$, which has been used in scattering experiments before~\cite{Tkac2014,Tkac2015b,Gao2019}. However, the scattering images recorded with this REMPI scheme are blurred because the total photon energy far exceeds the ionization threshold. This excess energy is transferred to the ion-electron pair as kinetic energy. The recoil velocity of the ion is \SI{\sim17}{m/s}, spoiling the images. Therefore, we developed (C), a novel 1+1$'$ REMPI scheme which offers superior resolution \cite{Kuijpers:JPCA128:10993}. A single photon of \SI{160}{nm} is used to directly excite the $B \; {}^1 \! E'' \leftarrow X \; {}^1 \! A_1'$ transition, after which a photon of \SI{448}{nm} excites the ND$_3$ molecule to a Rydberg state above the IE that autoionizes with near-zero recoil. 

The \SI{160}{nm} photons were generated through four-wave mixing in Xe, which acts as a nonlinear medium. More specifically, Difference Frequency Mixing (DFM) could be applied~\cite{Hilbig1983,Miyazaki1989}, where a two-photon transition of Xe was used to achieve higher conversion efficiency. \Cref{fig:exp_REMPI_schemes} schematically shows the DFM scheme. Two photons of \SI{250}{nm} plus a photon of \SI{615}{nm} mix to generate a fourth high energy photon. The \SI{250}{nm} light is near-resonant with a two-photon transition in Xe while the wavelength of the \SI{615}{nm} light can be chosen freely to tune the wavelength of the produced \SI{160}{nm} radiation. 

All three detection schemes were performed using pulsed dye laser systems pumped by a single Nd:YAG laser (Continuum PowerLite DLS8000) with a pulse width of \SI{8}{ns}~(FWHM). The Nd:YAG laser was frequency doubled and tripled to yield beams with a wavelength of \SI{532}{nm} and \SI{355}{nm} which were separated from each other and the \SI{1064}{nm} fundamental. For schemes (A,B), only the \SI{532}{nm} beam was used at a pulse energy of \SI{200}{mJ} or lower to pump a single dye laser (Liop-Tec LiopStar), operated using a solution of DCM in ethanol (\SI{0.34}{g/L} in oscillator cell, $6\times$ more dilute in Bethune cell). To detect H$_2$ with scheme (A) this dye laser produced a fundamental wavelength of \SI{603}{nm} which was frequency tripled to yield a beam of \SI{201}{nm}, \SI{0.2}{mJ} per pulse. This beam was focused into the machine by a lens with a focal length of $f=\SI{75}{cm}$. A low power is retrieved because DCM does not lase efficiently at \SI{603}{nm}, but a sufficiently strong H$_2$ signal could be observed nonetheless. To detect ND$_3$ with scheme (B) this dye laser produced a fundamental wavelength of \num{620} to \SI{640}{nm} which was frequency doubled to yield a beam of \num{310} to \SI{320}{nm}, \SI{12}{mJ} per pulse. This beam was focused into the machine by the same $f=\SI{75}{cm}$ lens. The position of the lens along the laser beam was optimized to maximize the ND$_3$ signal, achieving an optimal balance between a high power density and a large detection volume.

To detect ND$_3$ through scheme (C), the \SI{355}{nm} output of the Nd:YAG laser was maximized to \SI{280}{mJ} per pulse, leaving \SI{100}{mJ} per pulse of \SI{532}{nm}. The latter was used to pump the same dye laser as used for schemes (A,B), which produced a fundamental wavelength of \num{610} to \SI{620}{nm}, \SI{15}{mJ} per pulse. This beam is referred to as the `red' beam. The \SI{355}{nm} beam was split by a $70/30$ beam splitter. One beam of \SI{200}{mJ} per pulse was used to pump a second dye laser (Fine Adjustment Pulsare), operated using a solution of Coumarin 503 in ethanol (\SI{0.4}{g/L} in oscillator cell, $3\times$ more dilute in Bethune cell). This laser produced a fundamental wavelength of \SI{500}{nm} and was frequency doubled to yield a beam of \SI{250}{nm}, \SI{6}{mJ} per pulse. The coumarin dye degraded during operation, leading to a power loss of around \SI{30}{\percent} after measuring continuously at \SI{10}{\hertz} for \SI{8}{\hour}. This beam is referred to as the `UV' beam. Finally, the \SI{80}{mJ} beam of \SI{355}{nm} was used to pump a third dye laser (Liop-Tec LiopStar), operated using a solution of coumarin 440 in ethanol (\SI{0.25}{g/L} in oscillator cell, $3\times$ more dilute in Bethune cell). This laser produced a fundamental wavelength of \SI{448}{nm}, \SI{8}{mJ} per pulse. For this laser, degradation of the dye led to a power loss of around \SI{50}{\percent} after measuring for \SI{8}{\hour}. This beam is referred to as the `blue' beam.

The red and UV beams were brought to overlap with a dichroic mirror and tightly focused into a Xe cell by an $f=\SI{25}{cm}$ lens. To correct for the slight difference in the refractive indices for both wavelengths, the red laser was pre-focused by an $f=\SI{3}{m}$ lens placed before the merging dichroic, \SI{45}{cm} upstream of the main lens. The cylindrical Xe gas cell had a length of \SI{25}{cm} and an inner diameter of \SI{24}{mm}. It was filled with \SI{25}{\milli\bar} of Xe and sealed on both ends by rubber gaskets and $\diameter0.5$" MgF$_2$ windows. At the combined focus of the two beams, DFM led to the generation of copropagating laser radiation at a wavelength of \SI{160}{nm}. This wavelength belongs to the Vacuum UltraViolet (VUV) part of the electromagnetic spectrum between \num{100} and \SI{200}{nm}, and is readily absorbed by the oxygen in air. Therefore, upon exiting the Xe cell, all three colors entered a nitrogen over-pressure region sustained by a flexible polymer boundary. In this region three \SI{157}{nm} F$_2$ excimer laser mirrors were used as dichroic mirrors to separate the VUV beam from the other two colors, which were directed at a beam dump. The final two mirrors could be used to align the VUV beam towards the setup. At this stage, the VUV radiation diverged from its point of creation. The beam was collimated and slightly focussed towards the setup by placing a MgF$_2$, $f=\SI{15}{cm}$ lens approximately \SI{22}{cm} downstream of the focus in the Xe cell. The VUV beam entered the detection chamber through a $\diameter1$" MgF$_2$ window, perpendicular to the Stark decelerator axis. Finally, the blue beam was focused into the detection region using a cylindrical $f=\SI{45}{cm}$ lens placed \SI{55}{cm} before the point of detection. It entered the detection chamber at an angle of \SI{135}{\degree} with respect to the Stark decelerator axis.

The wavelengths of all three dye lasers could be monitored using a wavemeter (HighFinesse, WS6-600) equipped with an 8-channel fiber switch to continuously cycle between all lasers. Either a small portion of the fundamental or the reflection from one of the lenses was focused into a fiber optic cable, which guided the light towards the wavemeter.

\subsection{Velocity map imaging}

To detect the ions formed by REMPI, a Velocity Map Imaging (VMI)~\cite{Eppink:RSI68:3477} lens was used to extract the ions upwards, out of the plane spanned by the two molecular beams. The VMI lens used here has been described in detail before~\cite{Plomp2020}. Briefly, it consists of 16 stacked cylindrical electrodes (see \cref{fig:exp_setup}) with an inner diameter of \SI{85}{mm}, a height of \SI{26}{mm} and a gap of \SI{2}{mm} between them. This design ensures a large aperture and excellent shielding from external fields. The first cylinder was closed at the bottom and had a height of only \SI{14}{mm}. Both molecular beams as well as the REMPI lasers enter through $\diameter\SI{5}{mm}$ holes in the side of the second electrode. The two entrance holes of the molecular beams were squeezed to $\diameter\SI{3}{mm}$ by plastic inserts. Only the first and eighth electrode were connected to a power supply, with typical applied voltages of \num{2000} and \SI{1470}{V} respectively. The final electrode was grounded. Consecutive electrodes were linked in-vacuum by resistors to linearly distribute the potential and create two homogeneous field regions and an electrostatic lens at their boundary. The first and second electrode were connected by a potentiometer outside vacuum. This way, the fields at the point of ionization could be fine tuned to optimize the resolution of the VMI lens. Typically, the second lens was placed at a voltage of \SI{1993}{V}.

Compared to the conventional three-plate VMI design~\cite{Eppink:RSI68:3477} the added electrodes distribute the first homogeneous field region over a longer distance, lowering the extraction field. This can have several benefits~\cite{Townsend:RSI74:2530,Lin:RSI74:2495}, two of which are relevant in particular. First, with a large extraction field there is a large potential difference over the region where ions are produced. As a result, ions created at different positions along the field will follow slightly different trajectories. These so-called chromatic abberations can limit the resolution of the VMI lens. Second, a large extraction field will mix the molecular parity states due to the Stark effect. This spoils the state-specificity of the REMPI scheme, causing some ND$_3$ ($1_1^-$) to be detected in the ($1_1^+$) product channel as background. With the current design the extraction field was \SI{20}{V/cm}, limiting the parity mixing ratio to $\sim\num{2e-5}$. It is noted that reducing the fields by simply lowering the applied voltages would not work, as the ions need to be accelerated to sufficiently high energies to trigger the final stage of the detection, which is discussed in the next section.


\subsection{Detection and data acquisition}
\label{subsec:det_and_data_acq}
The VMI lens focussed the molecules at a detector placed \SI{1.134}{m} above the plane of the molecular and laser beams. The $\diameter\SI{4}{cm}$ detector (Photonis APD) consisted of a V-stacked Multi-Channel Plate (MCP) with a \SI{5}{\micro m} pore size and \SI{6}{\micro m} center-to center spacing, coupled to a P43 phosphor screen with a high conversion efficiency and a \SI{3}{ms} \SI{99}{\percent} decay time. Ions impinging on the MCP generated an avalanche of electrons towards the back of the detector. These electrons were accelerated towards the phosphor screen to excite it on impact, creating a fluorescent spot. Typically, the front plate of the MCP was grounded, while \SI{1.6}{kV} was applied to the back plate. The phosphor screen was placed at a voltage of \SI{4.6}{kV}.

Data could be recorded in two ways. First, the current over the back plate of the MCP could be monitored and read out with a four-channel digitizer (Picoscope 5444A) connected to a PC. This yielded a mass spectrum, in which ND$_3^+$ arrived roughly \SI{14}{\us} after the ionization pulse. Second, the phosphor screen was recorded by an iDS CMOS camera (UI-1240SE-M-GL) connected to a PC for further analysis. In this mode, mass-resolution could be retrieved by mass-gating the MCP detector with a GHTS30 Behlke switch to detect only the ND$_3^+$. This way mass- and position-dependent data could be collected to yield the highest sensitivity. During a DCS measurement, typically around one ion per shot was detected, whose central position could be determined by event counting and centroiding algorithms. Although this camera-based measurement was most sensitive, the digitizer-based approach had different advantages, especially while recording large signal levels. It involved less steps, and as such behaved more linear over a larger range of signal intensities. For the highest signal intensities, the voltages on the detector could be decreased to protect the detector against burn-in. The limiting factor in the sensitivity of this approach was the susceptibility of the analogue signal to electromagnetic noise.

The experiment was controlled and signals were read out digitally by the custom-built program FullDAQ \cite{Cremers2019}, created using the LabVIEW graphical programming environment. This program analyzed the data recorded by the camera and digitizer. Additionally, it controlled the wavelength of all three dye lasers and the output of a 24-channel delay generator (Spincore PulseBlasterUSB) capable of sending TTL trigger pulses with \SI{10}{ns} time resolution from each channel. Pulse changes should either be simultaneous or at least \SI{50}{ns} apart across the 24 channels. Timings could be defined relative to one another, allowing for an elaborate yet intuitive sequence of triggers. These were used to trigger the valves, flash-lamps and Q-switch of the YAG laser, massgate, digitizer and camera as well as to control the switching scheme of the Stark decelerator. Moreover, the switching scheme of the Stark decelerator together with up to five related timings could be automatically read from a so-called sequence list to produce ammonia packets with different velocities in rapid succession.

To perform an ICS measurement, this feature was used to toggle every 20 shots between a background and signal measurement while iterating over different velocities in a back-and-forth way to scan the collision energy. The velocity of the secondary beam was kept fixed, while its timing was detuned to perform the background measurement. With typically 20 velocities per sequence list, one back-and-forth scan including background measurements could be completed in around 160 seconds. Repeating this scan many times yielded an accurate measure of the intensity of scattered molecules, robust against long term drifts in either the sources or lasers. The measured intensity was corrected for the intensity of the incident beam of ND$_3$ ($1_1^-$) molecules by measuring its density at each velocity several times a day. The normalized collision product intensity $I_\text{n}$ is defined for each velocity as

\begin{equation}
	I_\text{n} = \frac{I_\text{s} - I_\text{bg}}{I_\text{b}},
	\label{eq:normalized_intensity}
\end{equation}

\noindent with $I_\text{s}$ the measured intensity of collision products, $I_\text{bg}$ the intensity of the background and $I_\text{b}$ the intensity of the incident beam.

Typically, the ICS measurement was split up according to the velocities allowed by a single seed gas mixture, as shown in \cref{fig:beam_densities}. Different velocity ranges were combined by scaling the corresponding normalized intensities vertically to minimize the error-weighted root-mean-square deviation among them. To improve the reliability of this approach it was ensured that each velocity appeared in at least two ranges, covering the full range twice. The result of this measurement is the normalized intensity as a function of the velocity of the ND$_3$ packet, $I_n(v_1)$. To convert this data into a measurement of the energy-dependent cross section, $\sigma(E)$, both an accurate energy calibration and a way to correct the normalized intensities for density-to-flux effects are required. These two topics will be discussed in section \ref{subsec:energy_calibration}.

\section{Calibration methods}
\label{sec:exp_calibration}
A sound set of calibration methods is required to know the conditions under which experiments are performed and how to convert the observed quantities into physical ones. The state-purity of the incoming molecular beams was tested by recording REMPI spectra. The VMI detector and beam crossing angle could be calibrated by directly imaging the incoming beams, where the Stark decelerator could be used to generate an accurate set of velocity markers directly on the detector. The average velocity of the secondary beam could be calibrated by recording collision images at several different velocities of the primary beam. Finally, the longitudinal velocity spread of the secondary beam could be derived from the time-of-flight profile of the incoming beam. These steps are detailed below. Some of them have been described before \cite{Vogels:NatChem10:435,Shuai2020,Jongh2020}.

\subsection{State purity}
\label{subsec:beam_purity}
In order to verify the state-purity of both incoming beams, 2+1 REMPI spectra were recorded at the peak of each time-of-flight distribution.

To record spectra of ND$_3$ the wavelength of a frequency doubled dye laser was scanned around \SI{321}{nm} and \SI{317}{nm} in steps of \SI{5e-4}{nm} to cover the rotational transitions $B(\nu_2=4,j'_{k'}) \leftarrow X(j_k^+)$ and $B(\nu_2=5,j'_{k'}) \leftarrow X(j_k^-)$, respectively. Each datapoint was averaged for $20$ shots while monitoring the current over the back plate of the MCP. The spectra were recorded under similar conditions to the scattering experiments, with a laser power of \SI{12}{mJ} focused towards the detection volume by an $f=\SI{750}{mm}$ lens. A neon carrier gas was used to create the ND$_3$ beam. As a consequence of the high velocity, the Stark decelerator will be least effective at filtering the low-field-seeking states. So, these measurements reflect a worst-case scenario and place an lower limit on the purity of the ND$_3$ beam.

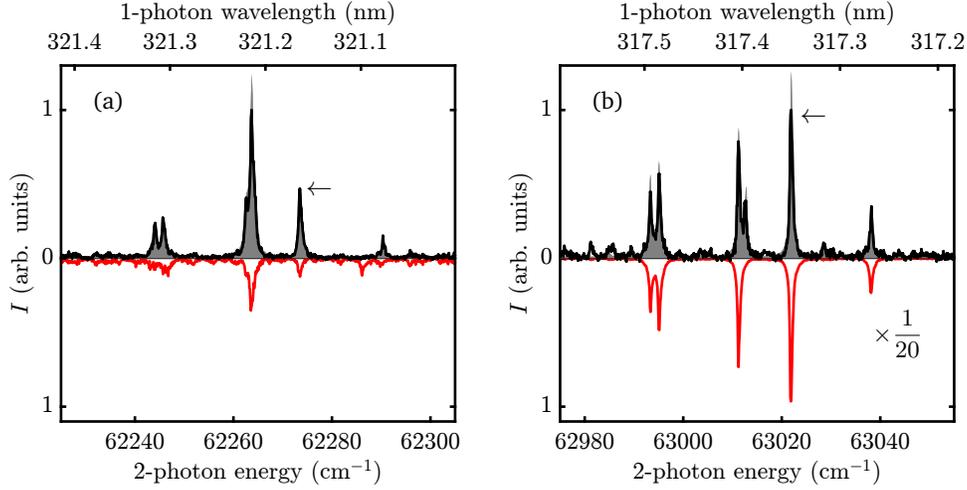
\begin{figure}[t!]
	\centering
	\begin{overpic}{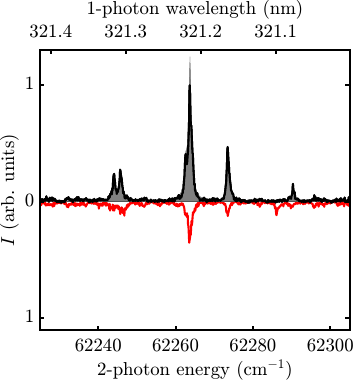}
		\put(17,78) {\small(a)}
		\put(61,60){$\leftarrow$}
	\end{overpic}
	\hspace{0.5cm}
	\begin{overpic}{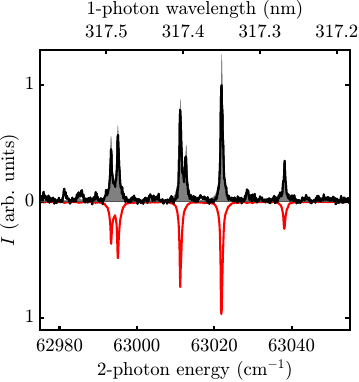}
		\put(17,78) {\small(b)}
		\put(60,75){$\leftarrow$}
		\put(75,30){\small$\displaystyle\times\frac{1}{20}$}
	\end{overpic}
	\caption{Rotational 2+1 REMPI spectra of ND$_3$ covering (a) $B(\nu_2=4,j'_{k'}) \leftarrow X(j_k^+)$ and (b) $B(\nu_2=5,j'_{k'}) \leftarrow X(j_k^-)$. Intensity $I$ as a function of 2-photon energy and 1-photon wavelength. Experimental spectra recorded with the Stark decelerator turned off (on) are shown as black (red) lines. The shaded gray area shows the spectra calculated by PGOPHER~\cite{Western:pgopher,Western2017} with molecular parameters as found in \cite{Ashfold:CPL138:201} for a rotational temperature of \SI{5}{K}. The horizontal arrows mark the main peaks used for detecting ND$_3$. Note that the red line in panel (b) has been scaled down by a factor of 20.}
	\label{fig:exp_ND3_spectrum}
\end{figure}

With the Stark decelerator disabled, thermal spectra matching a rotational temperature of $T_\text{rot}=\SI{5}{K}$ were observed for both parities as shown in the top traces of \cref{fig:exp_ND3_spectrum}. The simulated spectra were generated in PGOPHER \cite{Western:pgopher} with molecular parameters as found in \cite{Ashfold:CPL138:201}. In the scattering experiments only the strongest transitions for $1_1^+$ and $1_1^-$ were used, which both excite through $j'_{k'}=3_2$. These are assigned in \cref{fig:exp_ND3_spectrum} by small arrows. The signal-to-noise ratio of these experimental spectra is rather poor since these spectra are recorded after the molecular beam experienced free flight over \SI{3}{m}. In this scenario, the fraction of ND$_3$ molecules in the beam that occupy the $1_1^+$ and $1_1^-$ states is practically equal.

With the Stark decelerator turned on, \SI{\pm5}{kV} applied to its electrodes, and set to guide molecules at \SI{860}{m/s}, the spectra are significantly altered. As shown in the bottom traces of \cref{fig:exp_ND3_spectrum}, the signal intensity for transitions belonging to low-field-seeking states like the $1_1^-$ state are amplified by a factor of 20, while the high-field-seeking states are suppressed by a factor of four. During a scattering experiment, the ratio between the incoming and background population is therefore a factor 80 at these settings. States without a notable Stark effect are not affected. This includes the $M=0$ component of the $1_1^+$ states, which remain visible in the spectrum. Only the transition at \SI{62287}{\per \cm} is seen to gain intensity. This line has been observed before, and was interpreted as originating from the $1_1^-$ state, but transitioning through a different excited state~\cite{Veldhoven:thesis:2006,Tkac2014}. The spectra are slightly power broadened, as should be expected under these high power conditions. However, the broadening does not cause problematic overlap with surrounding transitions.

\begin{figure}[t!]
	\centering
	\begin{overpic}{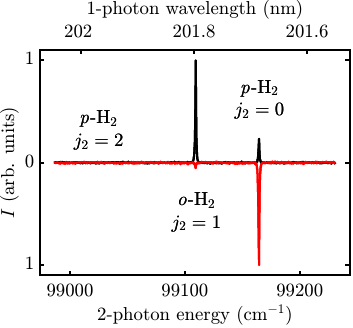}
		\put(17,68) {\small(a)}
		\put(27,40) {$\uparrow$}
	\end{overpic}
	\hspace{0.5cm}
	\begin{overpic}{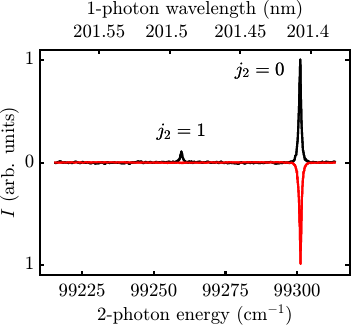}
		\put(17,68) {\small(b)}
	\end{overpic}
	\caption{Experimental 2+1 REMPI spectra of the hydrogen beam. (a) Spectra on the natural (black) and \emph{ortho}-to-\emph{para}-converted (red, inverted) H$_2$ beam at a valve temperature of \SI{35}{K}. Note that for the converted beam, mainly the $j_2=0$ rotational ground state is populated. (b) Spectra on the HD beam at a valve temperature of \SI{100}{K} (black) and \SI{40}{K} (red, inverted). Note that at \SI{40}{K}, only the $j_2=0$ rotational ground state is populated.}
	\label{fig:exp_H2_spectrum}
\end{figure}

Similarly, we characterized the hydrogen beam by inducing the $E,F \leftarrow X$ transition through 2+1 REMPI \cite{Rinnen1991}. \Cref{fig:exp_H2_spectrum} shows the rotational spectra of H$_2$ and HD, which were recorded by scanning the wavelength of a frequency tripled dye laser around \SI{201}{nm}. In panel (a), the spectrum of the natural H$_2$ beam at a valve temperature of \SI{35}{K} is shown along the top. The $j_2=1$ transition belonging to \emph{ortho}-H$_2$ is strongest since \SI{75}{\percent} of the H$_2$ molecules has an \emph{ortho} nuclear spin configuration prior to the expansion. Still, the beam is rotationally cold as the $j_2=2$ transition around \SI{99025}{\per\cm} is absent. Next, the spectrum of the converted H$_2$ beam is shown along the bottom. In this spectrum, the $j_2=1$ transition is severely suppressed and nearly all molecules reside in the $j_2=0$ rotational ground state belonging to \emph{p}-H$_2$. From the natural abundance of \emph{o}- to \emph{p}-H$_2$ (3:1) and the area under the curves in both plots, we calculate that the converted beam contains less than \SI{5}{\percent} \emph{o}-H$_2$. In panel (b), the spectrum of the HD beam at a valve temperature of \SI{100}{K} is shown along the top. Both the $j_2=0$ and $j_2=1$ transitions are visible, so both rotational states are present in the beam. When the valve temperature is further decreased to \SI{40}{K}, the setting used throughout this work, the spectrum shown along the bottom of panel (b) is obtained. Here only the $j_2=0$ transition is observed, confirming that a state-pure beam of HD molecules is generated.

\newpage
\subsection{Detector calibration and beam crossing angle}

\begin{figure}[b!]
	\centering
	\includegraphics{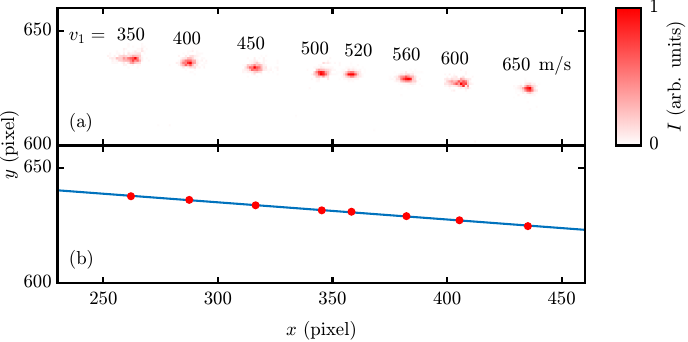}
	\caption{Detector calibration by imaging the ND$_3$ ($1_1^-$) packets guided by the Stark decelerator at eight different velocities. (a) Combined image, where each velocity produces a sharp beamspot on the detector. (b) Center-of-mass of each of the beamspots (red dots) with a linear fit through them (blue line). }
	\label{fig:beamspots}
\end{figure}

To calibrate the entire detection stack, which consists of the VMI lens, MCP, phosphor screen and camera, the exceptionally well-controlled packets emerging from the Stark decelerator can be used as a reference. \Cref{fig:beamspots}(a) shows the image obtained when directly imaging ND$_3$ ($1_1^-$) guided through the Stark decelerator (operated at \SI{\pm10}{kV}) at eight different velocities $v_1$ between \num{350} and \SI{650}{m/s}. For this measurement, the low-recoil 1+1$'$ REMPI scheme was used. A separate image was recorded for each velocity, yielding a single sharp beamspot on the detector. \Cref{fig:beamspots}(a) shows the sum of these images. The beamspots fall along a straight line at a position proportional to the velocity of the packet. This reveals both the propagation direction of the Stark decelerated package as well as the scale of these images in terms of velocity. The longitudinal velocity spread of the detected molecules is around \SI{5}{m/s}~(FWHM), as is evident from the spacing between the two beamspots at \num{500} and \SI{520}{m/s}.

We determined the center-of-mass of each beamspot, as shown in \cref{fig:beamspots}(b), and performed a linear fit through the ($x,y$) coordinates as a function of $v_1$. Based on these results, the angle between the ND$_3$ propagation direction and camera $x$-axis is \SI{4.29 \pm 0.01}{\degree}. The image scaling factor $f_\text{VMI}$ is \SI{1.71 \pm 0.04}{m/s/pixel}. All errors reported use a \SI{95}{\percent} Confidence Interval (CI), unless mentioned otherwise. While recording scattering images, a calibration like this was performed daily with at least four velocities.

The geometric angle between the two crossed molecular beams was determined previously for this setup by imaging the secondary beam in a similar way, as described in detail in~\cite{Jongh2020}. A line of beamspots could be created by tuning the beam velocity through the temperature of the cryogenic valve. This process was repeated for neat beams of H$_2$ and O$_2$ and yielded a beam crossing angle of \SI{5.2 \pm 0.1}{\degree}. This calibration was not repeated here, as the experimental setup has remained unchanged.

\subsection{Energy calibration}
\label{subsec:energy_calibration}
Accurately calibrating the energy at which collisions take place is vital to the interpretation of the recorded data. Especially when comparing our measurements to state-of-the-art theoretical predictions, it is important that the collision energy is known with confidence. When we assume all particles in a packet travel at the same velocity, $v_1$ and $v_2$ for the two packets, all collisions take place at the same energy

\begin{align}
	E &= \frac{1}{2} \mu v_\text{rel}^2 \label{eq:simple_Ecol} \ , \quad \text{where} \\
	v_\text{rel}^2 &= v_1^2 + v_2^2 - 2 v_1 v_2 \cos{\beta} \ , \label{eq:vrel}
\end{align}

\noindent with $\mu$ the reduced mass of the two colliding species, $\beta$ the beam crossing angle and $v_\text{rel}$ the relative velocity. The main uncertainty in determining the collision energy is the velocity of the secondary beam.

For realistic packets the velocity spreads and velocity sorting that takes place as a consequence can lead to deviations from \cref{eq:simple_Ecol}. In that case \cref{eq:simple_Ecol} is only valid per event, and the average detected collision energy is a function of the phase spaces of both packets and their overlap with each other and the detection laser. For now, we will ignore these effects, and focus on determining a value for $v_2$. 

\begin{figure}[p!]
	\centering
	\begin{overpic}[grid=0,tics=10,scale=0.8]{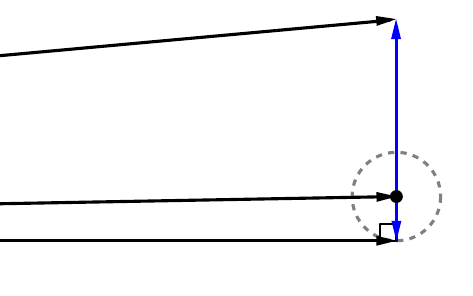}
		\put(86,7) {\small$\vec{u}_1$}
		\put(86,53) {\small$\vec{u}_2$}
		\put(60,5) {\small$\vec{v}_1$}
		\put(60,50) {\small$\vec{v}_2$}
		\put(60,25) {\small$\vec{v}_\text{COM}$}
		\put(77,15) {\small$\alpha$}
	\end{overpic}
	\caption{Newton diagram for the lowest collision energy given a beam-crossing angle of \SI{5.35}{\degree} and fixed velocity of the secondary beam $\vec{v}_2$. $\vec{v}_1$ is the primary beam velocity, $\vec{v}_\text{COM}$ the Center-Of-Mass (COM) velocity and $\vec{u}_1$, $\vec{u}_2$ are the velocity of the primary and secondary beam in the COM frame, such that $v_\text{rel}=\abs{\vec{u}_1} + \abs{\vec{u}_2}$. Conservation of energy and momentum demands that the scattering products end up on a circle concentric with the COM. $\alpha$ is the orientation angle, which is \SI{90}{\degree} at the turnaround point.}
	\label{fig:newton_diagram_turnaround}
\end{figure}

To calibrate the velocity of the secondary beam, the VMI detector can be used to record scattering images at low collision energies. Two independent features of these scattering images can be used to calibrate: the orientation angle $\alpha$ as defined in \cref{fig:newton_diagram_turnaround} and the image radius $r$. Since both properties are hard to determine accurately from a single image, several images were recorded at different ND$_3$ velocities, $v_1$. The velocity of the hydrogen beam, $v_2$, could be fitted to this data. To illustrate, \cref{fig:newton_diagram_turnaround} depicts the velocity vector diagram (Newton diagram) where $r$ is minimal and $\alpha=\SI{90}{\degree}$. This point is called the turnaround point, at which the collision energy is minimal. Changing $v_1$ in either direction increases the collision energy, probing the same physics under different kinematic conditions. 

\begin{figure}[p!]
	\centering
	\includegraphics[width=\textwidth]{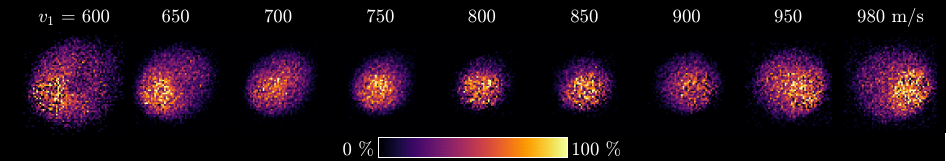}
	\caption{Raw scattering images for the ND$_3$+HD system, as a function of the velocity $v_1$ of the packet of ND$_3$ molecules emerging from the decelerator. These images were recorded with the 2+1 REMPI scheme and were used to calibrate $v_2$.}
	\label{fig:calibration_images}
\end{figure}

\begin{figure}[p!]
	\centering
	\begin{overpic}{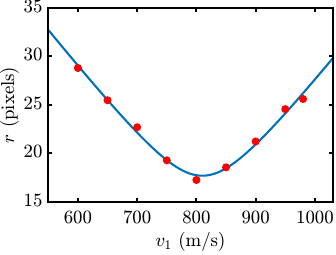}
		\put(20,20) {\small(a)}
	\end{overpic}
	\hspace{0.5cm}
	\begin{overpic}{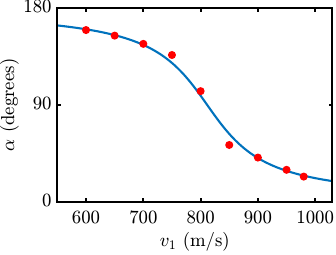}
		\put(20,20) {\small(b)}
	\end{overpic}
	\caption{Calibration of the ND$_3$+HD experiment based on the images in \cref{fig:calibration_images}. (a) Calibration based on the radius $r$ of the scattering images. (b) Calibration based on the orientation angle $\alpha$ of the scattering images. Each dot represents the data extracted from one image, while the solid blue line shows the least-squares fit through the data.}
	\label{fig:exp_calib}
\end{figure}

\begin{table}[p!]
	\centering
	\begin{tabular}{ r | c c c }
		Calibrated $v_2$, \SI{95}{\percent} CI    & radius calib. & orientation calib. & combined \\ \hline
		H$_2$ beam (m/s) & 865.7 $\pm$ 7.4 & 862.7 $\pm$ 7.1 & 864.2 $\pm$ 5.1 \\  
		HD beam (m/s) & 813.5 $\pm$ 4.8 & 815.0 $\pm$ 8.1 & 813.9 $\pm$ 4.0    
	\end{tabular}
	\caption{Calibrated values for $v_2$ based on the radius and orientation of a set of scattering images. The combined values were obtained using inverse-variance weighted statistics.}
	\label{tab:v2_calibration}
\end{table}

A series of calibration images was taken for both ND$_3$-H$_2$ and ND$_3$-HD. \Cref{fig:calibration_images} shows the results for ND$_3$-HD as an example. Similar to the ICS measurements, half the time was spent recording background signal by detuning the secondary beam in time. This background has been subtracted for the images shown here. The extracted radii $r$ together with their least-squares fit are shown in \cref{fig:exp_calib}(a). $r$ was found by an edge detection followed by a circle finding algorithm. The fit for $r$ follows the formula

\begin{equation}
	r = v_\text{rel} \cdot \frac{m_2}{m_1+m_2} \cdot f_\text{VMI}^{-1} + C \ ,
\end{equation}

\noindent with $v_\text{rel}$ as in \cref{eq:vrel}, $m_1$ and $m_2$ the masses of the two colliding species, $f_\text{VMI}$ as determined in the previous section and $C$ a constant fitparameter. $C$ is needed since the circle finding algorithm detects the outer edge of each scattering image, which is displaced from the true radius.

The extracted orientation angles $\alpha$ are shown in \cref{fig:exp_calib}(b), again with their least-squares fit. $\alpha$ was obtained from the line connecting the center of the aforementioned circles with the beamspots present in each background measurement (not shown). This beamspot overlapped with the forward direction and resulted from a small amount of parity mixing. The fit for $\alpha$ follows the formula

\begin{equation}
	\alpha= 90\si{\degree} - \frac{180\si{\degree}}{\pi} \cdot \atan \left( \frac{v_1/v_2-\cos{\beta}}{\sin{\beta}} \right) \ .
\end{equation}

\noindent For both fits, we assume an effective beam crossing angle of $\beta=\SI{5.35}{\degree}$, as determined from simulations of detectable collision products from the two overlapping beams. Both methods of calibrating the secondary velocity were consistent for the two systems studied here. An overview of the calibrated values for $v_2$ obtained by both methods can be found in \cref{tab:v2_calibration}.

\begin{figure}[h!]
	\centering
	\begin{overpic}{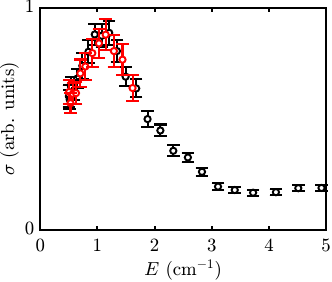}
		\put(15,20) {\small(a)  ND$_3$-H$_2$}
	\end{overpic}
	\hspace{0.5cm}
	\begin{overpic}{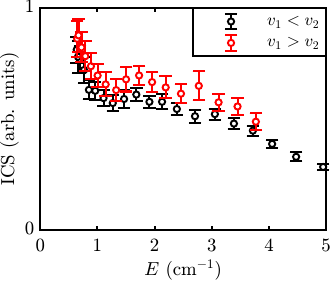}
		\put(15,20) {\small(b)  ND$_3$-HD}
	\end{overpic}
	\caption{Energy calibration by recording the relative ICS as a function of the collision energy $E$ while scanning through the turnaround point for (a) ND$_3$-H$_2$ and (b) ND$_3$-HD. The same structure can be observed on both sides of the turnaround point ($v_1<v_2$ versus $v_1>v_2$), at the same energy. This verifies that the collision energy has been calibrated correctly. Vertical error bars represent statistical uncertainties, calculated as the standard deviation of the mean over hundreds of samples (\cref{subsec:det_and_data_acq}). All error bars represent a \SI{95}{\percent} confidence interval.
	}
	\label{fig:turnaround_calibration}	
\end{figure}

In addition to using the orientation angle and radius of the scattering images to calibrate the velocity of the secondary beam, a third energy calibration is possible using the ICS measurements. When scanning the collision energy by tuning $v_1$ with the Stark decelerator, one can cross the turnaround point. Since the turnaround point marks the minimum collision energy, the same resonance structures can be measured on either side. This is shown in \cref{fig:turnaround_calibration} for both ND$_3$-H$_2$ and ND$_3$-HD. For ND$_3$-HD, the two sides of the turnaround point do not overlap perfectly in intensity, but they do in energy. This points towards a slight inaccuracy in our Flux-to-Density correction (see \cref{sec:sim_methods}), but a correct assignment of the collision energy. All three calibration methods outlined here work by scanning the kinematic conditions to tune the collision energy around the turnaround point. We thus expect that the calibration is most accurate for the lowest collision energies.

\subsection{Hydrogen beam calibration}
\label{subsec:hydrogen_beam}
With the velocity of the secondary beam calibrated most experimental parameters are known. But to better describe the dynamics between the two colliding packets, the velocity spread should be known as well. Because of this spread, the front of the packet will contain molecules that are faster on average, while the slower molecules tend to lag behind. This is important as the Stark packet intersects different parts of the secondary beam, and thus encounters different velocities, depending on the experimental settings.

By measuring the time of flight (TOF) of the hydrogen packet, the longitudinal velocity spread could be determined. To retrieve its value from the measurement, the secondary beam was modeled as an ideal Gaussian packet with no transverse velocity. In 1D, the phase-space density of such a packet, $\rho$, at time $t$ is given by

\begin{equation}
	\label{eq:1Dgaussianpacket}
	\rho(x,v,t)=N \cdot \exp(-\frac{(v-v_c)^2}{2\sigma_v^2}) \cdot \exp(-\frac{(x-v(t-t_0))^2}{2(v\sigma_t)^2}),
\end{equation}

\noindent with $x$ the position and $v$ the velocity in the phase-space, and $t_0$ the starting time. $t_0$ accounts for the mechanical delay of the valve in the experiment with respect to the valve trigger pulse. Furthermore, $N$ is a normalization factor, $v_c$ and $\sigma_v$ are the mean velocity and its standard deviation, and $\sigma_t$ relates to the valve opening duration. In this model, the TOF intensity at a given point $x$ is just the integral over all velocities. The local average velocity can similarly be calculated as an expectation value.

\begin{figure}[h!]
	\centering
	\begin{overpic}{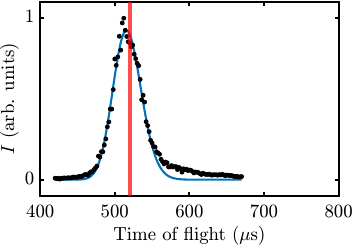}
		\put(70,60) {\small(a)  H$_2$}
	\end{overpic}
	\hspace{0.5cm}
	\begin{overpic}{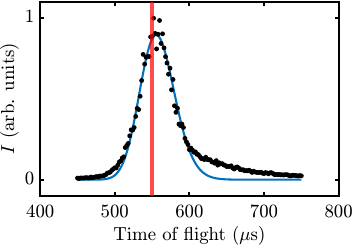}
		\put(70,60) {\small(b)  HD}
	\end{overpic}
	\caption{Intensity $I$ as function of the time of flight for the (a) H$_2$ and (b) HD molecular beams. The experimental intensity (black dots) was modeled as an ideal Gaussian packet in 1D (blue curve) according to \cref{eq:1Dgaussianpacket}. The vertical red bar shows the arrival time used throughout the scattering experiments.}
	\label{fig:H2_TOF_cal}
\end{figure}

The experimentally recorded TOF profiles for both the H$_2$ and HD beam are shown in \cref{fig:H2_TOF_cal}, together with their least-squares fits. According to the fit, both beams have a velocity spread of around \SI{10}{\percent}~(FWHM). For either fit $N$, $\sigma_v$ and $t_0$ were varied while keeping $v_c$, $x$ and $\sigma_t$ fixed. $x=\SI{410}{mm}$ on this setup and $\sigma_t$ was chosen to match a \SI{15}{\us}~(FWHM) opening-time of the valve. $v_c$ was picked such that the local average velocity at the time of arrival was given by $v_2$ as determined in the previous section. Because the arrival time and peak of the TOF profile did not match exactly, $v_c$ and $v_2$ differed by a few \si{m/s}. With these beam parameters known, the phase space of the secondary beam can be accurately modeled.

\section{Simulation methods}
\label{sec:sim_methods}

For the scattering experiments presented here, the experimental results and theoretical predictions could not be compared directly since the finite spreads in the experiment lead to a convolution of the underlying observables. This can distort the imaged DCSs~\cite{Zastrow2015} and limits the resolution of the recorded ICSs. In addition, as the collision energy is tuned, the overlap between the two beams, as well as the overlap between the scattered molecules and the detection laser changes. This problem is usually referred to as the flux-to-density effect, and can distort the detected signal when probing the ICS. To quantify these effects, the entire experiment was modeled in a Monte-Carlo simulation, taking the theoretical predictions for the ICS and DCSs as an input. The experimental and simulated images could then directly be compared, while measurements of the ICS could be corrected.

In this section, the simulation of crossed beam scattering experiments is discussed in detail. The program, which was written in Fortran, is an extension to the previously published numerical trajectory simulation for generating scattering images~\cite{Zastrow:NatChem6:216,Zastrow2015}. The program has been modified to more accurately model the overlap between two beams, i.e. select which pairs of particles form viable collision partners. This new approach is not specific to any beam source and can use the phase space generated by non-Gaussian sources such as Stark or Zeeman decelerators and (curved) hexapoles or magnetic guides as an input to both beams. Beyond simulating scattering images, the program is now suited to model ICS measurements too. 
The approach outlined here has also been applied recently in a merged-beam scattering experiment between Stark-decelerated NO with a packet of ND$_3$ prepared by a curved hexapole~\cite{Tang2023}.

\subsection{Phase spaces}
An accurate simulation of the collision dynamics starts with an accurate simulation of the incoming molecular packets. To this end the trajectories of many individual, randomly picked point particles were simulated throughout the experiment. Between $10^5$ and $10^7$ particles were required to densely populate the six-dimensional phase spaces crossing each other in the interaction region. 

The simulated particles were initialized at the skimmer as a Gaussian packet. Conditions before the skimmer are chaotic, and were therefore excluded. Transversely, the particles were distributed uniformly over the skimmer opening while the velocities were sampled from a Gaussian distribution. The particles were created at different times with a Gaussian distribution corresponding to the valve opening duration. The longitudinal velocity of the particles was sampled from a Gaussian distribution with a velocity spread of around \SI{10}{\percent}~(FWHM) of the average velocity, matching the parameters found in \cref{sec:exp_calibration}. From here the particles experienced free-flight, modeled by linear trajectories.

Next the particles belonging to the primary beam entered the Stark decelerator. The electric fields inside the decelerator were obtained by solving the Laplace equation in SIMION~\cite{Simion8.1:2012,Dahl2000}. Combined with the known Stark shift of ND$_3$, the force experienced by the particles could be calculated. The equations of motion are determined exclusively by this force, and can be expressed as two coupled first order ordinary differential equations for each dimension, yielding six in total. Each particle was propagated through the Stark decelerator by integrating the differential equations numerically using the Runge-Kutta-Fehlberg method~\cite{Fehlberg1970}. If a particle hit one of the electrode surfaces, it was removed from the simulation. Propagating $10^6$ particles through the Stark decelerator took around 30 minutes (Intel Core i7-8700 CPU, single core).

\begin{figure}[b!]
	\centering
	\includegraphics{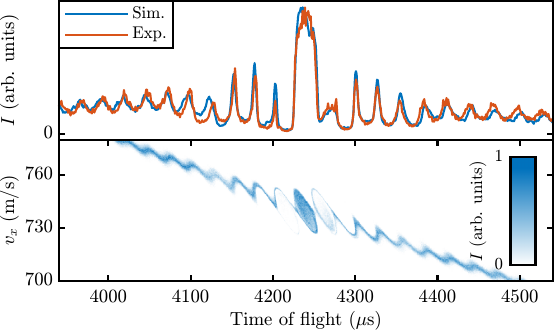}
	\caption{Molecular beam emerging from the Stark decelerator. Molecules with a velocity of \SI{740}{m/s} were selected from a beam of \SI{2}{\percent} ND$_3$ in a Ne/Ar mixture. (a) Simulated and experimental TOF profile. (b) Simulated ($t,v_x$) phase space at plane of detection.
	}
	\label{fig:Stark_phase_space}
\end{figure}

Packets emerging from the Stark decelerator could be simulated with extraordinary accuracy, thanks to the high level of control exerted by the Stark decelerator. This has been underlined by previous studies on the Stark decelerator using the same trajectory simulations~\cite{Meerakker:PRA71:053409,Meerakker:PRA73:023401,Scharfenberg:PRA79:023410}. The TOF profile obtained after the Stark decelerator when guiding an ND$_3$ packet at \SI{740}{m/s} is shown in \cref{fig:Stark_phase_space}(a), together with the simulated profile. \Cref{fig:Stark_phase_space}(b) displays the ($t,v_x$) phase space at the plane of detection. The periodic potential of the Stark decelerator has maintained a narrow packet surrounding the synchronous molecule. Particles outside this packet were not in sync with the pulse sequence, resulting in the observed `wiggles'.

We now have two phase spaces propagating towards the interaction region. Next, a way of simulating the interaction between the two packets is required.

\subsection{Classical collision sampling}
\label{subsec:CCS}
The interaction between the two packets was modeled using a method we call Classical Collision Sampling (CCS). This approach is both physically accurate and computationally efficient. It finds suitable pairs of collision partners from the input phase spaces as follows.

For two randomly picked particles, let the vectors representing their individual positions as a function of time $t$ be $\vec{x}_1(t)$, $\vec{x}_2(t)$. The distance between these two particles is 

\begin{equation}
	D(t)=|\vec{x}_1(t)-\vec{x}_2(t)|.
\end{equation}

\noindent Next, the possible collision time can be found by minimizing $D(t)$. This results in the time $t_{ca}$ and distance $D_{ca}$ of closest approach: 

\begin{align}
t_{ca} = t : \min_{t\in\mathbb{R}} D(t) \\
D_{ca} = D(t_{ca}).
\end{align}

\noindent Note that $D_{ca}$ is equal to the classical impact parameter.

Realistically two particles will only collide if they pass close to one another. For a typical cross section of \SI{100}{\angstrom\squared}, this will only occur when the distance between the two particles is around \SI{10}{\angstrom}, or one nanometer. As an approximation this constraint can be relaxed. Instead, we let pairs of particles collide if they pass below a certain limit, $D_{ca}<D_{lim}$. Effectively, this assumes that two particles passing within $D_{lim}$ are representative of two particles that pass much closer. This is a valid approximation when the phase-space density is uniform on the scale of $D_{lim}$. We varied $D_{lim}$ between \num{10} and \SI{0.01}{mm}, and found that variables like the mean collision energy converged below \SI{\sim0.5}{mm}. Throughout all simulations, we chose $D_{lim}=\SI{0.1}{mm}$. 

For particles in free flight $t_{ca}$ can be found analytically by solving the linear equation $\frac{dD^2}{dt}=0$. Note that $D^2(t)$ is minimized instead of $D(t)$ to omit a square root, which only works because $D(t)$ is strictly positive. This yields

\begin{equation}
	t_{ca} = -\frac{\Delta\vec{x}_0\cdot\Delta\vec{v}}{(\Delta\vec{v})^2},
	\label{eq:t_ca_freeflight}
\end{equation}

\noindent with the ordinary dot product, $\Delta\vec{x}_0$ the difference in position of the two particles at $t=0$ and $\Delta\vec{v}$ the difference in their velocity. In this case it is extremely efficient to check whether two particles collide. Many pairs can be evaluated and rejected quickly until $D_{ca}<D_{lim}$ is satisfied for one of them.

Now, with two viable collision partners and their collision time identified, all parameters necessary to describe the collision itself and complete the simulation are known, as has been described before~\cite{Zastrow:NatChem6:216,Zastrow2015}. For example, the COM velocity and collision energy follow from the velocity vectors of both particles. Similarly, the Newton sphere could be determined, describing all allowed final velocities. Based on the collision energy and theoretical predictions, the DCS and ICS could be assigned for each collision event. Prior to the simulation, the theoretical DCS was evaluated on a grid of deflection angles and collision energies. The resulting matrix was read by the simulation program as an input. For each collision event in the simulation the DCS and ICS were then determined by performing a linear interpolation between the two closest collision energies.

Formally the collision probability is dictated by the ICS, while the collision results in a delocalized wave covering the entire surface of the Newton sphere with a probability density given by the DCS. In this model, the Newton sphere is instead populated with individual particles that each carry the collision-energy-dependent DCS predicted by theory as a weight. The scattering probability is automatically taken into account, as the value of the ICS is retrieved by summing up all the weights, integrating the DCS. After this, each product particle is propagated to find its position when the detection laser fires. If the position lies within the laser detection volume, the weight of the particle will be added to the simulated image.

Typically, we used $10^6$ particles from the phase space of the primary beam. For each of those, we picked $10^4$ collision partners from the phase space of the secondary beam. Of these, only a fraction satisfied the condition $D_{ca}<D_{lim}$ to yield a collision. For every collision event, we then populated the resulting Newton sphere with around $2\times10^5$ particles.

The flux-to-density correction factor was finally obtained from a comparison of the numerically simulated detected number of molecules and the input ICS, and used to correct the experimentally obtained scattering signal intensities.

\section{Theoretical methods}
\subsection{Reference potential energy surface}\label{section_ref_pes}
\begin{figure}[h]
	\centering
	\includegraphics[width=.55\textwidth]{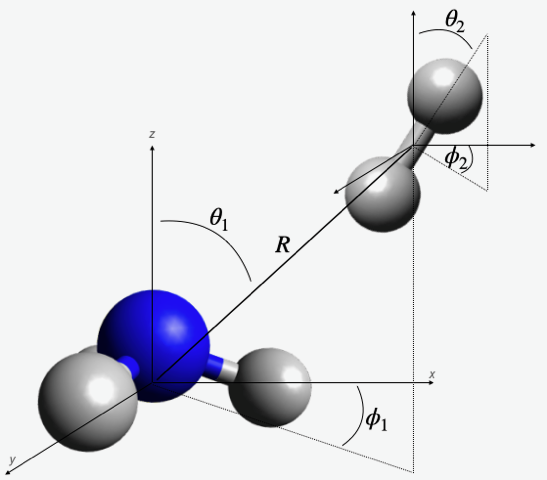}
	\caption{Coordinate system used in Ref.~\cite{Maret2009}. The origin is placed at the center of mass of NH$_3$. The molecule images were drawn in Avogadro~\cite{Hanwell2012}.}
	\label{fig_coords}
\end{figure}

The reference PES for NH$_3$-H$_2$ is that of Maret et al. \cite{Maret2009}, which has been used in a number of studies of NH$_3$-H$_2$ collisions, including deuterated isotopologues of ammonia \cite{Daniel2014, Daniel2016, Ma2015, Tkac2015b, Bouhafs2017, Gao2019, Demes2023, Loreau2023}. This five-dimensional PES depends on the coordinates $R$ (the length of the vector \textbf{R} connecting the centres of mass of NH$_3$ and H$_2$), $\theta_1$ (the angle between \textbf{R} and the $C_3$ axis of NH$_3$), $\phi_1$ (the angle of rotation of this vector around the $C_3$ axis), and  $(\theta_2, \phi_2)$, which are the polar and azimuthal angles used to describe the orientation of H$_2$ relative to NH$_3$.
Both molecules are thus assumed to be rigid rotors. The N-H bond length of NH$_3$ was fixed at $r_{\rm NH}=\SI{1.9512}{\bohr}$, and the HNH angle at \SI{107.38}{\degree}. The geometry of H$_2$ was fixed at the vibrationally-averaged value, $R_{\rm HH}=\SI{1.4488}{\bohr}$. The coordinate system is illustrated in \cref{fig_coords}.

The PES was constructed by computing the interaction energy on a grid on 29 distances in the range 3-15~\si{\bohr}. At each distance, the angular variables were sampled randomly. In total, 89,000 points were calculated at the CCSD(T) level of theory with the aug-cc-pVDZ (aVDZ for short) basis set. These points were corrected by means of 29,000 points computed at the same level of theory but with a larger aug-cc-pVTZ  basis set, based on which a complete basis set extrapolation (CBS) correction was performed. Such two-point extrapolation schemes are however known to be of limited accuracy. The resulting energies were fitted to an expansion in spherical harmonics of the form
\begin{equation}\label{PES_expansion}
	V(R,\theta_1,\phi_1,\theta_2,\phi_2) = \sum_{l_1 \mu_1 l_2 l} v_{l_1 \mu_1 l_2 l}(R) \, t_{l_1 \mu_1 l_2 l }(\theta_1,\phi_1,\theta_2,\phi_2),
\end{equation}
where the angular functions $t_{l_1 \mu_1 l_2 l}$ are given in Ref.~\cite{Phillips1994}. The radial functions were fitted using cubic splines and extrapolated to large $R$ with the appropriate $R^{-n}$ behaviour.
The fit in \cref{PES_expansion} contains 120 terms corresponding to keeping only the largest terms in \cref{PES_expansion} with maximum values of $l_1=11$ and $l_2=4$.

In order to treat ND$_3$-H$_2$ and ND$_3$-HD collisions, a coordinate transformation was performed to account for the change in the position of the centers of mass. This assumes that all isotopologues are described by the same PES, which is the case within the Born-Oppenheimer approximation. The PES was then re-expanded keeping terms up to $l_1=11$ and $l_2=4$ for ND$_3$-H$_2$ and $l_1=11$ and $l_2=6$ for the more anisotropic PES of ND$_3$-HD. For ND$_3$-HD the largest 801 terms in the expansion were retained.

\subsection{Scattering of ND$_3$ by H$_2$ and HD}

Quantum scattering calculations were then performed using a in-house scattering code in the body-fixed frame, which has previously been used to study ND$_3$-D$_2$ collisions \cite{Gao2019}. The close coupling scattering equations were solved on a dense grid of 400 energies in the range 0.01-25~\si{\per\cm} and the corresponding integral and differential cross sections were calculated at each collision energy. Convergence tests were performed for  various parameters. For ND$_3$-H$_2$ the radial grid consisted of 270 regularly spaced points in the range 4-60~\si{\bohr}, while for ND$_3$-HD the number of points was 322. The rotational basis included all levels with $j\leq 6$ (for ND$_3$) and $j_2\leq 4$ (for H$_2$/HD). In particular for transitions induced by HD, including the level $j_2=4$ of HD was found to be important to obtain fully converged cross sections.
The rotational constant of H$_2$ was taken as $B=\SI{59.340}{\per\cm}$, that of HD as $B=\SI{44.662}{\per\cm}$,
while the rotational constants of ND$_3$ were chosen as $A=\SI{5.1432}{\per\cm}$ and $C=\SI{3.1015}{\per\cm}$.
The symmetry group used in the calculations is the permutation-inversion group $G_{24}$ for ND$_3$-H$_2$ and $G_{12}$ for ND$_3$-HD, the latter being isomorphic to $D_{3h}$.

The inversion splitting in the ground umbrella vibrational state is \SI{0.0530}{\per\cm}. Since the PES does not describe the inversion of ND$_3$, the umbrella motion was treated with a two-state model in which the inversion-tunnelling states are taken as linear combinations of the two rigid equilibrium structures. This model was shown to be in excellent agreement with results obtained by treating the umbrella motion of ammonia explicitly for the scattering of NH$_3$ with rare gas atoms.\cite{Gubbels:JCP136:074301, Loreau2015, Loreau2024}
Partial waves with total angular momenta up to $\mathcal{J} = 16$ were considered, and the convergence of the cross sections was checked with respect to all the parameters discussed above.

\begin{figure}[bt]
	\centering
	\includegraphics[scale=1.1]{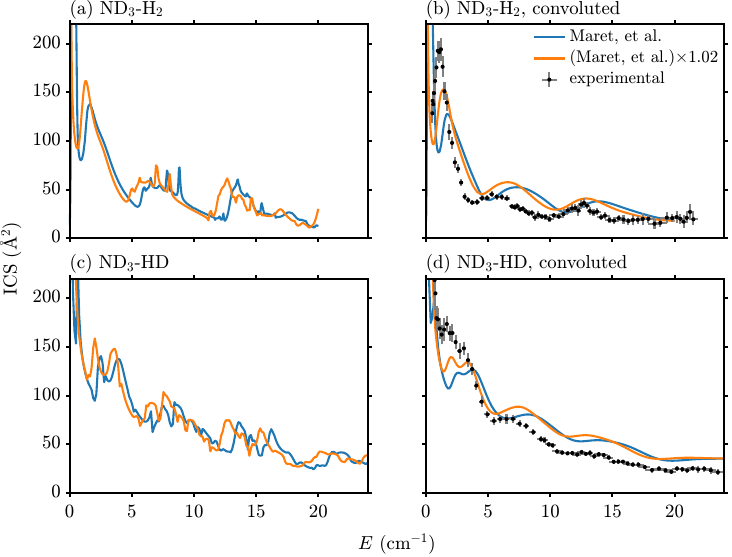}
	\caption{Cross sections for the $j_k^\pm = 1_1^- \rightarrow 1_1^+$ transition in ND$_3$ induced by collision with H$_2$ (top panels) or HD (bottom panels). Panels (a) and (c) display the results obtained with the reference PES of Maret et al. \cite{Maret2009} as well as the impact of multiplying the whole PES by a constant scaling of 1.02. Panels (b) and (d) show a comparison between the theoretical cross sections convoluted with the experimental spread and the experimental cross sections. Horizontal experimental error bars reflect the energy calibration uncertainty, computed by propagating the uncertainty in $v_2$ (\cref{tab:v2_calibration}). Vertical error bars represent statistical uncertainties, calculated as the standard deviation of the mean over hundreds of samples (\cref{subsec:det_and_data_acq}). All error bars represent a \SI{95}{\percent} confidence interval.}
	\label{fig_MaretPES}
\end{figure}

The results of these calculations are shown in \cref{fig_MaretPES} for the $j_k^\pm = 1_1^- \rightarrow 1_1^+$ transition. For ND$_3$-H$_2$, while theory (convoluted by assuming an experimental Gaussian energy spread with \SI{10}{\percent} FWHM) predicts a series of resonances in the range of energy studied, the energies and strengths of the resonance peaks in the ICS does not reproduce the experimental data. All resonances occur at higher energies than observed in the experimental ICSs. Scattering calculations were then performed on a PES scaled by a constant multiplying factor in the range $1.005-1.04$. This increases the depth of the PES by $0.5-4$~\si{\percent}, thereby shifting the position of the resonances to lower collision energies. We found that by scaling the PES by a factor 1.02, the resonances are slightly better reproduced, although it proved impossible to accurately reproduce the position and magnitude of all resonances simultaneously with a single scaling factor, as shown in \cref{fig_MaretPES}.

Given that the scattering calculations are fully converged, it is likely that the PES of \cite{Maret2009} is not accurate enough to match the experimental accuracy at low collision energy. Possible explanations for this observation include (i) the level of theory not being high enough, (ii) the basis sets used not being large enough to perform a CBS extrapolation, (iii) the fit of the PES not being accurate enough; (iv) the large value chosen for the N-H bond (\SI{1.9512}{\bohr}, compared to values \SI{\sim1.92}{\bohr} used in other works on NH$_3$-rare gas scattering \cite{Gubbels:JCP136:074301, Loreau2015, Loreau2024} as derived from the inertia moments of NH$_3$ or ND$_3$); and (v) vibrational effects, e.g., an effect of the umbrella motion, which must be described explicitly instead through a PES and a Hamiltonian that both depend on the umbrella coordinate. Below we examine all of these effects to gain insight into their relative importance.

\subsection{Impact of the N-H bond distance}
The effect of reducing the N-H distance is expected to reduce the strength of the interaction between NH$_3$ and H$_2$, which is confirmed by our tests performed at the CCSD(T) level with various basis sets. The depth of the PES was found to vary by approximately \SI{0.8}{\percent} at the CCSD(T)/aVQZ level when increasing the N-H bond length from a value of \SI{1.92}{\bohr} to \SI{1.95}{\bohr}, while the correction is smaller in the long range and larger in the short-range, repulsive region, and also depends on the orientation of the monomers. It is seen that the magnitude of this effect cannot be large enough to explain the discrepancy between theory and experiment. In addition, using a N-H bond length shorter than that used in Ref. \cite{Maret2009} leads to a shallower PES, which shifts resonances towards higher collision energies, thereby worsening the discrepancy between theory and experiment.
Finally, we note that an accurate value of the vibrationally-averaged bond length, $r_{NH}=\SI{1.946}{\bohr}$, can be extracted from high level quantum chemistry calculations \cite{Huang2008, Lee_priv_comm}, which is close to the value employed in Ref. \cite{Maret2009}.

\subsection{Impact of the umbrella motion}
The impact of the umbrella motion in NH$_3$-rare gas scattering was examined in previous works, see e.g. \cite{Gubbels:JCP136:074301, Loreau2015,Loreau2024, Tkac2015a} by comparing results obtained with a two-state model that considers that ammonia exists as a superposition of two equilibrium structures with a model that describes the inversion motion explicitly. Such calculations require two ingredients: a PES that includes a dependence on the umbrella coordinate $\rho$, and an extension of the rigid-rotor Hamiltonian and close-coupling equations to include an explicit dependence on $\rho$. The latter is realized by including a kinetic energy operator for the motion along the umbrella coordinate, as well as a double-well potential that describes the inversion. The theory and application for NH$_3$-rare gas collisions can be found in the above-mentioned papers.
It was found that for NH$_3$-He, the impact of the umbrella motion is negligible \cite{Gubbels:JCP136:074301, Loreau2015} down to energies of \SI{1}{\per\cm}. Unsurprisingly, this roughly corresponds to the inversion splitting (\SI{0.79}{\per\cm}), where an effect of the umbrella motion can be expected. On the other hand, for NH$_3$-Ne and NH$_3$-Ar collisions differences between the two models persist up to energies of about \SI{10}{\per\cm}, indicating that this effect might be important for NH$_3$-H$_2$ in the context of the present measurements \cite{Loreau2015}. The effect is however expected to be smaller for ND$_3$, given that the inversion splitting is \SI{0.053}{\per\cm}, about 15 times smaller than for NH$_3$.

\subsubsection{A new PES with inversion}
As a first step, we constructed a new PES that includes the inversion coordinate $\rho$. This leads to a six-dimensional PES depending on the five coordinates defined in \cref{section_ref_pes} and the umbrella angle $\rho$. The H-H distance in the H$_2$ molecule was fixed at $R_{\rm HH}=\SI{1.4488}{\bohr}$. The PES was expanded in angular functions defined in Eqs.~(6) and (7) of Ref.~\cite{Avoird2011} depending on the body-fixed Euler angles $(\alpha_A,\beta_A,\gamma_A,\beta_B)$. These angular functions differ from the expansion functions in \cref{PES_expansion}; the expansion coefficients $v_{L_A,K_A,L_B,L}(R,\rho)$ are related to the coefficients $v_{l_1	\mu_1 l_2 l}(R)$ in \cref{PES_expansion} by a transformation given in Eqs.~14 and 15 of Ref.~\cite{Avoird2011}. A grid of 28 distances $R$ was used in the range from \SI{4.2}{\bohr} to \SI{30}{\bohr}, with a step size that increases from \SI{0.3}{\bohr} at short range to \SI{3}{\bohr} in the long range region. Taking into account the symmetry of the problem, 1056 different orientations were considered for each value of $R$ and for 10 values of the angle $\rho$ between \SI{51.48}{\degree} and \SI{128.52}{\degree}, leading to a total of 295,680 \textit{ab initio} points. These were calculated at the CCSD(T)-F12a/aVTZ level of theory.

For all values of $R$ and $\rho$, the potential was expanded in the angular functions depending on the Euler angles $(\alpha_A,\beta_A, \gamma_A, \beta_B)$. The expansion coefficients $v_{L_A,K_A,L_B,L}(R,\rho)$ were calculated for each $R$ and $\rho$ by integration over the four angular coordinates using Gauss-Legendre
quadrature for $\beta_A$ (with 12 points) and $\beta_B$ (with 5 points) and Gauss-Chebyshev quadrature for $\alpha_A$ (with 5 points) and $\gamma_A$ (with 7 points). The $\rho$ dependence of the coefficients $v_{L_A,K_A,L_B,L}(R,\rho)$ was taken into account by means of a ten-term polynomial fit in ($\rho-\rho_{\rm eq}$), where $\rho_{\rm eq}$ is the equilibrium value of the umbrella coordinate. The resulting $R$ dependent coefficients were then interpolated with the reproducing kernel Hilbert space (RKHS) method \cite{rkhs96}. The smoothness parameter $n$ of the RKHS interpolation was set to 2 and the parameter $m$ which determines the long range behavior of the potential was chosen to depend on $L_A,L_B,L$. For terms with $L_A+L_B=L$ containing electrostatic multipole-multipole interactions $m$ was set to $L$ so that the potential correctly decays as $R^{-(L+1)}$. For the other terms $m$ was set to 5 and 6 for terms with even and odd $L_A+L_B$, respectively, which makes the potential decay as $R^{-6}$ and $R^{-7}$. To confirm the quality of the RKHS method when extrapolating over the long-range part of the PES, we performed additional ab initio calculations. For 6 different molecular orientations, we computed points at radii between \SI{25}{\bohr} and \SI{60}{\bohr}. Comparing these points to the extrapolated values, we found excellent agreement, with the error at large distances being on the order of \SI{1e-3}{\per\cm}.

\subsubsection{Scattering with inversion}

In a second step, the scattering code was updated to allow for the explicit treatment of the umbrella motion. The procedure is similar to the one explained in previous papers \cite{Gubbels:JCP136:074301, Loreau2015,Loreau2024}, to which we refer the interested reader. The wavefunctions are obtained on a grid of umbrella angles. We performed the calculations by including the first two wavefunctions (corresponding to the $+$ and $-$ state of the ground vibrational state) and carried out tests with four wavefunctions which showed that the results are converged.

\begin{figure}[h]
	\centering
	\includegraphics[scale=1.1]{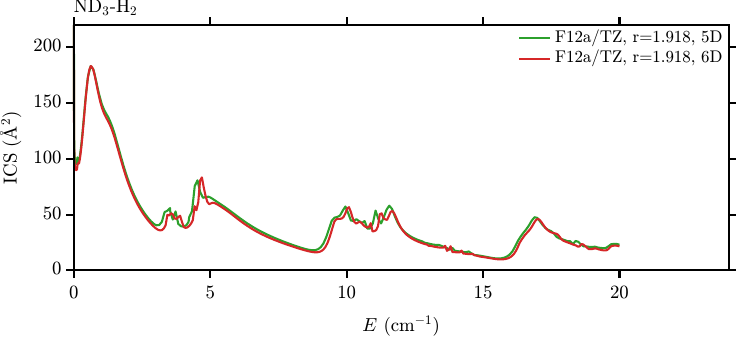}
	\caption{Impact of the umbrella motion on the $j_k^\pm=1_1^- \rightarrow 1_1^+$ cross section.}
	\label{fig_umbrella}
\end{figure}

The cross section for the $j_k^\pm = 1_1^- \rightarrow 1_1^+$ cross section is shown in \cref{fig_umbrella} and compared to scattering using the to-state model on the PES with $\rho$ fixed at its equilibrium value. The effect of the explicit umbrella motion is to shift the position of the resonances towards slightly higher collision energies. It thus appears that the widely-used two-state model with rigid NH$_3$ is valid in the range of collision energies explored here.

\subsection{Impact of the level of theory and basis set}\label{section_basis_set}
Additional tests were carried out for several geometries.
In these tests we compared the CCSD(T) and CCSD(T)-F12 calculations with a series of Dunning basis sets aVnZ on all atoms, with $n=2$ to 6. In addition, we also performed calculations with a basis set consisting of the aVnZ basis set supplemented by a set of ($3s3p2d2f1g$) diffuse midbond functions \cite{Cybulski1999} that was shown to give accurate results for NH$_3$-Ar \cite{Loreau2014b}.
Finally, the complete basis set extrapolation was applied to the CCSD(T)/aVnZ results by extrapolating separately the Hartree-Fock energy and the correlation energy. The CBS extrapolation is not unique as it can be performed based on two, three, or four basis sets. All tests were carried out for a N-H distance fixed the equilibrium value of \SI{1.918}{\bohr}.

The results are shown for three geometries in \cref{table_geom1,table_geom2,table_geom3}. All energies are given in \si{\per\cm}. The results from the three best CBS extrapolations are reported (based on T-Q-5, Q-5-6, and T-Q-5-6 results), hence a range of values. The extrapolation based on two terms only (e.g., DZ+TZ as in Ref. \cite{Maret2009}) resulted in energies that differed from the three- and four-point CBS by more than \SI{2}{\percent} and could not be trusted.


\begin{table}[h!]
	\centering
	\caption{\label{table_geom1}
		Geometry 1: $\theta_1 = \phi_1 = \theta_2 = \phi_2=0$, $R=\SI{6.12}{\bohr}$ (H$_2$ aligned with $C_3$ axis of NH$_3$, on the N atom side, corresponding to the global minimum of the PES). Energies given in \si{\per\cm}.}
	\begin{tabular}[t]{ccccc}
		$n$	& CCSD(T)/aVnZ	& CCSD(T)-F12a/aVnZ	& CCSD(T)-F12b/aVnZ		& CCSD(T)/aVnZ+mb 			\\ \hline
		2	& $-207.08$	& $-250.64$	& $-246.42$		& $-257.65$		\\
		3	& $-255.42$	& $-263.89$	& $-261.24$		& $-263.53$		\\
		4	& $-264.37$	& $-268.01$	& $-266.42$		& $-266.78$		\\
		5	& $-266.80$	& 			& 				& $-267.84$		\\
		6	& $-267.93$	& 			& 				& 		\\
		CBS	& $-269.17 - -270.44$	& &				& 		\\
	\end{tabular}
\end{table}

\begin{table}[h!]
	\centering
	\caption{\label{table_geom2}
		Geometry 2: $\theta_1=90, \phi_1=60, \theta_2 = \phi_2=0, R=\SI{6.14}{\bohr}$. (H$_2$ parallel to $C_3$ axis of NH$_3$, next to the center of mass, between two H atoms of NH$_3$). Energies given in \si{\per\cm}.
	}
	\begin{tabular}[t]{ccccc}
		$n$	& CCSD(T)/aVnZ	& CCSD(T)-F12a/aVnZ	& CCSD(T)-F12b/aVnZ		& CCSD(T)/aVnZ+mb 			\\ \hline
		2	& $-82.22	$	& $-101.99$	& $-98.87	$	& $-125.19$		\\
		3	& $-112.07$	& $-115.59$	& $-113.62$	& $-120.83$		\\
		4	& $-118.47$	& $-119.86$	& $-118.84$	& $-120.74$		\\
		5	& $-119.88$	& 			& 			& $-120.81$		\\
		6	& $-120.56$	& 			& 			& 		\\
		CBS	& $-121.06 - -122.22$	& &			& 		\\
	\end{tabular}
\end{table}

\begin{table}[h!]
	\centering
	\caption{\label{table_geom3}
		Geometry 3: $\theta_1=180, \phi_1= \theta_2 = \phi_2=0, R=\SI{6.76}{\bohr}$ (H$_2$ aligned with the $C_3$ axis of NH$_3$, on the H atoms side). Energies given in \si{\per\cm}.
	}
	\begin{tabular}[t]{ccccc}
		$n$	& CCSD(T)/aVnZ	& CCSD(T)-F12a/aVnZ	& CCSD(T)-F12b/aVnZ		& CCSD(T)/aVnZ+mb 			\\ \hline
		2	& $-48.71$	& $-57.03$	& $-55.15$		& $-76.21$		\\
		3	& $-69.13$	& $-70.42$	& $-69.11$		& $-74.04$		\\
		4	& $-72.71$	& $-73.82$	& $-73.09$		& $-74.41$		\\
		5	& $-74.00$	& 		& 			& $-74.64$		\\
		6	& $-74.50$	& 		& 			& 		\\
		CBS	& $-74.33 - -75.50$	& &			& 		\\
	\end{tabular}
\end{table}

\begin{figure}[t]
	\centering
	\includegraphics[scale=1.1]{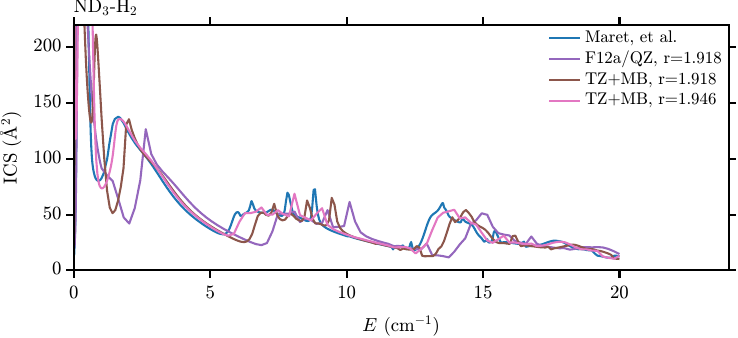}
	\caption{ICS for the ND$_3(1_1^-) + $H$_2 \rightarrow $ ND$_3(1_1^+) + $H$_2$ with different PESs. See text for details.}
	\label{fig_comp_new_PES}
\end{figure}

Based on these results, we conclude that the combinations CCSD(T)-F12a/aVQZ and CCSD(T)/aVTZ+mb both give results that are in good agreement with the CBS results, although the level of agreement depends on the geometry.

Two new five-dimensional PESs (for rigid NH$_3$ with a N-H distance set at the equilibrium value, \SI{1.918}{\bohr}) were constructed for these two combinations using the methods discussed above. For each PES, the same grid of angles and distances was used, resulting in 29,568 unique geometries. The ICS for the transition $j_k^\pm=1_1^- \rightarrow 1_1^+$ in ND$_3$-H$_2$ collisions is shown in \cref{fig_comp_new_PES} and compared with the ICS obtained with the Maret et al. PES. The behaviour of the cross section is similar with all three PESs. However, both the CCSD(T)/aVTZ+mb PES as well as the CCSD(T)-F12a/aVQZ PES predict resonances that are shifted towards higher energies compared to the results obtained with the PES from Ref. \cite{Maret2009}. We note that the use of the CCSD(T)/aVTZ+mb PES leads to a better agreement with experimental data than the CCSD(T)-F12a/aVQZ PES. Overall It is again clear that these new PESs are unable to fully explain the experimental position of the resonances. A third new PES was constructed with the CCSD(T) method combined with the aVTZ+mb basis set for a N-H distance set at the vibrationally-averaged value of \SI{1.946}{\bohr} \cite{Huang2008, Lee_priv_comm}. The result is also shown on \cref{fig_comp_new_PES} and is in good agreement with the ICSs calculated based on the PES from Ref. \cite{Maret2009}.

\subsection{Impact of quadruple excitations in CCSDT(Q)}

Lastly, we investigated the impact of further corrections in the coupled cluster expansion, namely the impact of including explicit triple excitations (CCSDT) and including quadruple excitations at the perturbative level [CCSDT(Q)]. In a recent work on NO-He collisions \cite{Jongh2020}, the effect of perturbative quadruples was investigated over the whole PES at the aVDZ level. However, due to the much larger number of geometries required to obtain a complete PES (29,568 for NH$_3$-H$_2$ for a five-dimensional PES instead of 912 for the two-dimensional PES of NO-He) and the associated increase in computational time, it proved impossible to construct a full PES with these corrections.

\begin{figure}[p]
	\centering
	\includegraphics[,width=.6\textwidth]{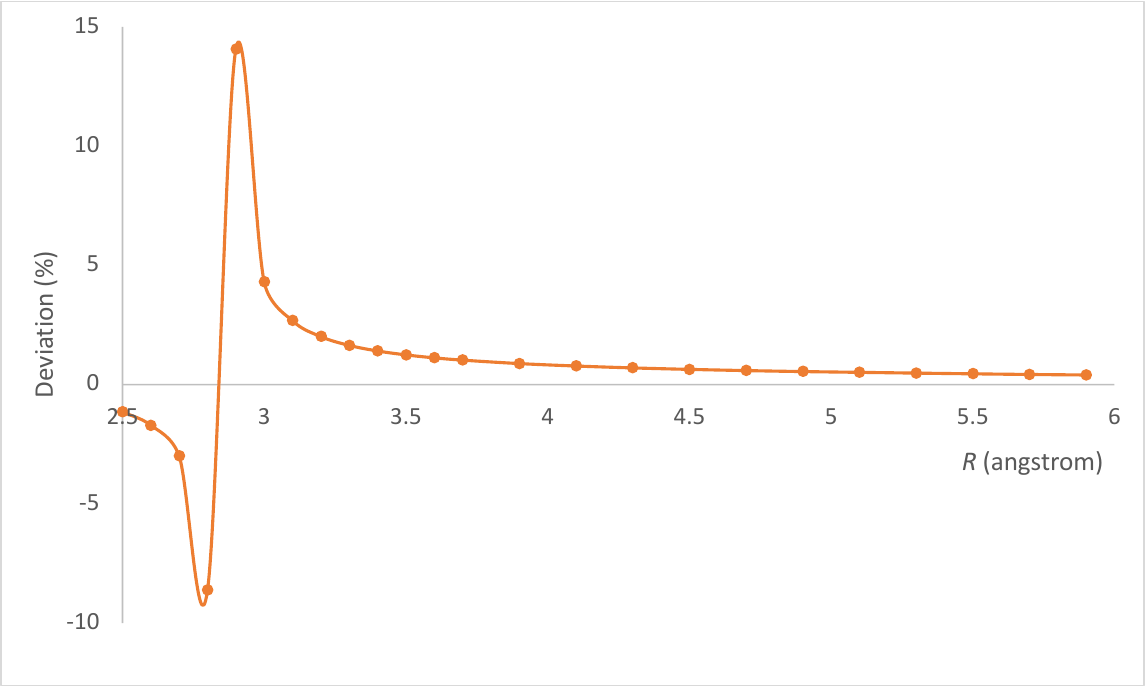}
	\caption{Deviation between CCSDT(Q) and CCSD(T) results as function of the intermolecular distance $R$, in \si{\percent}, for the orientation $\theta_1=0, \phi_1=0, \theta_2=0, \phi_2=0$.}
	\label{fig_CCSDTQ_percent}
	
	\includegraphics[,width=.6\textwidth]{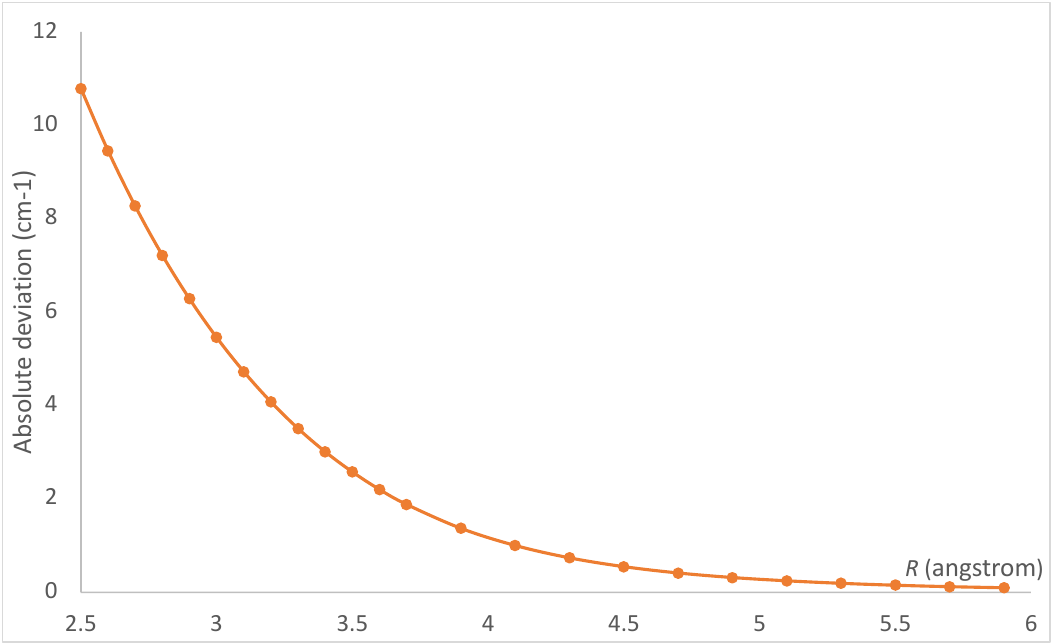}
	\caption{Same as \cref{fig_CCSDTQ_percent}, but absolute deviation.}
	\label{fig_CCSDTQ_absolute}
\end{figure}

Tests calculations were performed with the aVDZ basis set and a N-H distance of \SI{1.946}{\bohr} for selected geometries using the MRCC program interfaced with MOLPRO \cite{MRCC}. The results are reported in \cref{table_CCSDTQ} for the three geometries used in \cref{section_basis_set}. The effect of including full triple and perturbative quadruple excitations is seen to be non-negligible for these geometries. We then investigated the dependence of this correction on the intermolecular distance $R$. A representative example is shown in \cref{fig_CCSDTQ_percent,fig_CCSDTQ_absolute}. This illustrates the impact of the CCSDT(Q) correction compared to CCSD(T) results as function of $R$ for the orientation $\theta_1=0, \phi_1=0, \theta_2=0, \phi_2=0$, corresponding to H$_2$ aligned with the $C_3$ axis of NH$_3$, on the N atom side. The deviation, expressed in \si{\percent} of the interaction energy, varies strongly with distance. It is the largest when the interaction potential is close to zero and changes sign when the potential becomes repulsive, which is strongly dependent on the orientation. The deviation decreases with increasing distance, with typical corrections smaller than \SI{0.5}{\percent} for $R>\SI{6}{\AA}$.

\begin{table}[p]
	\centering
	\caption{\label{table_CCSDTQ}
		Comparison of CCSD(T), CCSDT, and CCSDT(Q) energies for three NH$_3$-H$_2$ geometries with aVDZ basis sets on all atoms. The energy is given in units of \si{\per\cm}. The deviation (in \si{\percent}) refers to the difference between CCSD(T) and CCSDT(Q) energies.
	}
	\begin{tabular}[t]{cccc}
		& Geometry 1	& Geometry 2	& Geometry 3 	\\ \hline
		CCSD(T)		& -207.08		& -82.22		& -48.71		\\
		CCSDT		& -210.30		& -84.01		& -50.81		\\
		CCSDT(Q)	& -210.93		& -84.48		& -51.12		\\
		Deviation (\si{\percent}) 		& 1.86		& 4.95		& 2.74		\\
	\end{tabular}
\end{table}

\begin{figure}[h!]
	\centering
	\includegraphics[scale=1.1]{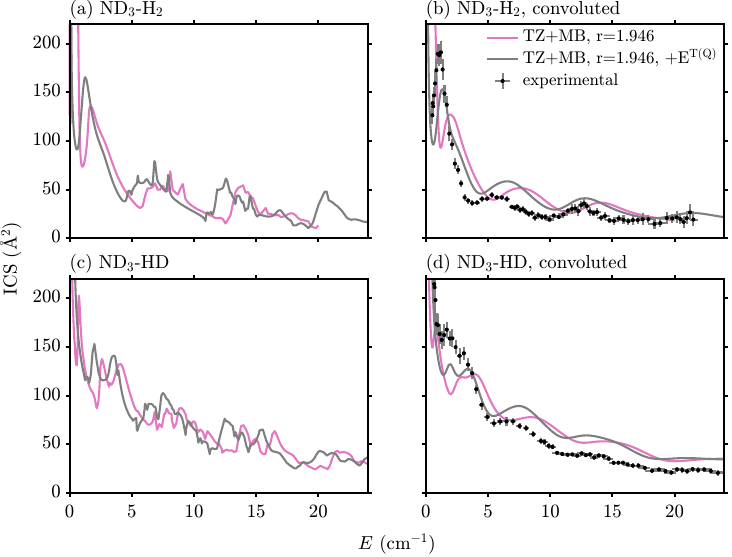}
	\caption{Cross sections for the $j_k^\pm = 1_1^- \rightarrow 1_1^+$ transition in ND$_3$ induced by collision with H$_2$ (top panels) or HD (bottom panels) with a PES computed at the CCSD(T)/aVTZ+mb, with or without the correction included to account for the full triple and perturbative quadruple excitations. The left panels display the theoretical ICSs while the right panels show the convoluted ICSs compared to the experimental data. Horizontal experimental error bars reflect the energy calibration uncertainty, computed by propagating the uncertainty in $v_2$ (\cref{tab:v2_calibration}). Vertical error bars represent statistical uncertainties, calculated as the standard deviation of the mean over hundreds of samples (\cref{subsec:det_and_data_acq}). All error bars represent a \SI{95}{\percent} confidence interval.}
	\label{Fig_CCSDTQ_ICS}
\end{figure}

\Cref{fig_CCSDTQ_percent} shows the absolute deviation between the CCSDT(Q) and CCSD(T) interaction potential in \si{\per\cm} for the same orientation. The magnitude of the impact of quadruple excitations is inversely proportional to the distance. By performing similar calculations for four other orientations of H$_2$, we found that the results shown in \cref{fig_CCSDTQ_absolute} were almost identical for all orientations. This suggests to apply a unique radial correction to the whole PES, averaged over the orientations of NH$_3$ and H$_2$. This correction was fitted to an analytical expression, $E_{\textrm{CCSDT(Q)-CCSD(T)}}=360\exp(-1.43R)$, with $R$ in \si{\bohr} and the energy in \si{\per\cm}. This correction was then added to our best PES, namely the one constructed with the combination CCSD(T)/aVTZ+mb.
The cross sections computed with this PES are presented in \cref{Fig_CCSDTQ_ICS} and compared to the ICSs obtained on the PES without the CCSDT(Q) correction. These data show a better agreement between theory and experiment for both ND$_3$-H$_2$ and ND$_3$-HD. Interestingly, we note that using this PES we recover almost exactly the results obtained with the reference PES of Ref. \cite{Maret2009} scaled by \SI{2}{\percent}, this despite the fact that the impact of the CCSDT(Q) correction does not correspond to a simple scaling law.

\subsection{Behaviour of the cross section at low energy}

\begin{figure}[t]
	\centering
	\includegraphics[scale=1.1]{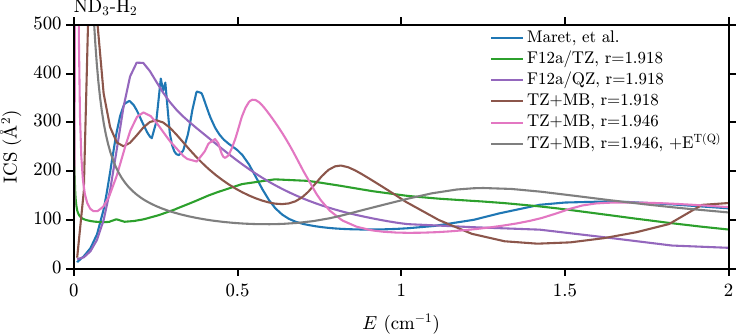}
	\caption{Low energy behavior of the $j_k^\pm = 1_1^- \rightarrow 1_1^+$ transition in ND$_3$ induced by H$_2$ collisions at low energy with the different PESs used in this work.}
	\label{fig_low_energy}
\end{figure}

At very low collision energies (\SI{<2}{\per\cm}) the various PES lead to dramatically different cross sections for the $j_k^\pm=1_1^- \rightarrow 1_1^+$. This is illustrated in \cref{fig_low_energy}, where six different PESs are compared.
While all PESs lead to the prediction of one or multiple resonances below collision energies of \SI{2}{\per\cm}, the most accurate PES (CCSD(T)/aVTZ+mb + CCSDT(Q) correction) predicts a single resonance, with a position (at $E=\SI{1.25}{\per\cm}$) in excellent agreement with experiment.

\subsection{Partial wave analysis}
To fully characterize the resonances a partial wave analysis can be performed. This will be done here for ND$_3$-H$_2$ only, as the observed resonances are most pronounced for this system. There are two conserved quantities while scattering, the total angular momentum $\mathcal{J}$, which is obtained by coupling all the angular momenta

\begin{equation}
	\bm{\mathcal{J}} = \vec{j} + \vec{j_2} + \bm{\ell},
	\label{eq:angular_momentum}
\end{equation}

\noindent and parity~\cite{Green1980}
\begin{equation}
	\mathcal{P} = p\cdot(-1)^{k+\ell},
	\label{eq:parity}
\end{equation}

\noindent where $\ell$ is the angular momentum quantum number of the partial wave, $j$, $k$ and $p$ are the quantum numbers of ND$_3$, denoted before as $j_k^p$, and $j_2$ is the rotational angular momentum of H$_2$. Fig.~2(a,c) of the main text shows the contribution of each total angular momentum state $\mathcal{J}$ to the ICS as a function of collision energy. This reveals which $\mathcal{J}$ corresponds to each resonance. In principle, these contributions should also be split per $\mathcal{P}$. However, in all cases both parities contribute near-equally, so their contribution was summed here. This is due to the small inversion splitting of ND$_3$, which causes the energy differences between states to be nearly independent of the overall parity. The ten most prominent resonances are listed in \cref{tab:res_ND3_H2}, with the corresponding value for $\mathcal{J}$.

\begin{figure}[p!]
	\centering
	\includegraphics{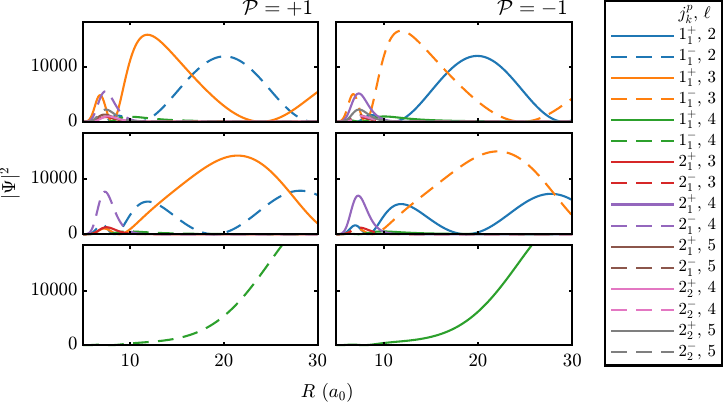}
	\caption{Calculated wave function for ND$_3$-H$_2$ scattering with $\mathcal{J}=3$ at $E=\SI{1.23}{\per\cm}$, the position of a resonance. Squared amplitude $\abs{\Psi}^2$ as function of the distance $R$ in Bohr radii, \si{\bohr}, between the two scattering partners. There are three scattering states (top to bottom) for each parity $\mathcal{P}$, one for each final state ($1_1^p,\ell\in\{5,6,7\}$), as discussed in the text.
	}
	\label{fig:wavef}
\end{figure}

\begin{table}[p!]
	\begin{center}
		\begin{threeparttable}
			\begin{tabular}{*7{r}}
				\hline\hline \rule{0pt}{2ex}
				$E$ (cm$^{-1}$) & $\mathcal{J}$ & $j_k^p{}_\text{,res}\bm{{}^\ast}$ & $\ell_\text{res}$ & $\ell_\text{in}\bm{{}^\ast\emph{}}$ & $\ell_\text{out}\bm{{}^\ast}$ & Type \\
				\hline \rule{0pt}{2ex}
				$1.23$      & $3$ & $2_1^-$ & $4$ & $2,4$ & $3  $ & Feshbach \\
				$1.75$      & $2$ & $2_1^-$ & $4$ & $2  $ & $1,3$ & Feshbach \\
				$5.46$      & $7$ & $2_1^+$ & $5$ & $6,8$ & $7  $ & Feshbach \\
				$6.15$      & $4$ & $2_2^+$ & $6$ & $4  $ & $3,5$ & Feshbach \\
				${}^\dag$   & $4$ & $2_1^+$ & $5$ & $4  $ & $3,5$ & Feshbach \\
				$6.81$      & $5$ & $2_2^+$ & $6$ & $4,6$ & $5  $ & Feshbach \\
				$7.87$      & $8$ & $1_1^-$ & $8$ & $8  $ & $7,9$ & Shape \\
				$7.97$      & $7$ & $1_1^-$ & $8$ & $6,8$ & $7  $ & Shape \\
				$12.10$     & $7$ & $2_1^-$ & $6$ & $6,8$ & $7  $ & Feshbach \\
				$12.60$     & $8$ & $2_1^-$ & $6$ & $8  $ & $7,9$ & Feshbach \\
				$14.47$     & $6$ & $2_2^-$ & $7$ & $6  $ & $5,7$ & Feshbach-Shape \\
				\hline\hline
			\end{tabular}
			\begin{tablenotes}
				\item[$\bm{\ast}$] These values are for $\mathcal{P}=+1$. For $\mathcal{P}=-1$, $p$ changes sign and the assignment of $\ell_\text{in}$ and $\ell_\text{out}$ is swapped.
				\item[$\dag$] Two resonant channels were present at $E=\SI{6.15}{\per\cm}$.
			\end{tablenotes}
		\end{threeparttable}
		\caption{Characteristics of the ten most prominent resonances appearing in the ICS for ND$_3$-H$_2$ scattering.
		} 
		\label{tab:res_ND3_H2}
	\end{center}
\end{table}

To assign these resonances we calculated the scattering wave functions at the corresponding energies, similar to earlier work on NH$_3$-H$_2$~\cite{Ma2015} and NO-He~\cite{Jongh2020}. \Cref{fig:wavef} shows one of these calculated wave functions as an example, for the resonance $E=\SI{1.23}{\per\cm}$, $\mathcal{J}=3$. Note that only the ingoing $1_1^-$ state, together with $\mathcal{P}$ and $\mathcal{J}$, is used as a boundary condition to compute these wavefunctions. As such, the wavefunction can extend over all open final states. These final states are allowed combinations of $\ell$ and $j_k^p$. At low energies only $1_1^p$ contributes (one per $\mathcal{P}$), with $\ell \in \{\mathcal{J}-1,\mathcal{J},\mathcal{J}+1\}$. Hence, the wavefunction consists of three components for collision energies below the threshold for $2_2^p$ (and 8 components above the $2_2^p$ threshold). There are three panels per parity in \cref{fig:wavef}, one for each of these components. However, instead of each panel corresponding to a single final state, they instead represent so-called scattering states. These scattering states are constructed such that the wavefunction can be plotted in terms of real, properly normalized components, but cannot straightforwardly be expressed in terms of the final states anymore.

\Cref{eq:angular_momentum,eq:parity} hold throughout the scattering process, but in particular for the ingoing, resonant and outgoing channels, which we denote with subscripts (e.g. $\ell_\text{in}$, $\ell_\text{res}$, $\ell_\text{out}$). Since only collisions in which ND$_3$ transfers from $j_k^p{}_\text{,in}=1_1^-$ to $j_k^p{}_\text{,out}=1_1^+$ are detected here, $\ell$ should scatter from $\ell_\text{in}$ even to $\ell_\text{out}$ odd or vice versa by \cref{eq:parity}. \Cref{eq:angular_momentum} furthermore implies $\ell_\text{in/out}=\left\{ \mathcal{J}-1,\mathcal{J},\mathcal{J}+1 \right\}$. As an example, again for the $\mathcal{J}=3$ resonance at $E=\SI{1.23}{\per\cm}$ this means $\ell_\text{in/out}=\left\{ 2,3,4 \right\}$. Picking one of the parities, say $\mathcal{P}=+1$, further restricts $\ell_\text{in}=\left\{ 2,4 \right\}$, $\ell_\text{out}=3$.

\Cref{fig:wavef} directly reveals the character of this resonance, as the resonant channel is the one with the strongest amplitude at close range, $R\sim\SI{8}{\bohr}$. So, for $\mathcal{P}=+1$ this particular resonance can be attributed to $j_k^p{}_\text{,res} = 2_1^-$, $\ell_\text{res}=4$, which makes it a Feshbach resonance (the $2_1^-$ state is asymptotically closed). The resonance has a weak amplitude and appears as a broad feature in the ICS in Fig.~2(a) of the main text, indicating the corresponding quasi-bound state is short-lived. At long range, $R\sim\SI{30}{\bohr}$, only the open channels contribute with values for $\ell$ satisfying \cref{eq:angular_momentum}. Both parities behave identically, except for the swapped inversion parity label and small differences in the amplitude.

\begin{figure}[p!]
	\centering
	\includegraphics{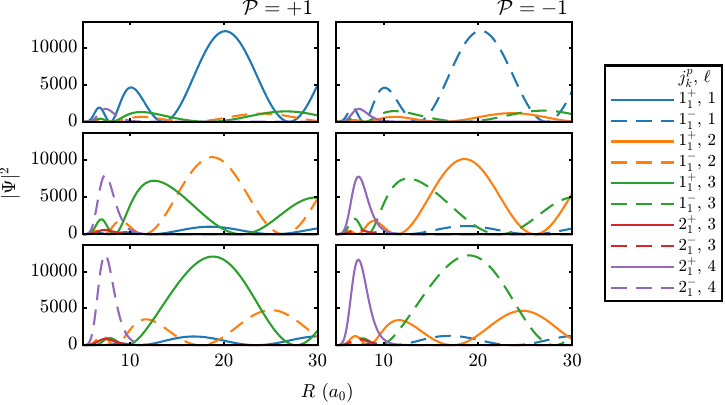}
	\caption{Calculated wave function for ND$_3$-H$_2$ scattering with $\mathcal{J}=2$ at $E=\SI{1.75}{\per\cm}$, the position of a resonance. As in \cref{fig:wavef}.
	}
	\label{fig:wavef_1.75}
\end{figure}

\begin{figure}[p!]
	\centering
	\includegraphics{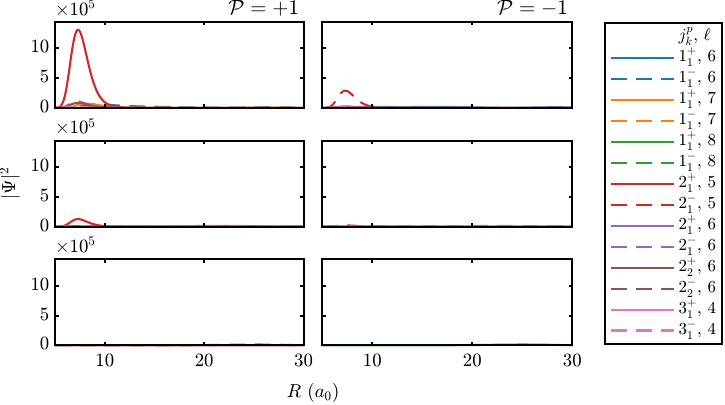}
	\caption{Calculated wave function for ND$_3$-H$_2$ scattering with $\mathcal{J}=7$ at $E=\SI{5.46}{\per\cm}$, the position of a resonance. As in \cref{fig:wavef}.
	}
	\label{fig:wavef_5.46}
\end{figure}

\begin{figure}[p!]
	\centering
	\includegraphics{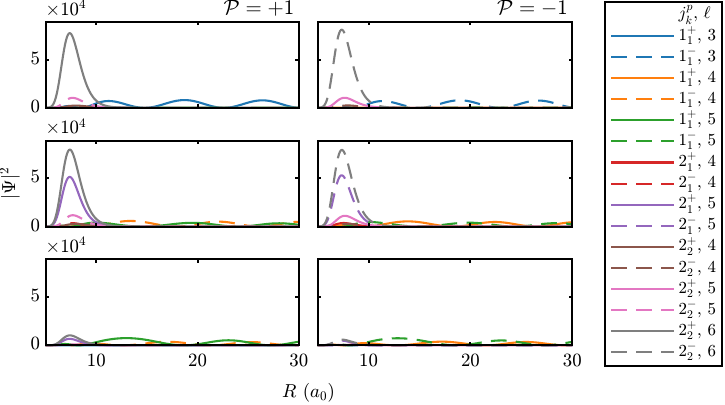}
	\caption{Calculated wave function for ND$_3$-H$_2$ scattering with $\mathcal{J}=4$ at $E=\SI{6.15}{\per\cm}$, the position of a resonance. As in \cref{fig:wavef}.
	}
	\label{fig:wavef_6.15}
\end{figure}

\begin{figure}[p!]
	\centering
	\includegraphics{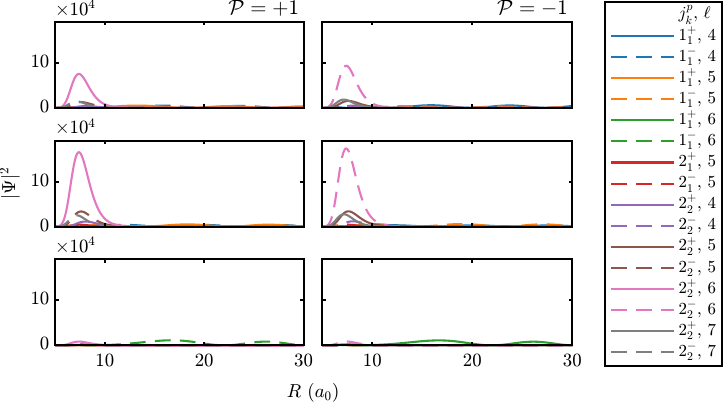}
	\caption{Calculated wave function for ND$_3$-H$_2$ scattering with $\mathcal{J}=5$ at $E=\SI{6.81}{\per\cm}$, the position of a resonance. As in \cref{fig:wavef}.
	}
	\label{fig:wavef_6.81}
\end{figure}

\begin{figure}[p!]
	\centering
	\includegraphics{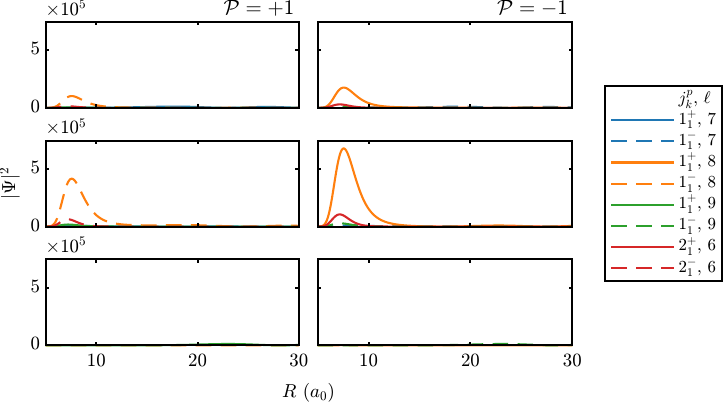}
	\caption{Calculated wave function for ND$_3$-H$_2$ scattering with $\mathcal{J}=8$ at $E=\SI{7.87}{\per\cm}$, the position of a resonance. As in \cref{fig:wavef}.
	}
	\label{fig:wavef_7.87}
\end{figure}

\begin{figure}[p!]
	\centering
	\includegraphics{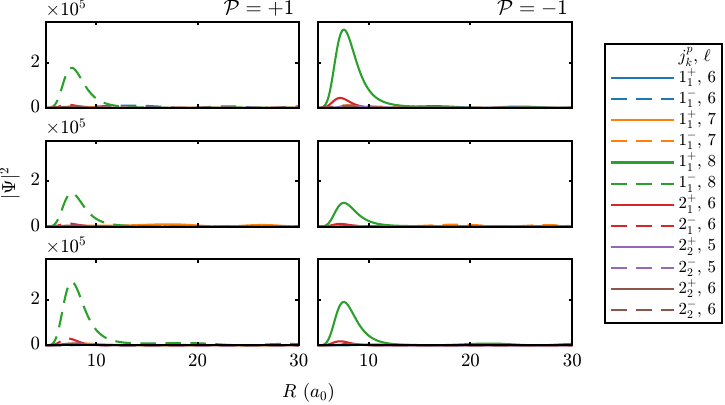}
	\caption{Calculated wave function for ND$_3$-H$_2$ scattering with $\mathcal{J}=7$ at $E=\SI{7.97}{\per\cm}$, the position of a resonance. As in \cref{fig:wavef}.
	}
	\label{fig:wavef_7.97}
\end{figure}

\begin{figure}[p!]
	\centering
	\includegraphics{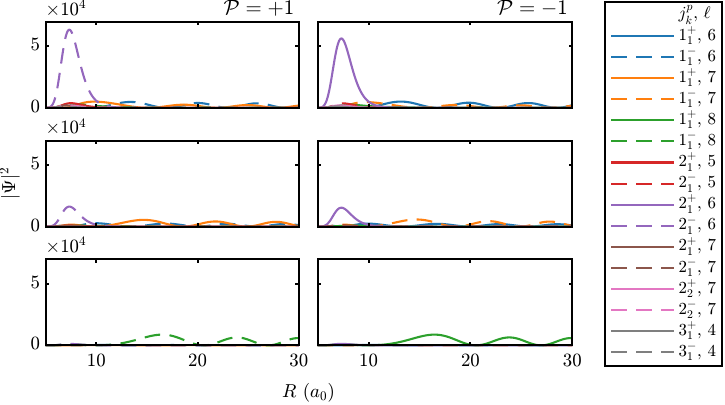}
	\caption{Calculated wave function for ND$_3$-H$_2$ scattering with $\mathcal{J}=7$ at $E=\SI{12.10}{\per\cm}$, the position of a resonance. As in \cref{fig:wavef}.
	}
	\label{fig:wavef_12.10}
\end{figure}

\begin{figure}[p!]
	\centering
	\includegraphics{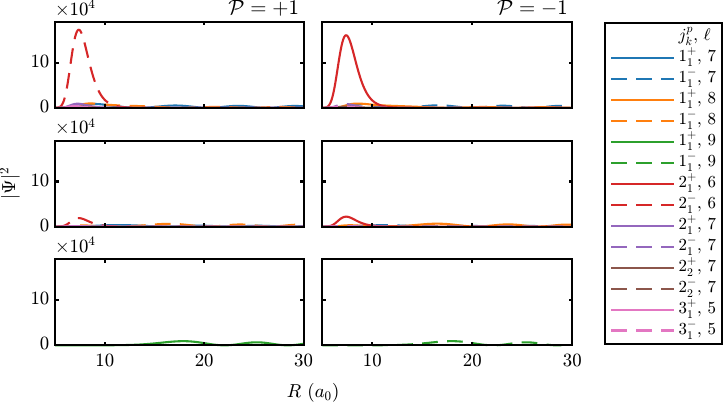}
	\caption{Calculated wave function for ND$_3$-H$_2$ scattering with $\mathcal{J}=8$ at $E=\SI{12.60}{\per\cm}$, the position of a resonance. As in \cref{fig:wavef}.
	}
	\label{fig:wavef_12.60}
\end{figure}

\begin{figure}[p!]
	\centering
	\includegraphics{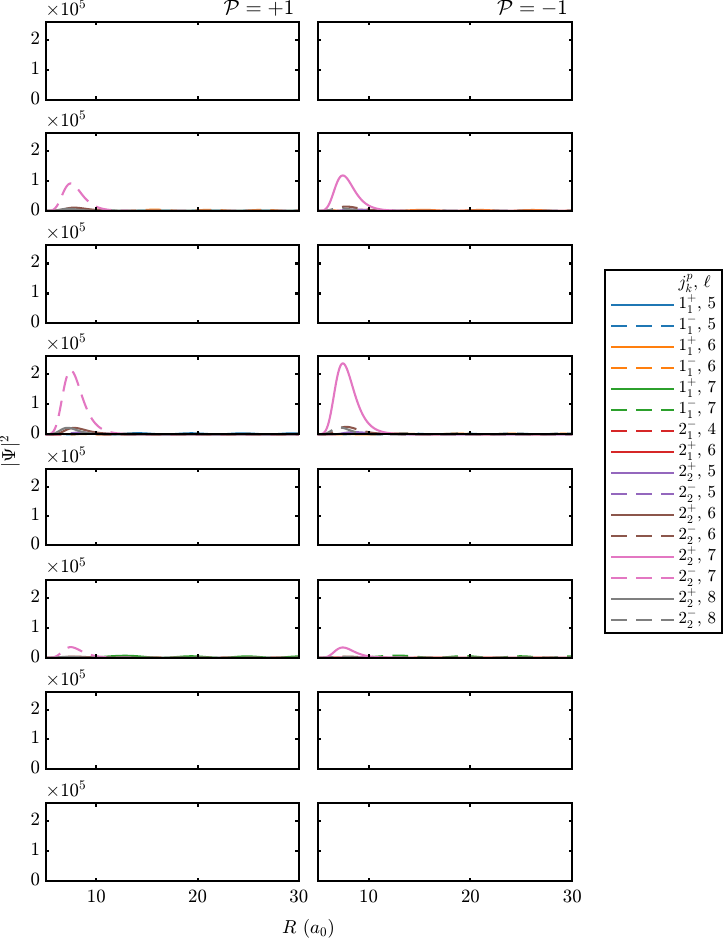}
	\caption{Calculated wave function for ND$_3$-H$_2$ scattering with $\mathcal{J}=6$ at $E=\SI{14.47}{\per\cm}$, the position of a resonance. As in \cref{fig:wavef}. There are 8 scattering states (top to bottom) for each parity $\mathcal{P}$, one for each final state ($1_1^p,\ell\in\{5,6,7\}$ and $2_2^p,\ell\in\{4,5,6,7,8\}$), as discussed in the text. 
	}
	\label{fig:wavef_14.47}
\end{figure}

By following the same reasoning, \cref{tab:res_ND3_H2} could be completed, fully characterizing the ten most prominent resonances featuring in the ICS for ND$_3$-H$_2$ scattering. Only the two resonances at $E=\num{7.87}$ and \SI{7.97}{\per\cm} can be characterized as Shape resonances. The resonance at $E=\SI{14.47}{\per\cm}$ can be described as a combined Feshbach-shape resonance, because the $2_2^p$ channel opens just below this collision energy. All other resonances are Feshbach resonances. The remaining wavefunctions are shown in \cref{fig:wavef_1.75,fig:wavef_5.46,fig:wavef_6.15,fig:wavef_6.81,fig:wavef_7.87,fig:wavef_7.97,fig:wavef_12.10,fig:wavef_12.60,fig:wavef_14.47}.

\newpage
\renewcommand\refname{Supplementary references}
\bibliography{bibliography-ND3H2}
\bibliographystyle{naturemag}